\documentclass[aps,twocolumn,amsfonts,amssymb,amsmath,nofootinbib,preprintnumbers,eqsecnum,superscriptaddress]{revtex4-1}

\usepackage{epsfig}
\usepackage[latin1]{inputenc}
\usepackage{float}
\usepackage{graphicx}
\usepackage{cancel}
\usepackage{mathrsfs}
\usepackage{amssymb}
\usepackage{amsfonts}
\usepackage{amsmath}
\usepackage{slashed}
\usepackage{hyperref}
\usepackage[usenames,dvipsnames]{color}

\newcommand{\be}{\begin{equation}}
\newcommand{\ee}{\end{equation}}
\newcommand{\beq}{\begin{equation}}
\newcommand{\eeq}{\end{equation}}
\renewcommand\r[1]{(\ref{#1})}
\def\({\left(}  \def\){\right)}  \def\[{\left[}  \def\]{\right]}  \def\<{\langle}  \def\>{\rangle} 
\newcommand{\bew}{\begin{widetext}}
\newcommand{\ew}{\end{widetext}}
\newcommand\sW{{\mathcal W}}
\newcommand\sR{{\ensuremath{{\mathcal R}}}}
\newcommand\vev[1]{{\ensuremath{\left\langle{#1}\right\rangle}}}
\def\Tr{\mathop{\rm Tr}}

\begin{document}

\title{Momentum Broadening in Weakly Coupled Quark-Gluon Plasma\\
(with a view to finding the quasiparticles within liquid quark-gluon plasma)}

\author{Francesco D'Eramo}
\affiliation{Center for Theoretical Physics, Massachusetts Institute of Technology, Cambridge, MA, USA 02139}
\affiliation{Center for Theoretical Physics, Department of Physics, University of California, Berkeley, CA 94720}
\affiliation{Theoretical Physics Group, Lawrence Berkeley National Laboratory, Berkeley, CA 94720}

\author{Mindaugas Lekaveckas}
\affiliation{Center for Theoretical Physics, Massachusetts Institute of Technology, Cambridge, MA, USA 02139}
\author{Hong Liu}
\affiliation{Center for Theoretical Physics, Massachusetts Institute of Technology, Cambridge, MA, USA 02139}
\author{Krishna Rajagopal}
\affiliation{Center for Theoretical Physics, Massachusetts Institute of Technology, Cambridge, MA, USA 02139}
\affiliation{Physics Department, Theory Unit, CERN, CH-1211 Gen\`eve 23, Switzerland}


\preprint{MIT-CTP 4400}
\preprint{UCB-PTH-12/14}
\preprint{CERN-PH-TH/2012-303}

\begin{abstract}

We calculate $P(k_\perp)$, the probability distribution for an energetic parton 
that propagates for a distance $L$  through 
a medium without radiating to pick up transverse momentum $k_\perp$, for a medium consisting of weakly coupled quark-gluon plasma. We use full or HTL self-energies in appropriate regimes, resumming each in order to find the leading large-$L$ behavior.  
The jet quenching parameter $\hat q$ is the second moment of $P(k_\perp)$, and
we compare our results to other determinations of this quantity in the literature,
although we emphasize the importance of looking at $P(k_\perp)$ in its entirety.
We compare our results for $P(k_\perp)$ in weakly coupled quark-gluon plasma
to expectations from holographic calculations that
assume a plasma that is strongly coupled at all length scales.
We find that the shape of $P(k_\perp)$ at modest $k_\perp$ may not be 
very different in weakly coupled and strongly coupled plasmas,
but we find that $P(k_\perp)$ must be parametrically {\it larger} in a weakly coupled plasma than in a strongly coupled plasma --- at large enough $k_\perp$.  This
means that by looking for rare (but not exponentially rare) 
large-angle deflections of the jet resulting from a parton produced initially back-to-back with a hard photon, experimentalists can find the weakly coupled 
short-distance quark and gluon 
quasiparticles within the strongly coupled liquid quark-gluon plasma produced
in heavy ion collisions, much as Rutherford found nuclei within atoms or Friedman, Kendall and Taylor found quarks within nucleons.
\end{abstract}

\date{December 17, 2012}

\maketitle

\section{Introduction}

The droplets of quark-gluon plasma (QGP) produced in heavy-ion collisions can be studied 
via analyzing the ``shape'' (in various senses including almost literally) of the
explosion of hadrons it decays into and via analyzing their effects on
various internally generated ``probes''.   High energy partons, produced
in hard parton-parton scattering events occurring at the earliest moments
within a heavy ion collision, are particularly useful probes.
After their production they will propagate through as much as
$5-10 \, {\rm fm}$ of the hot and dense medium produced in the collision. 
In proton-proton collisions, the production rates and decay products (jets) of
such high energy partons are well measured and well understood, meaning
that in heavy ion collisions they constitute
well calibrated probes of the plasma.
The suite of observables that indicate the ways in which
high energy partons interact with the plasma, for example losing energy
to it, are collectively known as ``jet quenching''.
The 
suppression of the production rates for high-$p_T$ hadrons
in heavy ion collisions at RHIC
with respect to expectations based on scaling with the number of binary
nucleon-nucleon collisions was the earliest manifestation of
jet quenching to be discovered~\cite{Arsene:2003yk}.
Jet quenching manifests itself in many observables, which contain clues about how the fragmentation of a high energy parton is affected by the presence of the plasma and how the medium responds to the energy and momentum that a fragmenting parton transfers to it. 
Jet quenching has by now been seen in many ways, both
at RHIC and at the LHC.
At the LHC, jet energies are high enough that the jets can be detected calorimetrically event-by-event, and the phenomenon of jet quenching is manifest in single events with, say, a jet with an energy greater than 200~GeV back-to-back with a jet with an energy less than 100~GeV~\cite{Aad:2010bu,Aamodt:2010jd,Chatrchyan:2011sx,CMS:2012aa,Chatrchyan:2012nia,Chatrchyan:2012gt,Chatrchyan:2012gw,Collaboration:2012is}. 
(It is improbable that a pair of jets will be produced such that each travels the same distance through the plasma and loses the same amount of energy, so back-to-back pairs of jets with unbalanced energies are the norm.)   
The first measurements of heavy ion collisions at the LHC 
in which a hard parton (manifest in the final state as a jet) 
was produced back-to-back with a single hard photon have recently been 
reported~\cite{Chatrchyan:2012gt,QMPlenaryTalksOnGammaJetsatLHC},
and early results on energetic hadrons back-to-back with a single hard photon
in heavy ion collisions at RHIC have also been reported~\cite{QMTalkWithGammaHadrons}. 
Since the photon tells us 
the initial transverse momentum and direction of the parton that produced the jet,
analyzing sufficiently
large data sets of such events
can tell us how the plasma produced
in heavy ion collisions affects the energy, fragmentation, and direction of
hard partons plowing through it.

Theoretical analyses of how the energy and momenta of
hard partons are modified by passage through
weakly coupled quark-gluon plasma
have been developed by many authors~\cite{Gyulassy:1993hr,Baier:1996kr,Baier:1996sk,Zakharov:1997uu,Wiedemann:2000za,Gyulassy:2000fs,Gyulassy:2000er,Guo:2000nz,Wang:2001ifa,Arnold:2001ba,Salgado:2003gb,Majumder:2007hx} and are reviewed in Refs.~\cite{Baier:2000mf,Kovner:2003zj,Gyulassy:2003mc,Jacobs:2004qv,CasalderreySolana:2007zz,Accardi:2009qv,Wiedemann:2009sh,Majumder:2010qh}. 
In the limit of high parton energy, parton energy loss occurs dominantly
via the radiation of nearly collinear gluons, an effect that is distinct
from the changes in the direction of the momentum of the hard parton
via (repeated, soft) elastic collisions.  
The latter effect is often called
``transverse momentum broadening'', where the word ``transverse'' here and throughout the remainder of this paper means 
perpendicular to the original direction of the energetic parton, not perpendicular
to the beam direction. Transverse momentum broadening describes the accumulation
of changes to the direction of the momentum of the hard parton as it
propagates through a medium.
Transverse momentum broadening plays a central role in all the calculations of radiative energy loss, since the incoming and outgoing partons and the radiated gluons are all continually being jostled by the medium in which they find themselves. 

Analysis of an energetic parton propagating through a medium immediately involves
(at least) two well-separated energy scales,
$Q \gg T$, where $Q$ is the energy of the hard parton and $T$ 
refers to any of the soft scales that characterize the medium itself.  In the
case of weakly coupled quark-gluon plasma, ``$T$'' could refer to the temperature
$T$ itself or to even softer scales like $gT$ or $g^2T$, with $g$ the QCD coupling.
This separation of scales suggests the use of 
Effective Field Theory (EFT),
which takes advantage of a separation of energy scales to simplify
calculations and enhance predictive power by identifying
relevant degrees of freedom, simplifications, or new symmetries that
appear in the limit of large scale separation.
Our long-term goal is to use EFT techniques to develop
controlled theoretical calculations of the propagation of energetic
particles through hot and dense media.
The hierarchy $Q \gg T$ makes Soft Collinear Effective Theory (SCET)~\cite{Bauer:2000ew}, 
the natural EFT with which to describe hard partons in a medium.
(SCET was introduced in other contexts in
which both energetic (collinear) partons and soft degrees of freedom are relevant.) 
The work of Ref.~\cite{Idilbi:2008vm},  which builds upon
the earlier analysis of transverse momentum
broadening in Ref.~\cite{Majumder:2007hx}, was the first 
use of SCET to analyze partons in medium.  These authors looked at the transverse
momentum broadening of partons produced in deep inelastic scattering on large
nuclei.  In particular, they discovered that the transverse momentum broadening
of an energetic parton is induced by its interactions with the
gluons from the medium in a particular kinematic regime, known as
 ``Glauber gluons'', that were not present in the original formulation of SCET.
Upon choosing coordinate axes
such that
the three-momentum of the energetic parton is initially along the negative $z$-axis,
meaning that its initial four-momentum  is
\be
q_0 \equiv  \left(q_0^+, q_0^-, q_{0\,\perp}\right) = \left(0, Q, 0\right) \ .
\ee
(in light-cone coordinates defined by $q^{\pm} = \frac{1}{\sqrt{2}} (q^0 \pm q^3)$)
Glauber gluons are gluons whose momenta are parametrically of 
order~\cite{D'Eramo:2010ak}
\be
\left(\frac{T^2}{Q},\frac{T^2}{Q},T\right)\ .
\ee
After absorbing or emitting any number of Glauber gluons, the momentum of the energetic parton 
is parametrically of order $(T^2/Q,Q,T)$, meaning that it is off-shell only by
of order $T^2$ and so does not radiate either collinear
or soft gluons~\cite{D'Eramo:2010ak}.  
And yet, repeated absorption and
emission of Glauber gluons continually kicks the hard parton and can result
in significant transverse momentum broadening.
In Ref.~\cite{D'Eramo:2010ak}, three of us 
developed a SCET formulation
for the calculation of transverse momentum broadening
in any medium, accounting for arbitrarily many interactions between
the energetic parton and the Glauber modes from the medium.
SCET has also been used to study momentum broadening and collinear gluon radiation 
to first order in opacity in Ref.~\cite{Ovanesyan:2011xy}.
In this paper, we evaluate the general result of~\cite{D'Eramo:2010ak} for the 
specific case of a weakly coupled quark-gluon plasma, in QCD, in thermal equilibrium.

Transverse momentum broadening is described by $P(k_\perp)$, the probability
density that after propagating through the medium for a distance $L$
without radiating, the energetic parton has acquired momentum
transverse to its initial direction given by $k_\perp$. The probability
distribution is normalized as
\be
\int \frac{d^2 k_{\perp}}{(2\pi)^2} P (k_{\perp}) = 1\ .
\label{eq:ProbabilityNormalization}
\ee
$P(k_\perp)$ depends on the medium length $L$, but we will keep this dependence implicit
in our notation. 
$P(k_\perp)$ is obtained by summing over infinitely many Feynman diagrams, accounting for the interactions between the energetic parton with any number of Glauber gluons from the medium. 
In Ref.~\cite{D'Eramo:2010ak} this calculation was performed by treating the Glauber
gluons as background fields, analyzing the propagation of the energetic parton in the presence
of any background field configuration, and then averaging the result over all 
possible background field configurations.

The final result for the differential probability distribution reads~\cite{D'Eramo:2010ak}
\be
P(k_\perp) = \int d^2 x_{\perp} \, e^{-i k_{\perp} \cdot x_{\perp}}\,
 \mathcal{W}_{\mathcal{R}} (x_\perp) \ ,
\label{eq:provb}
\ee
where $\mathcal{R}$ is the $SU(N_c)$ representation to which the energetic particle belongs 
and $ \mathcal{W}_{\mathcal{R}} (x_\perp)$ is the expectation value of two light-like Wilson lines 
with spatial extent $L$ (and therefore length $L^-\equiv \sqrt{2}L$ along the light cone)
in representation $\mathcal{R}$ separated from each
other in the transverse plane by the 
vector $x_\perp$.\footnote{The expression  (\ref{eq:provb}) 
is gauge invariant only after the ends of the lightlike Wilson lines
are closed with transverse segments, completing a Wilson loop.
However, in the limit $L\gg 1/T$ the contribution of these transverse
segments is subleading in any covariant gauge, meaning
that the gauge invariant result can be obtained by evaluating (\ref{eq:provb}), which
includes only the two long lightlike Wilson lines, in 
any covariant gauge~\cite{D'Eramo:2010ak}.
In lightcone gauge, the expectation value of the lightlike lines
vanishes and the same result is obtained entirely from the
transverse segments~\cite{D'Eramo:2010ak,Benzke:2012sz}.
In the present paper, we shall work in a covariant gauge,
obtaining the gauge invariant result directly from (\ref{eq:provb}).}  The explicit definition of $ \mathcal{W}_{\mathcal{R}} (x_\perp)$ is given at the beginning of 
Sec.~\ref{sec:resumming}.
The expression (\ref{eq:provb}) was obtained previously using different methods
in Refs.~\cite{CasalderreySolana:2007zz,Liang:2008vz}.
The nature of the medium --- for example whether it is weakly coupled or strongly
coupled --- does not enter in the analysis of the propagation in any one background
field configuration, and therefore does not affect the 
expression (\ref{eq:provb})~\cite{D'Eramo:2010ak}.  
This distinction, or indeed any property of the medium, only becomes
relevant when one averages over all possible background field configurations,
which is to say when one {\it evaluates} the expectation value 
$ \mathcal{W}_{\mathcal{R}} (x_\perp)$.
If the medium of interest is in thermal equilibrium, the expectation
value $\mathcal{W}_{\mathcal{R}}(x_\perp)$ is a thermal average
that can be evaluated in equilibrium thermal field theory.
If the medium of interest is not in equilibrium, the expectation value
$\mathcal{W}_{\mathcal{R}}(x_\perp)$ is much harder to evaluate
but the expression (\ref{eq:provb}) remains correct.
The result (\ref{eq:provb}) is valid only in the 
high energy limit for the propagating parton. Specifically, it requires that
$Q \gg k_\perp^2 L$, as this is the criterion that
ensures that the trajectory of the hard parton
in position space
remains well-approximated as a straight line, even as the parton 
picks up transverse momentum $k_\perp$. 
In this limit, $P(k_\perp)$ is given by (\ref{eq:provb}) and is
therefore independent of the energy of the hard probe, depending only on the properties of the medium through the thermal expectation value $\mathcal{W}_{\mathcal{R}} (x_\perp)$. 
Transverse momentum broadening without radiation thus ``measures'' a field-theoretically well-defined property of the medium. 

In Ref.~\cite{D'Eramo:2010ak},
an explicit evaluation of 
the thermal average 
$\mathcal{W}_{\mathcal{R}} (x_\perp)$ and
hence $P(k_\perp)$ was provided only for the plasma of
${\cal N}=4$ supersymmetric Yang-Mills (SYM) theory in the large number
of colors ($N_c$) and strong coupling limit, in which this strongly coupled plasma has
a holographic description, allowing the calculation to be done via
gauge/gravity duality.\footnote{The result so obtained~\cite{D'Eramo:2010ak} confirms one obtained
first in Ref.~\cite{Liu:2006ug}, and eliminates certain subtleties in
the earlier derivation.}
In this plasma (and,  
quite likely in any plasma in the strong
coupling limit)
$P(k_\perp)$ is Gaussian in $k_\perp$ and transverse momentum broadening
can be understood as diffusion in $k_\perp$-space due to repeated, even
continuous, interactions
between the hard parton and the strongly coupled medium.

In this work, we shall consider a  QCD plasma in equilibrium at high enough temperatures
that physics at scales $\sim T$ is weakly coupled.  We shall evaluate
the thermal average 
$\mathcal{W}_{\mathcal{R}} (x_\perp)$ 
perturbatively, using standard methods from thermal field theory.
From this we then obtain $P(k_\perp)$ and hence  momentum broadening
in a weakly coupled quark-gluon plasma by applying (\ref{eq:provb}).
We shall present this calculation over the course of Sections~\ref{sec:resumming},
\ref{sec:propagator}
and \ref{sec:magnetic}. 
We set up  the general formalism and identify the leading order contribution in the 
gauge coupling $g$ (assumed $\ll1$) to the expression in~(\ref{eq:provb}).  
In doing so, we resum an infinite class of
diagrams which are enhanced by the medium length $L$.  For a thick medium, the resummation alters $P(k_\perp)$ at small $k_\perp$, as we will show explicitly. 
Once we have set up the formalism and identified the expression that we need
to evaluate, we find that $P(k_\perp)$
depends on the  retarded gluon propagator. In Sec.~\ref{sec:propagator},
we show how this propagator can be expressed in terms of self-energies,
which we compute.
We compute  the self-energies first using ordinary perturbation theory, for any value of the external momentum. Since our goal is to compute the probability distribution in~(\ref{eq:provb}), which manifestly depends on a gluon correlator in coordinate space, we need the gluon propagator for any value of the external momentum. Famously,  perturbation theory for non-Abelian gauge theories at finite temperature breaks down in the infrared~\cite{Linde:1980ts,Kalashnikov:1980tk}. In the case of the gluon propagator this happens when the external momentum is of order $g^2 T$, and we recover this pathology from our expression. We take care of the infrared problem by using the hard thermal loop (HTL) effective theory~\cite{Braaten:1989kk}, which
is valid for momenta of order $gT$ and below and which restores a consistent
perturbative expansion.
We work in the weak-coupling regime, where the hierarchy $g^2 T \ll g T$ guarantees that the infrared problem does not show up in our calculation. In Sec.~\ref{sec:magnetic} we discuss this in detail, and we explain how we use full or HTL retarded self-energies in the appropriate regimes, as well as how we match them at an intermediate scale. 
The reader only interested in our results, not in their derivation, 
can jump directly to Sec.~\ref{sec:results}. 
There we present our results, compare them to results at strong 
coupling~\cite{Liu:2006ug,D'Eramo:2010ak},
and compare them to other weak-coupling results in the literature
for $P(k_\perp)$ and its second moment, which is called the jet quenching
parameter $\hat q$~\cite{Majumder:2007hx,Arnold:2008zu,Arnold:2008vd,CaronHuot:2008ni,Majumder:2009cf}.

The most important qualitative feature of our result is that $P(k_\perp)\propto 1/k_\perp^4$
in the large $k_\perp$ regime.  Since this is parametrically larger than a Gaussian
at large $k_\perp$ this means that at sufficiently large $k_\perp$ $P(k_\perp)$ is greater
in a weakly coupled plasma than in a strongly coupled one.
We close Sec.~\ref{sec:results} by exploring this comparison, 
semi-quantitatively.
In Sec.~\ref{sec:conclusions} we conclude and look ahead.
We do not expect that our result for weakly coupled quark-gluon plasma
describes $P(k_\perp)$ for quark-gluon plasma produced in heavy ion
collisions correctly at all $k_\perp$.  Heavy ion collisions do not reach
the asymptotically high  temperatures at which $g\ll 1$ at scales of
order the temperature.  Instead, there are many indications
that the plasma produced in heavy ion collisions at both RHIC
and LHC is a strongly coupled liquid, with no well-defined quasiparticles
(i.e. no quasiparticles with mean free paths long compared to $1/T$).  
At small $k_\perp$, we therefore expect that its $P(k_\perp)$ is more
similar to that in the strongly coupled ${\cal N}=4$ SYM plasma~\cite{Liu:2006ug,D'Eramo:2010ak}
than to that we calculate in this paper.
However, QCD is asymptotically free meaning that at short enough distance scales
the strongly coupled liquid must be described by weakly interacting quark
and gluon quasiparticles.  
We do not incorporate the running of $g$ with $k_\perp$ in our calculation. Nevertheless we expect that, because $g$ {\it does} run, $P(k_\perp)$ for the strongly coupled liquid produced in heavy ion collisions {\it is} 
well described at large enough $k_\perp$ by the result of our weakly coupled calculation
of $P(k_\perp)$. This means that if $P(k_\perp)$ can be measured over a sufficiently
wide range of $k_\perp$ it could yield insights into how quark-gluon plasma
with liquid-like properties at length scales of order $1/T$ 
emerges from a weakly coupled gauge theory at short distances.

\section{Setting up the formalism}
\label{sec:resumming}

\begin{figure}[t]
\begin{center}
\includegraphics[scale=0.43]{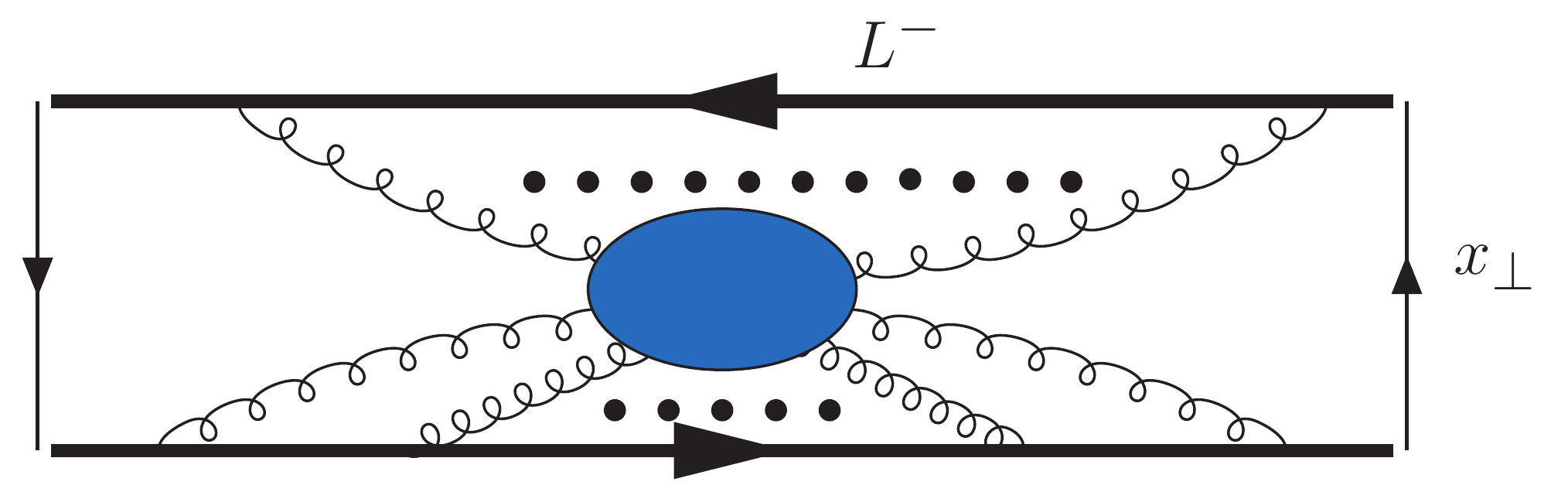}
\end{center}
\caption{Diagrams contributing to the expectation value of the Wilson loop, $ \mathcal{W}_{\mathcal{R}} (x_\perp)$. The length of each 
of the long light-like sides of the loop is $L^- = \sqrt{2} L$.
The light-like Wilson lines are separated in the transverse direction
by a distance $x_\perp$.}
\label{fig:Wloop}
\end{figure}

In this section, we derive an expression valid to leading order in the QCD coupling
constant, and hence in a weakly coupled QCD plasma, relating
 the general result for $P(k_\perp)$,  derived in Ref.~\cite{D'Eramo:2010ak} 
 and given in (\ref{eq:provb}), 
 to the retarded gluon propagator.
We see in (\ref{eq:provb}) that the probability distribution $P(k_\perp)$ that
describes momentum broadening is the Fourier transform of  ${\cal W}_{\cal R}(x_\perp)$,
the expectation value of  two light-like Wilson lines in representation ${\cal R}$ separated
from each other by a distance $x_\perp$ in the transverse plane. We begin by defining
  ${\cal W}_{\cal R}(x_\perp)$ explicitly:
\be \label{Wils}
\sW_{\sR} (x_\perp)  \equiv \frac{1}{d\left(\mathcal{R}\right)} \vev{\Tr \left[ W_{\sR}^\dagger [0,x_\perp] \, W_{\sR} [0,0] \right]} \ ,
\ee 
where each Wilson line along the light-cone is in turn defined by
\be
W_{\sR}\left[y^+, y_{\perp}\right] \equiv P \left\{ \exp \left[i g \int_{0}^{L^-} d y^{-}\, A_{\sR}^+ (y^+, y^-,y_{\perp})\right] \right\} \ .
\ee
Next, we Taylor expand~(\ref{Wils}), and in doing so we define
\be
\mathcal{W}_{\mathcal{R}}(x_\perp)
\equiv  1 \, + \, \sum_{j=2}^{\infty} \mathcal{W}^{(j)}_{\mathcal{R}}(x_\perp) =  1 +  \mathcal{W}^{(2)}_{\mathcal{R}} + \mathcal{W}^{(3)}_{\mathcal{R}}  + \ldots \ ,
\label{eq:Wseries}
\ee
where $\mathcal{W}^{(j)}_{\mathcal{R}}$ denotes the contribution 
in which $j$ gluon fields are evaluated at points on the light-like Wilson lines
or, equivalently, the contribution from those diagrams (see Fig.~\ref{fig:Wloop}) 
in which gluon propagators
end at $j$ vertices on the light-like Wilson lines.
We have dropped the explicit denotation of the $x_\perp$-dependence on the
right-hand side of (\ref{eq:Wseries}), and will do so at many points below.
Each term on the right-hand side of (\ref{eq:Wseries}) is itself a series,
\be
 \mathcal{W}^{(j)}_{\mathcal{R}}
 = \sum_{k=0}^j W_{\mathcal{R}}^{(k, j-k)}
 \label{eq:DoubleSeries}
 \ee
 where the contribution from a diagram with $k$ gluon vertices on the light-like
 line at the perpendicular position $x_\perp$ and $j-k $
 gluon vertices on the light-like line at the origin of the perpendicular plane is given by 
\begin{widetext}
 \be
 W_{\mathcal{R}}^{(k,j-k)}
 =\frac{(-i)^k \,i^{j-k} \, g^j}{d(\mathcal{R})\, k!\, (j-k)!}
 \int_0^{L^-} \hspace{-0.1in}
 dy_1^- \ldots 
 dy_j^-
\left \langle {\rm Tr} \left[ \mathcal{P} \left\{ 
 A_{\mathcal{R}}^+(y_1^-,x_\perp) \ldots 
 A_{\mathcal{R}}^+(y_k^-,x_\perp)
 A_{\mathcal{R}}^+(y_{k+1}^-,0_\perp) \ldots 
 A_{\mathcal{R}}^+(y_j^-,0_\perp)
 \right\} \right] \right\rangle\,.
 \label{eq:ExplicitDefn}
 \ee
 \end{widetext}
Here, $g$ is the $SU(N_c)$ gauge coupling constant, $\mathcal{P}$ stands
for path ordering of the gluon fields and the trace is taken over  $SU(N_c)$
color indices.
We need not specify the $y^+$ coordinates at which
the gluon fields in (\ref{eq:ExplicitDefn}) are evaluated because they
are evaluated on the light-cone described by varying $y^-$ at fixed $y^+$, and we can 
use the translational
invariance of the medium to set all of the $y^+$ coordinates to $0$. 
The gluon fields in (\ref{eq:ExplicitDefn}) can each
be written as the product of an operator and a group
matrix: $A_{\mathcal{R}}^+=A^{a+}t_{\mathcal{R}}^a$.
Note, finally, that since the expectation value of a single gluon field vanishes, there
is no $j=1$ contribution in the expansion (\ref{eq:Wseries}).
We have now specified the $\mathcal{W}_{\mathcal{R}}$ of (\ref{Wils})
fully explicitly.
The probability distribution $P(k_\perp)$
that describes momentum broadening 
is then given by the Fourier transform of $\mathcal{W}_{\mathcal{R}}$, 
as in~(\ref{eq:provb}).

Both the gluon operators $A^{a+}$ and the group matrices $t^a_{\mathcal{R}}$ within $\mathcal{W}_{\mathcal{R}} $ are ordered along the path as indicated by arrows in Fig.~\ref{fig:Wloop}, in contrast with the time ordered operators in a conventional Wilson loop~\cite{D'Eramo:2010ak}. 
Hence, the expectation value $\mathcal{W}_{\mathcal{R}}$ should be described by a Schwinger-Keldysh contour 
with one light-like Wilson line on the ${\rm Im} \, t = 0$ segment of the contour and the other one on the ${\rm Im} \, t = - i  \epsilon$ segment. The infinitesimal displacement in imaginary time ensures that the operators from the two lines are ordered such that all operators from one line come before any operators from the other. The Schwinger-Keldysh formalism relevant to our calculation is reviewed in Appendix~\ref{app:RTF}. A typical diagram contributing to $\mathcal{W}_{\mathcal{R}} (x_\perp)$ is shown in Fig.~\ref{fig:Wloop}.

\begin{figure*}[t]
 \begin{center}
\includegraphics[scale=0.6]{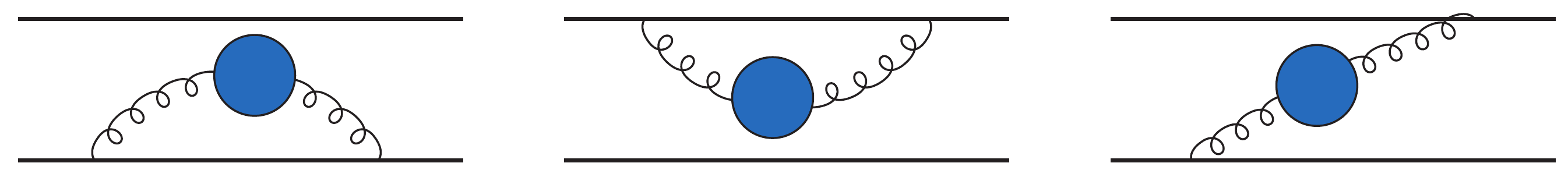}
\end{center}
\caption{Diagrams contributing to $\mathcal{W}^{(2)}_{\mathcal{R}}$. Here, the blue blobs stand for full interacting gluon two-point  Green functions.}
\label{fig:W2diag}
\end{figure*}

We are looking for the leading order contribution to $\mathcal{W}_{\mathcal{R}}$, leading 
in the $g \ll 1$ limit. The first non-trivial and non-vanishing contribution appears for $j=2$, and 
is given by the diagrams in Fig.~\ref{fig:W2diag}. Upon writing 
each gluon field as 
$A_{\mathcal{R}}^+ =  A^{a \,+} \,  t_{\mathcal{R}}^a$, the propagator that arises in $\mathcal{W}^{(2)}_{\mathcal{R}}$ reads
\be
\begin{split}
& \left\langle {\rm Tr} \left[ A_{\mathcal{R}}^+ (y_1^-, y_{1 \perp}) A_{\mathcal{R}}^+ (y_2^-, y_{2 \perp}) \right] \right\rangle =  \\
& {\rm Tr} \left[ t^a_{\mathcal{R}} t^b_{\mathcal{R}} \right] \,  \left\langle A^{a\,+} (y_1^-, y_{1 \perp}) A^{b\,+} (y_2^-, y_{2 \perp})  \right\rangle  \ ,
\label{eq:gluonprop}
\end{split}
\ee
where $y_{1\perp}$ and $y_{2\perp}$ can each be either $0_\perp$ or $x_\perp$.
The expectation value 
on the right-hand side of (\ref{eq:gluonprop}) 
is diagonal in the color indices, making it convenient to define
\be
\begin{split}
& \left\langle A^{a\,+} (y_1^-, y_{1 \perp}) A^{b\,+} (y_2^-, y_{2 \perp})  \right\rangle \equiv  \\
&\qquad\qquad\qquad \delta^{ab} \, D^{>}(y_1^- - y_2^- , y_{1 \perp} - y_{2 \perp}) \ , 
\end{split}
\ee
where the Wightman propagator $D^{>}$ has no color 
indices.\footnote{It would be more precise to call the propagator $D^{>\,++}$, since we only have the $+$ component of the gluon fields. We omit the $++$ for notational simplicity. However, we keep the symbol $>$ to remind ourselves that this is the Wightman propagator. 
We shall later relate $D^>$ to the retarded propagator.} We can now write
\be
\begin{split}
& \left\langle {\rm Tr} \left[ A_{\mathcal{R}}^+ (y_1^-, y_{1 \perp}) A_{\mathcal{R}}^+ (y_2^-, y_{2 \perp}) \right] \right\rangle = \\ &
\qquad\qquad\qquad d(\mathcal{R}) \, C_{\mathcal{R}} \, D^{>}(y_1^- - y_2^- , y_{1 \perp} - y_{2 \perp}) \ ,
\end{split}
\ee
where we have identified the quadratic Casimir factor $C_{\mathcal{R}}$ for the $SU(N_c)$ representation $\mathcal{R}$ via the relation
\be
\delta^{ab} \, t^a_{\mathcal{R}} t^b_{\mathcal{R}} = C_{\mathcal{R}} \, \mathbf{I}_{\mathcal{R}} \ ,
\ee
where $\mathbf{I}_{\mathcal{R}}$ is the identity matrix for the representation $\mathcal{R}$. With
all these  definitions in place, $\mathcal{W}^{(2)}_{\mathcal{R}}$ now reads
\be
\begin{split}
\mathcal{W}^{(2)}_{\mathcal{R}} = - &\, g^2 \, C_{\mathcal{R}}  \int_{0}^{L^-} d y^{-}_1\int_{0}^{L^-} d y^{-}_2 \, \times \\ & 
\left[D^{>}(y^{-}_1 - y^{-}_2, 0_\perp) - D^{>}(y^{-}_1 - y^{-}_2, x_\perp) \right]  \ ,
\end{split}
\ee
where $L^- = \sqrt{2} L$. 
Finally, we perform a change of variables for the integrations over the $y^-$ coordinates,
defining
\be
Y^- \equiv \frac{y^{-}_1 + y^{-}_2 }{2}, \qquad \qquad y^{-} \equiv y^{-}_1 - y^{-}_2 \ .
\label{eq:com}
\ee
The integration over the ``center of mass'' coordinate $Y^-$ is straightforward, and we find
\be
\mathcal{W}^{(2)}_{\mathcal{R}} = - g^2 C_{\mathcal{R}} L^-  \int d y^{-} \left[D^{>}(y^{-}, 0_\perp) - D^{>}(y^{-}, x_\perp) \right]  \ .
\label{eq:W2}
\ee

In order to determine the leading gauge coupling dependence 
of $\mathcal{W}^{(2)}_{\mathcal{R}}$, we need to
determine how the gluon propagator depends on $g$. The tree level term
in the gluon propagator
is proportional to the metric tensor $g^{\mu\nu}$, and since $g^{++}=0$ this vanishes. The first non-vanishing contribution is found 
at next order in perturbation theory,
when we evaluate the one-loop propagator --- i.e. replacing the blue blobs
in Fig.~\ref{fig:W2diag} by single loops. 
It is straightforward to count the powers of $g$ when the external
momentum in the propagator (i.e. momentum in the
gluon lines shown explicitly in Fig.~\ref{fig:W2diag}) are greater than the 
scale $gT$. In this case, conventional perturbation theory is under control, the one-loop propagator 
is $\mathcal{O}(g^2)$ and the full expression in (\ref{eq:W2}) is $\mathcal{O}(g^4)$. 
For external momentum in the propagator
of the order $gT$ and lower, we shall see below that 
HTL resummation is necessary.  We shall complete the discussion
of the $g$-dependence of our results in Section~\ref{sec:results}.

Before we move on, we need to check that 
the $\mathcal{W}^{(j)}_{\mathcal{R}}$ for $j>2$ are suppressed
relative to $\mathcal{W}^{(2)}_{\mathcal{R}}$.
Each gluon field from the Wilson lines brings a factor of $g$ with it,
meaning that $\mathcal{W}^{(j)}_{\mathcal{R}}$ comes with a factor of 
$g^j$ from attaching $j$ gluons to the Wilson lines.
This, together with the fact that $\mathcal{W}^{(3)}_{\mathcal{R}}$ must
include a  three-gluon vertex that comes with another $g$, suggests
that all the $\mathcal{W}^{(j)}_{\mathcal{R}}$ for $j>2$ are suppressed relative
to $\mathcal{W}^{(2)}_{\mathcal{R}}$ by a factor of at least $g^2$.
The only way to avoid this conclusion would be if the tree-level
contribution to $\mathcal{W}^{(3)}_{\mathcal{R}}$ or 
$\mathcal{W}^{(4)}_{\mathcal{R}}$ were nonzero, since we saw that
$\mathcal{W}^{(2)}_{\mathcal{R}}$ vanishes at tree-level.  
However, because the three-point gluon vertex has the form $g^{\mu\nu} p^{\rho}$, where $p^{\rho}$ is one of the incoming gluon momenta, and
because $g^{++}=0$, the tree-level contribution to
$\mathcal{W}^{(3)}_{\mathcal{R}}$  vanishes.  It is also straightforward
to check that the tree-level contribution to 
$\mathcal{W}^{(4)}_{\mathcal{R}}$ vanishes.
We conclude that 
the contribution to the series in~(\ref{eq:Wseries}) that is leading order in powers
of $g$  is given 
by 
$\mathcal{W}^{(2)}_{\mathcal{R}}$, which is related to the gluon
propagator $D^>$ according to~(\ref{eq:W2}).

The formalism from Ref.~\cite{D'Eramo:2010ak} within which (\ref{eq:provb}) was derived
requires $LT\gg 1$.  If we were to require in addition that $g^2 C_{\mathcal{R}} L \, T  \ll 1$,
which is satisfied at weak enough coupling for any given $L$, we would have 
achieved our goal for this section having derived the relationship (\ref{eq:W2}).
However, we would prefer to have a result that is valid at large $L$ for any
given weak value of the coupling $g$.  
For this purpose the 
leading order contribution to $\mathcal{W}_{\mathcal{R}}$ given by
(\ref{eq:W2}) does not suffice because it is proportional to the length of the medium $L$,
meaning that
if we evaluate
 $P(k_\perp)$ by taking the Fourier transform of~(\ref{eq:W2}) as 
 prescribed in (\ref{eq:provb}), 
 we obtain a probability distribution that is proportional to 
 $L$.   This cannot be the correct result at large enough $L$ 
 for any fixed $g$.  In particular, we find that
 when
 $g^2 C_{\mathcal{R}} L \, T  \sim 1$ or greater, the perturbative expansion is not under control.
In Appendix~\ref{app:thinmed} we complete the discussion of 
a ``thin medium'', in which  
$g^2 C_{\mathcal{R}} L \, T  \ll 1$,
showing how in this circumstance
(\ref{eq:W2}) yields a correctly normalized probability distribution $P(k_\perp)$.
Here, we shall do better.

In order to find a perturbative expansion that is valid for any value of $L$ (and which
of course reduces to what we have already derived if $g^2 C_{\mathcal{R}} L \, T$ is $\ll 1$)
we need to consider the $L$-dependence of each diagram contributing 
in (\ref{eq:Wseries}). In a translation-invariant medium,
 each contribution to the series  (\ref{eq:Wseries}) is, at minimum, 
 proportional to $L$. This can easily  be seen by starting from the definition in~(\ref{eq:Wseries}) and changing the $y^-$ coordinates 
 to a new set including the center-of-mass coordinate $Y^-$, in analogy to what we have done for $\mathcal{W}^{(2)}_{\mathcal{R}}$ in (\ref{eq:com}). The gluon correlator cannot depend on the center of mass position, and therefore the integration along $Y^-$ is straightforward, 
 giving just a factor of $L$. If this were the complete story, we would (incorrectly) 
 conclude that power counting in $g$ and $L$ results in 
\be
\mathcal{W}^{(j)}_{\mathcal{R}} \propto g^{r_j} \, L \ , \qquad  r_j \geq j \ ,
\label{eq:Lwrong}
\ee
and
would further conclude that
 the previous perturbative expansion needs no modification. 
 In fact, the $L$ power counting in~(\ref{eq:Lwrong}) is incorrect
 because its ``derivation'' assumed that
 the diagram in Fig.~\ref{fig:Wloop} is connected. If this were the case, there 
 would be only one global translational invariance, and therefore one single center of mass integration. 
 Disconnected diagrams, however,
 always have additional translational invariances (corresponding to the
 freedom to translate disconnected pieces of the diagram independently)
that yield additional integrations over center-of-mass coordinates
that in turn result in the contribution of the diagram being
enhanced by additional powers of $L$.
Thus when $g^2 C_{\mathcal{R}} L \, T  \sim 1$ or greater we will need to resum 
a suitable set of disconnected diagrams, namely those whose contributions are the most
``length-enhanced''.
Disconnected diagrams can always be drawn for $j \geq 4$, 
 and the greatest number of translational invariances is reached
 in diagrams with the greatest number of disconnected pieces. 

\begin{figure}[t]
\begin{center}
\includegraphics[scale=0.45]{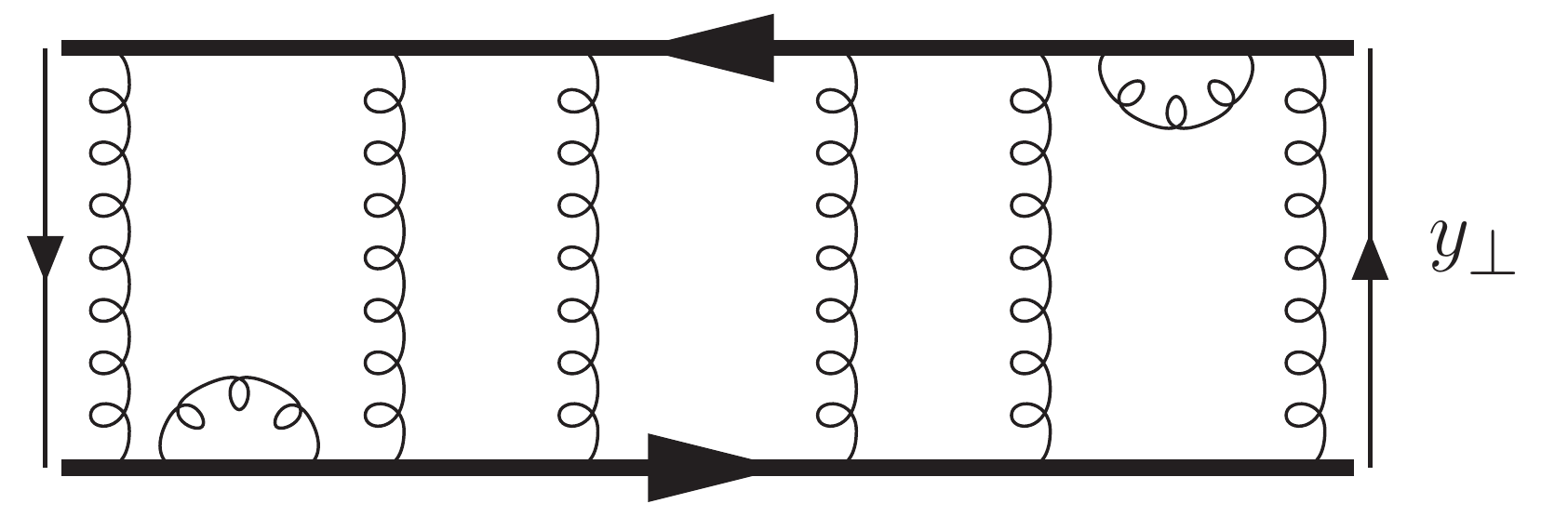}
\end{center}
\caption{A contraction of $n=j/2$ gluons, here with $n=8$, 
giving a contribution proportional to $L^n$. This
is  an example of a diagram whose
contribution is length-enhanced.}
\label{fig:WloopEN}
\end{figure}

\begin{figure}[t]
\begin{center}
\includegraphics[scale=0.45]{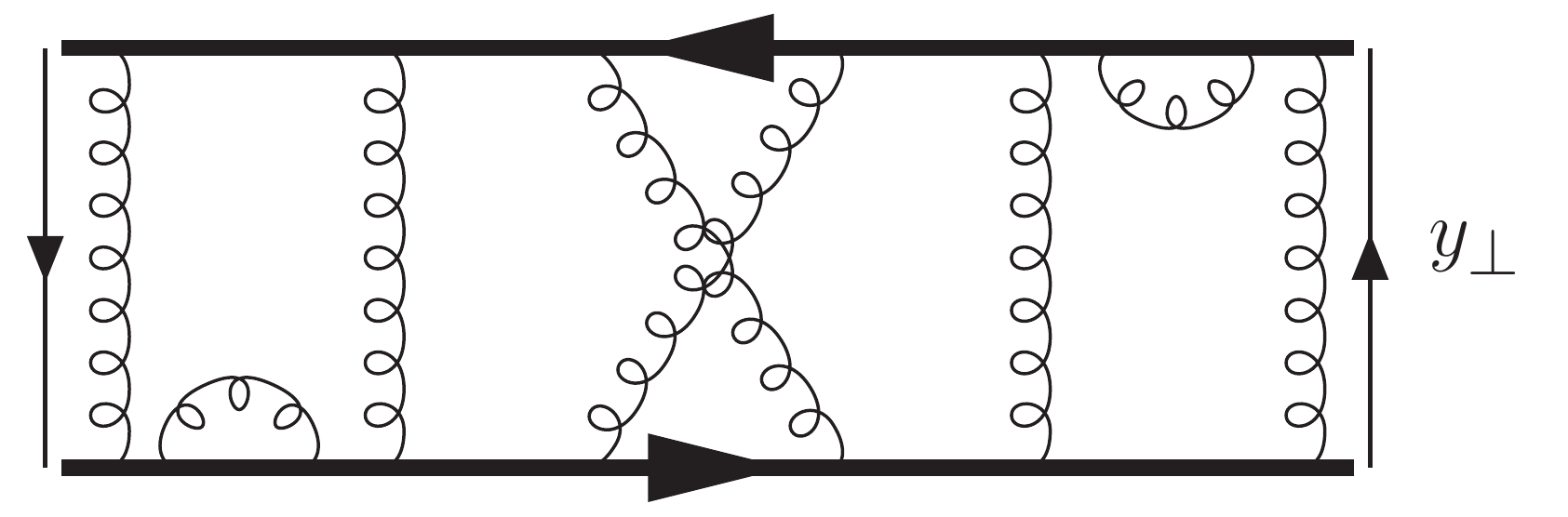}
\end{center}
\caption{A different contraction of $k=j/2=8$ gluons
giving a contribution which is proportional only to $L^7$, making it
less ``length-enhanced'' than that of Fig.~\ref{fig:WloopEN}. If 
we neglect powers of $g$ coming from within the propagators, 
the cross diagram in this figure gives a factor of $g^4 L$ which should be compared with the $g^4 L^2$ factor coming from two disconnected lines. This diagram and that of Fig.~\ref{fig:WloopEN} give contributions proportional to the
same power of $g$ but, among all such contributions, that 
from Fig.~\ref{fig:WloopEN} is one of those that comes with the
highest possible power of $L$ while that from this diagram is not.
So, the diagram in Fig.~\ref{fig:WloopEN} is one of the length-enhanced diagrams
that we resum while this diagram is not.}
\label{fig:WloopNO}
\end{figure}

From the cluster decomposition principle,  
we expect that when we include all disconnected diagrams, ${\cal W}_{\cal R} (x_\perp)$ can be  written in the exponentiated  form
\be \label{exPne}
{\cal W}_{\cal R} = \exp \left(\sum {\rm connected \; diagrams} \right) \ .
\ee
We emphasize that here connected diagrams include diagrams which could be disconnected were the gauge group Abelian. For example, consider the diagrams in Figs.~\ref{fig:WloopEN} and \ref{fig:WloopNO}.  The first is clearly disconnected. Naively,
the cross diagram within Fig.~\ref{fig:WloopNO} also 
appears disconnected in coordinate space. However,
the contractions in color space restrict the coordinate space integrations, and  the cross diagram 
should in fact be considered connected.
 Since all connected diagrams come with 
 precisely one factor of $L$, our earlier power counting of $g$ goes through, but now when
 applied to the exponent in~\eqref{exPne}. So, to leading order in weak coupling
 we now have
\be
P(k_\perp) = \int d^2 x_{\perp} \, e^{-i k_{\perp} \cdot x_{\perp}}\, \exp\left[ \mathcal{W}^{(2)}_{\mathcal{R}}  \right] \ .
\label{Pdef}
\ee
The proof of~\eqref{exPne}  is almost 
analogous to the textbook proof of the relation between the connected and disconnected diagrams in,  say, $\lambda \phi^4$ theory.  There are some additional complications involving
path ordering and group contractions.  We illustrate how these are resolved
in Appendix~\ref{app:exon}, by giving as an example a proof of the
exponentiation of $\mathcal{W}^{(2)}_{\mathcal{R}}$, namely (\ref{Pdef}).


We conclude that
whereas for
a thin medium in which $g^2 C_{\mathcal{R}} L \, T  \ll 1$
the leading contribution to $P(k_\perp)$ at weak coupling is given by 
(\ref{eq:provb}) with $W_{\mathcal{R}}$ replaced by just 
$\mathcal{W}^{(2)}_{\mathcal{R}}$,
if $g^2 C_{\mathcal{R}} L \, T$ is not small we must resum
all disconnected diagrams involving only $\mathcal{W}^{(2)}_{\mathcal{R}}$
as in Fig.~\ref{fig:WloopEN}, obtaining
\be
\exp\left[ \mathcal{W}^{(2)}_{\mathcal{R}}  \right]  = 1 \, + \, \sum_{n=1}^{\infty}  \frac{1}{n!} \,  \left(\mathcal{W}^{(2)}_{\mathcal{R}} \right)^n  \ 
\label{eq:W2en}
\ee
where $ \left(\mathcal{W}^{(2)}_{\mathcal{R}} \right)^n $ contains a ``length enhancement'' factor 
$L^n$.   By keeping only the leading order term in the exponent in (\ref{exPne}) we are resumming
those diagrams that have the highest possible power of $L$ for a given power of $g$.  This is
illustrated in Figs.~\ref{fig:WloopEN} and \ref{fig:WloopNO}.  The diagram in Fig.~\ref{fig:WloopEN}
is included in our resummation (\ref{Pdef}); it contributes to 
the $n=8$ term in the expansion (\ref{eq:W2en}).  The diagram in Fig.~\ref{fig:WloopNO}
arises in (\ref{exPne}) from a cross-term involving 6 powers of 
$\mathcal{W}^{(2)}_{\mathcal{R}}$
and one power of
$\mathcal{W}^{(4)}_{\mathcal{R}}$. It therefore
does not arise in (\ref{eq:W2en}) or (\ref{Pdef}).
It is not included in our resummation because, for its power of $g$,
it is less length-enhanced
than the diagram in Fig.~\ref{fig:WloopEN}.

The physical interpretation of 
resumming length enhanced diagrams is
that by doing so we are taking into account 
the possibility that the energetic parton scatters many times
over the course of propagating for a distance $L$ through the medium.
In Appendix~\ref{app:Boltz} we show
explicitly that the resummation we have performed is equivalent
to considering multiple scattering
by deriving and solving a Boltzmann equation for momentum broadening. 
The rederivation of our results in Appendix~\ref{app:Boltz} is also helpful
in making contact between our results and those in previous literature.
To that end, 
in Appendix~\ref{app:Boltz} we 
analyze the Boltzmann equation with a collision kernel that includes 
only one gluon exchange. 
The solution to this Boltzmann equation is identical to Eq.~(\ref{Pdef}), 
which we obtained by 
describing processes with one gluon exchange (via $\mathcal{W}^{(2)}_{\mathcal{R}}$, obtained by evaluating the Wilson line diagrams in Fig.~\ref{fig:W2diag}) and then exponentiating
in order to resum length-enhanced diagrams as we have just discussed. 
From our approach, we know that Eq.~(\ref{Pdef}) could be extended to higher
orders in the coupling by including further disconnected diagrams 
in (\ref{exPne}). Analogously,
the calculation in 
Appendix~\ref{app:Boltz} can immediately be generalized to analyze a Boltzmann equation
with more terms in its collision kernel, and to show that if the collision kernel includes all terms
to arbitrarily high orders in the coupling the result of the Boltzmann equation approach
would indeed agree with our more general expression (\ref{exPne}).

The expression for $\mathcal{W}^{(2)}_{\mathcal{R}}$ is given in (\ref{eq:W2}) as a function of the gluon propagator in coordinate space. Together with (\ref{eq:W2}), the result 
(\ref{Pdef}) provides an expression for the probability distribution
that describes transverse momentum broadening in a weakly coupled plasma of any
length $L$, thick or thin.
In Appendix~\ref{app:thinmed} we complete the analysis of a thin medium, where
$g^2 C_{\mathcal{R}} L \, T  \ll 1$, disconnected diagrams are not
length-enhanced,  and only the diagrams in Fig.~\ref{fig:W2diag} contribute.

We end this Section by rewriting
the expression (\ref{eq:W2}) for
$\mathcal{W}^{(2)}_{\mathcal{R}}$, which appears in
the final result (\ref{Pdef}), in Fourier space. We first introduce the Wightman propagator in momentum space
through
\be
D^{>}(X) = \int \frac{d^4 Q}{(2 \pi)^4} \, e^{- i Q \cdot X} \, D^{>}(Q) \ .
\ee
Here and below, we denote Lorentz four-vectors by an uppercase character, e.g.~$Q$, and the modulus of the three-vector by lowercase character, e.g. $q$.
Integrating over $y^-$ in~\eqref{eq:W2} (taking $L \to \infty$) yields
a delta function $\delta (q^+)$. 
Keeping in mind that  
the coordinate-space
gluon fields are evaluated on the negative light-cone $y^+ = 0$, we then find that
\begin{eqnarray}
\mathcal{W}^{(2)}_{\mathcal{R}}(x_\perp)&=& - g^2 C_{\mathcal{R}} L^-  \times\nonumber\\ 
&&\hspace{-0.5in}\int \frac{d q^- d^2 q_\perp}{(2 \pi)^3} \, \left[1 - e^{i q_{\perp} \cdot x_{\perp}} \right] D^{>}(q^-, q_\perp) .
\label{eq:W2Fourier}
\end{eqnarray}
Finally, the propagator $D^{>}$ can be written as~\cite{LeBellac:1996}
\be
D^>(Q) = \left[1 + f(q^0)\right] \, 2 \,{\rm Re}\, D_R(Q) \ ,
\label{D21def}
\ee
where $f(q^0)$ is the Bose-Einstein distribution function and $D_R(Q)$ is the 
retarded propagator. 
We see that in a weakly coupled plasma 
$P(k_\perp)$ depends only on the retarded gluon  propagator.
Our goal in the next Section will be to derive an
explicit expression for the retarded propagator $D_R(Q)$, from which $D^>(Q)$ can be obtained
using (\ref{D21def}), $\mathcal{W}^{(2)}_{\mathcal{R}}$ can then be
obtained using (\ref{eq:W2Fourier}), and the probability distribution describing
momentum broadening in a weakly coupled QCD plasma
 then follows
using (\ref{Pdef}).



\section{Retarded gluon propagator}
\label{sec:propagator}

In this section, we evaluate the real time expression for
the retarded gluon propagator  
using the Schwinger-Keldysh formalism, 
with the two long light-like segments of the contour separated infinitesimally
in the imaginary time direction.
We give a brief
review of the real-time field theory framework that we use in Appendix~\ref{app:RTF}. 
The retarded gluon propagator $D_{R\, \mu\nu}$ is obtained by solving the Dyson equation
\be
D^{-1}_{R\, \mu\nu}(Q) = (D^{{\rm free}}_{R\, \mu\nu}(Q))^{-1} + i \, \Pi_{R\, \mu\nu}(Q) \ ,
\label{eq:Dyson}
\ee
where $D^{{\rm free}}_{R\, \mu\nu}$ is the free retarded propagator and $\Pi_{R\, \mu\nu}$ is the retarded self-energy. In a generic covariant gauge, the former reads
\begin{equation}
(D^{{\rm free}}_{R\, \mu\nu}(Q))^{-1}=i \, Q^2 \left[ g_{\mu\nu} - \left(1 -\frac{1}{\xi}\right) \frac{Q_\mu Q_\nu}{Q^2} \right] \ ,
\label{eq:freeinverseDR}
\end{equation}
where $\xi$ is the gauge fixing parameter. In this Section, we compute the one-loop expression for the retarded self-energy. With the self-energy in hand 
we can solve~(\ref{eq:Dyson}) and extract the $++$ component of the retarded propagator, 
which is what we need in order to determine the probability distribution $P(k_\perp)$ using the
formalism that we set up in the previous Section.

Unlike at zero temperature, the medium breaks Lorentz invariance and  the self-energy tensor
therefore has four independent components, in principle. 
We work in Feynman gauge ($\xi = 1$), where the one-loop self-energy is transverse~\cite{Kajantie:1982xx,Kobes:1989up,Weldon:1996kb,Brandt:1997se}, $Q^\mu \Pi_{R\, \mu\nu}(Q) = 0$. In this 
case we only have two independent components:
\begin{equation}
\Pi_{R\, \mu\nu}(Q) =  \Pi^T_R(Q)\, P_{\mu\nu}^T  +  \Pi^L_R(Q)\, P_{\mu\nu}^L  \qquad (\xi = 1) \ ,
\label{eq:Pixiequal1}
\end{equation}
where the projectors $P_{\mu\nu}^T$ and $P_{\mu\nu}^L$ are defined in 
Appendix~\ref{app:projectors}. After we substitute the expressions (\ref{eq:freeinverseDR}) and (\ref{eq:Pixiequal1}) into (\ref{eq:Dyson}), we can invert the Dyson equation, obtaining
\begin{equation}
D_{R\, \mu\nu}(Q) = \frac{i \, P_{\mu\nu}^T}{Q^2 - \Pi^T_R(Q)}  + \frac{i\, P_{\mu\nu}^L}{Q^2 - \Pi^L_R(Q)}  - \frac{i \,K_{\mu\nu}}{Q^2} \ ,
\label{eq:DRasPi}
\end{equation}
where the projector $K_{\mu\nu}$ is also given in Appendix \ref{app:projectors}.

\begin{figure*}[t]
\includegraphics[scale=0.5]{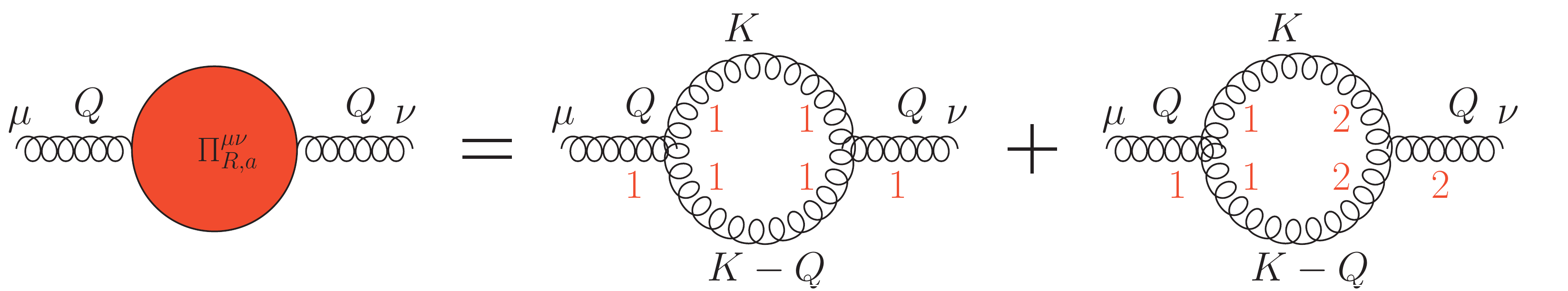} \\
\includegraphics[scale=0.5]{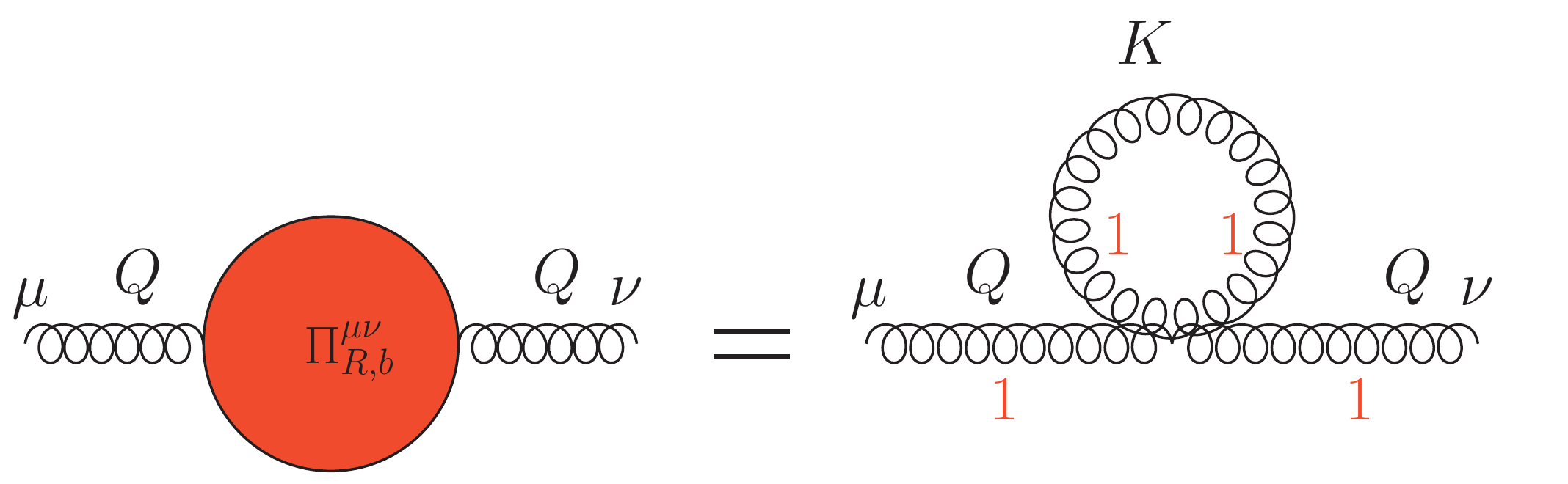} \\
\includegraphics[scale=0.5]{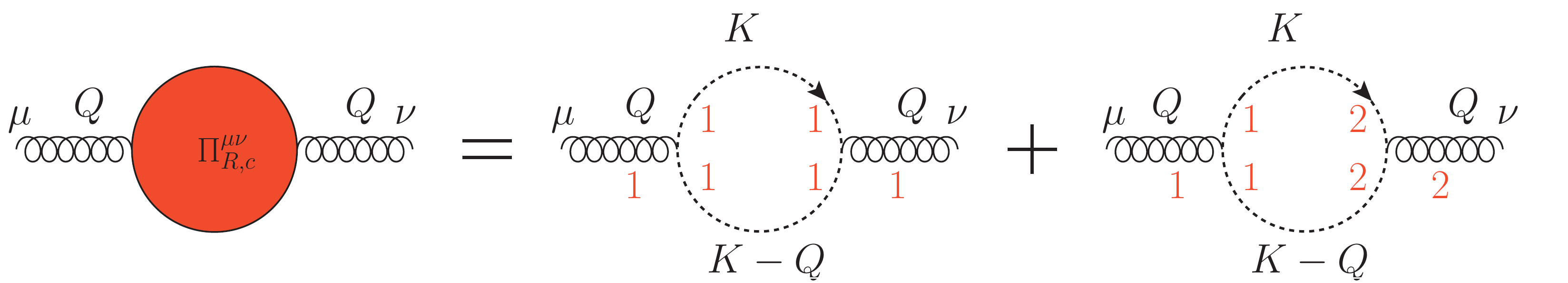} \\
\caption{Diagrams contributing to $\Pi^{\mu\nu}_{R, \, {\rm YM}}$, the contribution
to the retarded self-energy from the Yang-Mills sector. The red numbers denote entries of the Schwinger-Keldysh matrix propagator.}
\label{fig:PiRYM}
\end{figure*}

\begin{figure*}[t]
\includegraphics[scale=0.5]{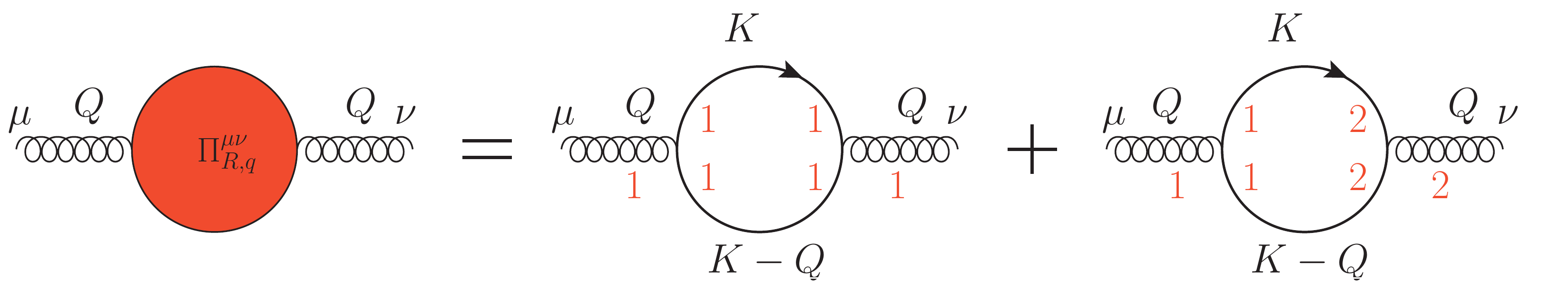}
\caption{Diagrams contributing to $\Pi^{\mu\nu}_{R, \, {\rm quarks}}$. Notation as in  Fig.~\ref{fig:PiRYM}.}
\label{fig:PiRq}
\end{figure*}

In what follows, we explicitly evaluate the full retarded gluon self-energy tensor $\Pi_{R\, \mu\nu}(Q)$ at one-loop, and then extract the components $\Pi^T_R(Q)$ and $\Pi^L_R(Q)$. We use the Schwinger-Keldysh formalism presented in Appendix \ref{app:RTF}, with the identification in (\ref{eq:PiRandPi1112}) between the retarded self-energy and the components of the self-energy matrix in the Schwinger-Keldysh formalism. There are two main contributions to $\Pi_{R\, \mu\nu}(Q)$, the Yang-Mills sector and the quarks, so we write
\be
\Pi^{\mu\nu}_R= \Pi^{\mu\nu}_{R, \, {\rm YM}} + \Pi^{\mu\nu}_{R, \, {\rm quarks}} \ .
\label{eq:fullPi}
\ee
In the Yang-Mills sector we have three contributions
\be
\Pi^{\mu\nu}_{R, \, {\rm YM}} = \Pi^{\mu\nu}_{R,a} +  \Pi^{\mu\nu}_{R,b} +  \Pi^{\mu\nu}_{R,c} \ ,
\label{eq:PiRYM}
\ee
corresponding to the different diagrams shown in Fig.~\ref{fig:PiRYM}: a) gluon loop with the three-point vertex; b) gluon loop with the four-point vertex; c) ghost loop. We start from the 
diagrams
for $\Pi^{\mu\nu}_{R,a}$ in Fig.~\ref{fig:PiRYM} and use the standard Feynman
rules for the three-gluon vertex to obtain the expression
\be
\begin{split}
\Pi^{\mu\nu}_{R,a} = &\, \frac{i}{4}  \, g^2 N_c
\int \frac{d^4 K}{(2\pi)^4} \, \\ & \left[D_{R}(K) D_{S}(K-Q)  +  D_{S}(K)  D_{A}(K-Q)  \right]  \, \times \\ & 
\left[- g^{\mu\nu} \left(5 Q^2 + 2 K^2 - 2 K \cdot Q \right)  + 2 Q^\mu Q^\nu \right. \\ & \left. + 5 Q^\mu K^\nu + 5 K^\mu Q^\nu -10 K^\mu K^\nu  \right]  \ .
\label{eq:PiRa}
\end{split}
\ee
The explicit expressions for the retarded, advanced and symmetric propagators $D_{R}$, $D_{A}$ and $D_{S}$ are given in Appendix~\ref{app:RTF}. We can 
combine the first two terms on the right-hand side of (\ref{eq:PiRa}) by
changing the loop integration variable $K \rightarrow  Q - K$ in one of them. The vertex factor is left unchanged, whereas for the propagators we have
\be
D_{R}(K) D_{S}(K-Q) \; \rightarrow \; D_{S}(K) D_{A}(K-Q)  \ .
\label{eq:chva}
\ee
The two contributions are then identical, and we obtain
\be
\begin{split}
\Pi^{\mu\nu}_{R,a} = & \frac{i}{2} \, g^2 N_c \int \frac{d^4 K}{(2\pi)^4} \, D_{S}(K)  D_{A}(K-Q)\\ & 
\left[- g^{\mu\nu} \left(5 Q^2 + 2 K^2 - 2 K \cdot Q \right)  + 2 Q^\mu Q^\nu \right. \\ & \left. + 5 Q^\mu K^\nu + 5 K^\mu Q^\nu -10 K^\mu K^\nu  \right]  \ .
\label{eq:PiRa2}
\end{split}
\ee
Next we consider the contribution $\Pi^{\mu\nu}_{R,b}$. In this case we have only one diagram, since interaction vertices cannot connect fields on different segments of the Schwinger-Keldysh contour, and it reads
\be
\Pi^{\mu\nu}_{R,b}  = g^2 \, N_c\, \int \frac{d^4 K}{(2\pi)^3}  \delta(K^2) \, n_B(k_{0}) \, (- 3\,  g^{\mu\nu}) \ .
\label{eq:PiRb}
\ee
Finally, since we are working in the Feynman gauge, there is also a contribution 
$\Pi^{\mu\nu}_{R,c}$
from ghosts in the loop. Its 
calculation is very similar to the one for the first contribution $\Pi^{\mu\nu}_{R,a}$, and in particular the two sub-diagrams in Fig.~\ref{fig:PiRq} are combined together by the same shift~(\ref{eq:chva}). After some algebra, the only difference with respect to the previous case is the interaction vertex, and we have
\be
\begin{split}
\Pi^{\mu\nu}_{R,c} = i \, g^2 N_c & \,
\int \frac{d^4 K}{(2\pi)^4} \,  D_{S}(K) D_{A}(K-Q)  \\ & \left[K^{\mu} K^\nu - Q^{\nu} K^\mu \right] \ .
\end{split}
\label{eq:PiRc}
\ee
Upon summing
(\ref{eq:PiRa2}), (\ref{eq:PiRb}) and (\ref{eq:PiRc}), we find
that the full gauge contribution to the retarded gluon self-energy is given by
\begin{widetext}
\be
\begin{split}
\Pi^{\mu\nu}_{R, \, {\rm YM}} = &\, g^2 N_c
\int \frac{d^4 K}{(2\pi)^3} \delta(K^2) \, n_B(k_{0}) \, \frac{1}{(Q - K)^2  - i \, {\rm sgn}(k_0 - q_0) \, \epsilon}  \, \times \\ & 
\left[g^{\mu\nu} \left(2 Q^2 + 4 Q \cdot K - K^2 \right) - 2 Q^\mu Q^\nu - 6 Q^\mu K^\nu - 2 K^\mu Q^\nu + 8 K^\mu K^\nu  \right]  \ .
\end{split}
\label{eq:PiRQCD}
\ee
\end{widetext}
We assume that there are $N_f$ quarks in the theory.
The fermion loop contribution is shown in  Fig.~\ref{fig:PiRq}, and its calculation proceeds analogously
to that for the gauge contribution. We find
\begin{widetext}
\be
\begin{split}
\Pi^{\mu\nu}_{R, \, {\rm quarks}} = & \,4 \, g^2 N_f  \int\frac{d^4 K}{(2 \pi)^3}  \delta(K^2) \, n_F(k_{0}) \, \frac{1}{(Q - K)^2  - i \, {\rm sgn}(k_0 - q_0) \, \epsilon}  \, \times \\ &
\left[g^{\mu\nu} \, ( Q \cdot K - K^2 ) -  Q^\mu K^\nu - K^\mu Q^\nu +  2 K^\mu K^\nu \right]  \ .
 \label{eq:PiRq}
\end{split}
\ee
\end{widetext}

The full retarded self-energy defined in~(\ref{eq:fullPi}) is given by the sum of the two results~(\ref{eq:PiRQCD}) and (\ref{eq:PiRq}). We use the 
transverse and longitudinal projectors defined in Appendix~\ref{app:projectors}
to
extract its components $\Pi^T_R(Q)$ and $\Pi^L_R(Q)$, defined by (\ref{eq:Pixiequal1}), 
in order to evaluate the expression for the retarded gluon propagator in (\ref{eq:DRasPi}). 
The longitudinal component is projected out as follows
\be
\Pi_R^L = P_{L\, \nu\mu} \, \Pi_R^{\mu\nu} = - \frac{U_\mu U_\nu}{N^2} \Pi_R^{\mu \nu} =  \frac{Q^2}{q^2} \Pi_R^{00} \ ,
\label{eq:PiLextracted}
\ee
and we get
\begin{widetext}
\be
\begin{split}
\Pi^{L}_{R, \, {\rm YM}} = & \, \frac{Q^2}{q^2}  \, g^2 N_c
\int \frac{d^4 K}{(2\pi)^3} \delta(K^2) \, n_B(k_{0}) \, \frac{2 Q^2 + 4 Q \cdot K - 2 q_0^2 - 8 q_0 k_0 + 8 k_0^2 }{(Q - K)^2  - i \, {\rm sgn}(k_0 - q_0) \, \epsilon}  \ ,\\
\Pi^{L}_{R, \, {\rm quarks}} = & \,  \frac{4 \, Q^2}{q^2}  \,  g^2 N_f  \int\frac{d^4 K}{(2 \pi)^3}  \delta(K^2) \, n_F(k_{0}) \, \frac{ k_0^2 + k^2 -  2 q_0 k_0 + Q \cdot K}{(Q-K)^2 - i \, {\rm sgn}(k_0 - q_0) \, \epsilon}  \ .
\label{PiL}
\end{split}
\ee
\end{widetext}
Likewise, the transverse component reads 
\be
\begin{split}
\Pi_R^T = & \, \frac{1}{2} \, P_{T\, \nu\mu} \, \Pi_R^{\mu\nu}  = - \frac{1}{2} \left[ \Pi_{R} + \Pi_R^L\right] \ , \\   \Pi_R \equiv & \, g_{\mu\nu} \Pi_R^{\mu\nu} \ ,
\label{eq:PiTextracted}
\end{split}
\ee
where 
the factor of $1/2$ arises because the projector $P_T$
defined in Appendix~\ref{app:projectors} has trace -2. In the second equality,
we have defined the trace of the retarded self-energy $\Pi_R$ and we have also identified the longitudinal self-energy just found above. Thus, once we know the longitudinal component, in order to get the transverse component we need only compute the trace of the self-energy and can 
then use~(\ref{eq:PiTextracted}). We find that the two contributions to the trace are given by
\begin{widetext}
\be
\begin{split}
\Pi_{R, \, {\rm YM}} =  & \, g^2 N_c \int \frac{d^4 K}{(2\pi)^3} \delta(K^2) \, n_B(k_{0}) \, \frac{6 Q^2 + 8 Q \cdot K + 4K^2}{(Q - K)^2  - i \, {\rm sgn}(k_0 - q_0) \, \epsilon}  \ , \\
\Pi_{R, \, {\rm quarks}} = & \,  g^2 N_f  \int\frac{d^4 K}{(2 \pi)^3}  \delta(K^2) \, n_F(k_{0}) \, \frac{8 \, Q \cdot K - 2K^2 }{(Q - K)^2  - i \, {\rm sgn}(k_0 - q_0) \, \epsilon}  \ ,
\end{split}
\label{eq:Piextracted}
\ee
\end{widetext}
for the pure gauge and quark contributions, respectively.

In order to obtain  an explicit expression for the retarded gluon propagator of Eq.~(\ref{eq:DRasPi}), 
we need  the longitudinal self-energy given in~(\ref{eq:PiLextracted}), and the transverse component obtained by combining~(\ref{eq:PiTextracted}) and~(\ref{eq:Piextracted}). The expressions we have
obtained so far are valid for any value of the gluon external momentum $Q$. However, as we have shown in~(\ref{eq:W2Fourier}), we only need the retarded gluon 
propagator evaluated on the negative light-cone, namely for $q^+ = 0$. For nonzero transverse momentum $q_\perp$ this corresponds to space-like external momentum $Q^2 = - q_\perp^2 < 0$, in which case the self-energy has a non-vanishing imaginary part. This is crucial to our analysis, since what enters the calculation of $P(k_\perp)$ 
is the real part of the retarded propagator, as shown in Eqs.~(\ref{eq:W2}), (\ref{Pdef}) and (\ref{D21def}). The retarded propagator in~(\ref{eq:DRasPi}) has a real part if and only if the self-energy has an imaginary part. Without this imaginary part, the probability distribution $P(k_\perp)$ in~(\ref{Pdef}) would just be a delta function centered at $k_\perp = 0$, and thus we would not have any momentum broadening.

In what follows, we sketch the extraction of the longitudinal component of the self-energy arising from loops involving gauge bosons and ghosts, and just state the final result for the transverse component. Starting from the explicit expression \r{PiL}, we first integrate over $k_0$, imposing the on-shell condition for the loop momentum via the delta function. 
We get two different contributions, for $k_0 = k$ and $k_0 = - k$. The integration over the spatial components of the loop momentum is performed in polar coordinates, with the polar axis defined by the direction of the spatial component $\vec q$ of the external momentum. The integration over the azimuthal angle $\phi$ is straightforward, giving just a $2 \pi$ factor. The polar angle $\theta$ satisfies $\cos\theta = \vec q \cdot \vec k$, and after we integrate over it we find
\begin{widetext}
\be
\begin{split}
\Pi^{L}_{R, \, {\rm YM}}  = & \, \frac{g^2 N_c T^2}{6} \frac{q_\perp^2}{q^2} + \\ &  
\frac{g^2 N_c}{8 \pi^2} \frac{q_\perp^2}{q^3} \int_0^\infty  dk \,  n_B(k)  \left[ \(2q^2  - (2k - q_0)^2\) \log \( \frac{q_\perp^2 + 2k(q_0 - q) + i \epsilon \text{ sgn}(k-q_0)}{q_\perp^2 + 2k(q_0 + q) + i \epsilon \text{ sgn}(k-q_0)} \) + \left(\begin{array}{c}q_0 \to -q_0 \\ \epsilon \to -\epsilon \end{array}\right) \] \ ,
\label{LYM1}
\end{split}
\ee
\end{widetext}
where we have used $\int_0^\infty dk \, k \, n_B(k) = \frac{\pi^2 T^2}{6}$.
The logarithms appearing in the expression (\ref{LYM1}) develop an imaginary part for $(q_\perp^2 \pm 2kq_0)^2 < (2kq)^2$. We expand the logarithms in the $\epsilon \rightarrow 0$ limit, obtaining a logarithm of the absolute value and a Heaviside step function for the real and imaginary part, respectively. We then identify the real and imaginary part of the longitudinal self-energy, obtaining
\begin{widetext}
\be
\begin{split}
{\rm Re} \, \Pi^L_{R, \, {\rm YM}} = & \, \frac{g^2 N_c T^2}{6} \frac{q_\perp^2}{q^2}  + \frac{g^2 N_c}{8 \pi^2} \frac{q_\perp^2}{q^3} \int_0^\infty  dk \,  n_B(k)
\left[ \(2q^2  - (2k - q_0)^2\) \log \left | \frac{q_\perp^2 + 2k(q_0 - q) }{q_\perp^2 + 2k(q_0 + q) } \right | + (q_0 \to -q_0)  \] \ , \\
{\rm Im}  \, \Pi^L_{R,\, {\rm YM}} = & \, \frac{g^2 N_c}{8 \pi} \frac{q_\perp^2}{q^3}  \[ \int_{\frac{q_0 + q}{2}}^\infty dk \, n_B(k) \(2q^2  - (2k - q_0)^2\) - (q_0 \to -q_0) \] \ .
\end{split}
\ee
\end{widetext}
The only integrations which are left are over the magnitude of the loop three-momentum. The integral for the real part can only be evaluated numerically, whereas the one for the imaginary part can be expressed in terms of the polylogarithmic functions $\text{Li}_\nu (z)$. The expressions for the real and imaginary part of the longitudinal self-energy coming from quark loop are evaluated analogously, with the main difference being the appearance of the 
Fermi-Dirac distribution thermal distribution function instead of the Bose-Einstein. After combining the Yang-Mills piece and the contribution from $N_f$ quarks we obtain
\begin{widetext}
\be
\begin{split}
{\rm Re} \, \Pi^L_{R} = & \,  \(N_c + \frac{N_f}{2} \)  \frac{g^2 T^2}{6} \frac{q_\perp^2}{q^2}  \, + \\ &
\frac{g^2}{8\pi^2}  \frac{q_\perp^2}{q^3}  \left [ \int_0^\infty dk \log \left | \frac{q_\perp^2 + 2k(q_0 - q)}{q_\perp^2 + 2k (q_0 + q)} \right |  \[ N_c n_B(k) q^2 - (N_f n_F(k) + N_c n_B(k)) (4k^2 - 4kq_0 - q_\perp^2)  \] + (q_0 \to -q_0) \right] \ , \\
{\rm Im} \, \Pi^L_{R} = & \, \frac{g^2 N_c}{24\pi} \frac{q_\perp^2}{q^3}  \left[5 q_0^3+8 q_0 T^2 \pi ^2+6 q_0 q_\perp^2 \right] 
-  \frac{g^2 T}{4\pi} \frac{q_\perp^2}{q^2} \left [ \frac{N_f}{2} \( T \, \text{Li}_2 \(-e^{\frac{q_0-q}{2T}} \) + \frac{2 T^2}{q}   \text{Li}_3 \(-e^{\frac{q_0-q}{2T}} \)  \) + \right.  \\
& - \left . N_c \left ( \frac{q}{2}  \log\left(1-e^{\frac{q-q_0}{2T}}\right)- 2 T \text{Li}_2\(e^{\frac{q-q_0}{2T}}\) + \frac{4T^2}{q} \text{Li}_3\(e^{\frac{q-q_0}{2T}}\) \right )  - (q_0 \to -q_0) \right. \bigg ],
\end{split}
\label{eq:ReImPiL}
\ee
\end{widetext}
By similar means, we calculate the trace of self-energy tensor
(\ref{eq:Piextracted})
which we then combine with the longitudinal components in~(\ref{eq:ReImPiL}) to obtain the transverse self-energy according to Eq.~(\ref{eq:PiTextracted}). We find
\begin{widetext}
\be
\begin{split}
{\rm Re} \, \Pi^T_{R} = & \, \(N_c + \frac{N_f}{2} \) \frac{g^2 T^2}{12} \(1 + \frac{q_0^2}{q^2} \) + \\ &
\frac{g^2 q_\perp^2}{16 \pi^2 q^3} \left [ \int_0^\infty  dk \log \left | \frac{q_\perp^2 + 2k(q_0 - q)}{q_\perp^2 + 2k(q_0 + q)}  \right | \[2 N_c n_B(k) q^2 +  (N_c n_B(k) + N_f n_F(k)) (q^2 + (2k - q_0)^2) \]  + (q_0 \to -q_0) \], \\
{\rm Im} \, \Pi^T_{R} = & \, \frac{g^2 N_c}{48\pi} \frac{q_\perp^2}{q^3} \[10 q_0^3 - 8 q_0 T^2 \pi^2 + 9 q_0 q_\perp^2 \] + \\ & \frac{g^2 T}{4\pi} \frac{q_\perp^2}{q^3} \left [ -\frac{N_f}{2} \( q^2 \log \(1 + e^\frac{q_0-q}{2T} \) + 2qT \text{Li}_2 \( -e^\frac{q_0-q}{2T} \) + 4T^2  \text{Li}_3 \( -e^\frac{q_0-q}{2T} \) \) +  \right.  \\
&\left. N_c \(  q^2  \log\(1 - e^{\frac{q - q_0}{2T}}\) - q T \text{Li}_2\(e^{\frac{q - q_0}{2T}}\) -2 T^2 \text{Li}_3\(e^{\frac{q - q_0}{2T}}\) \) - (q_0 \to -q_0) \right. \bigg ] \ .
\end{split}
\label{eq:ReImPiT}
\ee
\end{widetext}
Thus, we have obtained the components of the gluon
self-energy  in~(\ref{eq:ReImPiL}) and~(\ref{eq:ReImPiT}) by direct calculation in real-time field theory. In the imaginary time formalism, the gluon self-energies were first computed in 
Refs.~\cite{Kajantie:1982xx,Kalashnikov:1979cy}.
  
We end this section by performing a check of our calculation, a check
whose results we shall use in Section~\ref{sec:magnetic}. 
We recover the HTL self-energies~\cite{Braaten:1989kk}, which are valid for external momenta of order $g T$ or below. In this regime, the main contribution to the loop integral comes from hard loop momenta $k \sim T$, so the procedure corresponds to expanding the integrand in powers of $q / k$. Both the real and imaginary parts of the gluon self-energy that we have obtained above
can be computed analytically to leading order in this expansion, resulting in
\be
\begin{split}
\left. {\rm Re} \, \Pi^L_{R} \right|_{HTL} = &\, \frac{m_D^2 q_\perp^2}{q^2} \left ( 1 + \frac{q_0}{2q} \log \left ( \frac{q - q_0}{q + q_0} \right ) 
\right ) \ , \\
\left. {\rm Im} \, \Pi^L_{R} \right|_{HTL} = &\, \pi m_D^2 \( \frac{q_\perp^2 q_0}{2q^3} 
\) \ , \\
\left. \Pi^{T}_{R} \right|_{HTL} =  &\, \frac{m_D^2- \left. \Pi^L_{R} \right|_{HTL}}{2} \ ,
\end{split}
\label{FGHTL}
\ee
where $m_D^2$ is the Debye mass squared 
\be
m_D^2 = \frac{g^2 T^2}{3} (N_c + \frac{N_f}{2} ) \ .
\label{eq:mD}
\ee
As we will show explicitly in Section~\ref{sec:magnetic}, perturbation theory breaks down in the region where the external momentum
in a gluon propagator is of order $g^2 T$. We will fix this problem by using the HTL self-energies given in Eq.~(\ref{FGHTL}), which are well-behaved in the region where ordinary perturbation theory becomes problematic.


\section{Breakdown of perturbation theory and self-energy matching}
\label{sec:magnetic}

The purpose of this work is to evaluate the probability distribution
$P(k_\perp)$ in (\ref{Pdef}). In order to do that we have to evaluate the gluon propagator for $q^+=0$, and integrate it over $dq^- d^2 q_\perp$, as in Eq.~(\ref{eq:W2Fourier}). In particular, we have to integrate the gluon propagator over the 
region in momentum space where both $q^-$ and $q_\perp$ are of order
$g^2 T$ or smaller.  We shall begin this Section
with an explicit demonstration of the breakdown
of perturbation theory at the scale $g^2 T$ in our self-energy results~(\ref{eq:ReImPiL}) and~(\ref{eq:ReImPiT}), and then describe how we shall evade this difficulty. This problem in finite temperature non-abelian gauge theory
has been known for many years, since the early work of 
Refs.~\cite{Linde:1980ts,Kalashnikov:1980tk}. Here, we focus on the pure Yang-Mills contribution to the self-energy (setting $N_f = 0$), since the matter fermions are not responsible for the breakdown of perturbation theory in the infrared. We only consider the real part of the self-energies, since that
is where the problem arises.

We can find the infrared breakdown of perturbation theory
that occurs where both $q^-$ and $q_\perp$ are of order
$g^2 T$ by focussing on the slice through this
region where $q_0=0$ and $0< q_\perp < g^2 T$.
No problems arise in the longitudinal self-energy:
it is gauge independent~\cite{Kalashnikov:1980tk} and, 
upon taking the appropriate limit in~(\ref{eq:ReImPiL}), we find
\be
{\rm Re} \, \Pi^L_{R , \, YM}(q_0 = 0, q_\perp \to 0) \to \, m_D^2 = \frac{1}{3}g^2 N_c T^2   \ .
\ee
This longitudinal self-energy 
is responsible for screening the electric modes, giving them a screening mass $m_{{\rm el}}^2  = g^2 N_c T^2  / 3$. It does not cause any problems for perturbation theory. 
The problems arise in the transverse self-energy~(\ref{eq:ReImPiT}). 
In order to extract the infrared limit, we divide the loop integral $d k$
in (\ref{eq:ReImPiT})  into hard and soft regions. 
When $q_0=0$, $q_\perp < g^2 T$, and the loop integration variable is hard ($k \gtrsim T$ and
therefore $k\gg g^2T$) we find  that the integrand in the expression (\ref{eq:ReImPiT}) 
for $\Pi^T_{R, \, YM}$ vanishes.  
We return to this point below but, first, we push ahead into trouble
by attempting to evaluate the contribution to $\Pi^T_{R, \, YM}$ from
the region of the $d k$ integral in (\ref{eq:ReImPiT}) where $k\ll T$.
Here we are allowed to use the 
``soft approximation'' ($n_B(k) \sim T / k$) 
 in~(\ref{eq:ReImPiT}). We find\footnote{Our result is obtained in 
Feynman gauge ($\xi =1$). For a general covariant gauge the infrared behavior of the 
self-energy was first analyzed in~\cite{Kalashnikov:1980tk} using
the imaginary time formalism, finding
\be
{\rm Re} \, \Pi^T_{R , \, YM} (q_0 = 0, q_\perp \to 0) \to  -\frac{ 8 + (1+\xi)^2}{64} \, g^2 T N_c \, q_\perp \ , \notag
\ee
consistent with our result in the Feynman gauge. This contribution 
is gauge dependent, but it cannot be set to zero by any gauge choice.}
\be
\begin{split}
& {\rm Re} \, \Pi^T_{R , \, YM} (q_0 = 0, q_\perp \to 0) \to  \\ & 
\frac{3}{8 \pi^2} g^2 N_c q_\perp T \int_0^\infty \frac{dk}{k} \log \left | \frac{q_\perp- 2k}{q_\perp + 2k} \right | = -\frac{3}{16} g^2 T N_c q_\perp  \ .
\label{GaugeDepMagMass}
\end{split}
\ee
We immediately notice that for external momentum $q_\perp \sim g^2 T$, the
real part of the  transverse self-energy ${\rm Re} \, \Pi^T_{R , \, YM}$ is comparable 
to $q_\perp^2$.  This introduces an unphysical pole at a space-like momentum
of order $g^2 T$ 
in the propagator (\ref{eq:DRasPi}). It also invalidates the perturbative expansion
of the propagator (\ref{eq:DRasPi}), in
which $\Pi_R$ is supposed to be subleading compared to $Q^2$.
Clearly, 
perturbation theory cannot be trusted anymore at and 
below the scale $g^2 T$, and neither
can the result (\ref{GaugeDepMagMass}).

It is expected that a magnetic mass of order $g^2 T$ arises from nonperturbative effects.  
Even if this happens, though, 
perturbation theory still breaks down at the $g^6$ order, as shown in 
Ref.~\cite{Linde:1980ts} by an explicit example. (In contrast, neither perturbative nor non-perturbative effects generate a magnetic mass in an abelian gauge 
theory~\cite{Kalashnikov:1996qx}. The leading term for the transverse self-energy goes 
as $g^2 q_\perp^2$, as can be checked from our fermion loop result, and for this reason perturbation theory does not break down in the infrared limit.)

\begin{figure}[t]
\includegraphics[scale=0.57]{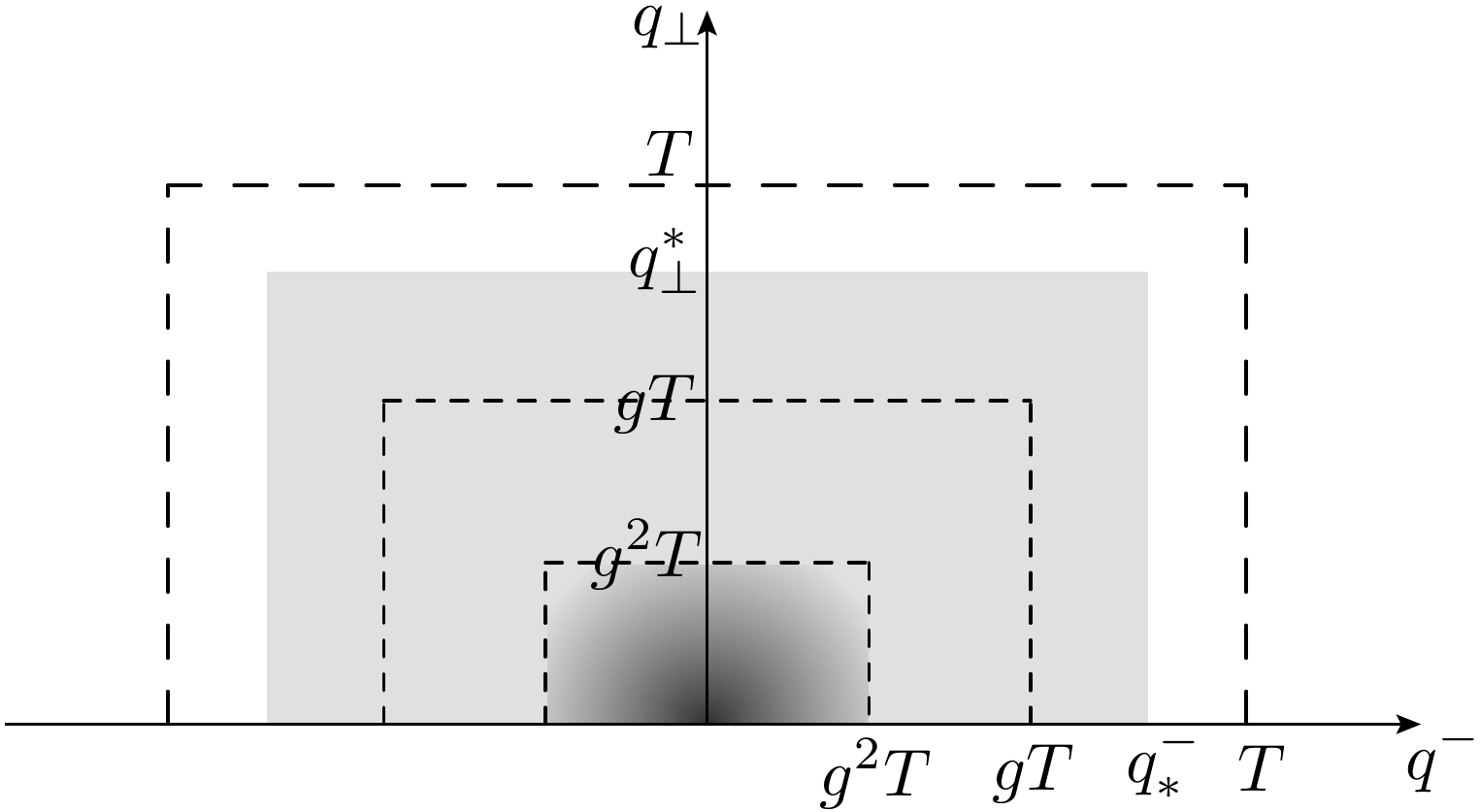}
\caption{Self-energy matching illustration. In the momentum region
shaded in grey, where  $|q^-| < |q^-_*|$ and $q_\perp < q_\perp^*$, we use HTL self-energies in the integration in~\r{eq:W2Fourier}, whereas we use full self-energies elsewhere, in the white
regions. 
The darker grey region, where momenta are ${\cal O}(g^2 T)$ is the dangerous
region where we {\it must} use HTL self-energies.   We do not want to do the matching
from grey to white anywhere near this darker region.  The HTL self-energies are
not valid once momenta are ${\cal O}(T)$, so the grey region must not extend this far.
We have checked explicitly that 
for $g \ll 1$ the numerical result for~\r{eq:W2Fourier} is insensitive to where we match
from grey to white, as long as $gT < |q^-_*| < T$ and $gT < k_\perp^* < T$.}
\label{IntegRegion}
\end{figure}

 To take care of this problem, we use the HTL self-energy~(\ref{FGHTL}) in the problematic region. In this approximation, the transverse component of the gluon self-energy is gauge independent~\cite{WeldonKlimov}
 and does not give rise to any additional pole at $q \neq 0$, 
 so we do not run into any infrared problems. 
 Along the $q_0=0$ slice that we analyzed above, the HTL self-energy
 is so well-behaved that it in fact vanishes, as we already saw above.
 Using the HTL self-energy in the $g^2 T$ momentum region avoids all infrared problems and
 gives us a well-behaved result at the leading order to which we are working, but of course it does not
 incorporate the effects of the magnetic mass of order $g^2T$, which is generated only nonperturbatively
 and so
 is absent in the HTL self-energy (\ref{FGHTL}).  In the high-temperature limit of QCD, the 
nonperturbative physics at momenta of order $g^2T$ is described by 
matching to a dimensionally reduced long-wavelength effective theory which turns out to be just 
Euclidean 3-dimensional $SU(N_c)$ gauge theory with the dimensionful coupling constant $g_E$
given by $g_E^2 = g^2 T$.
Following up on a suggestion by Caron-Huot~\cite{CaronHuot:2008ni}, Laine has very recently 
shown that the nonperturbative contributions to $P(k_\perp)$ can
be related 
to the static potential in this effective theory and he and others~\cite{Benzke:2012sz,Laine:2012ht} 
have used
this elegant observation to show that the nonperturbative contribution to $P(k_\perp)$ is
suppressed parametrically, contributing to $\hat q$ (the second moment
of $P(k_\perp)$) only at order $g^{6}T^3$,  and is further suppressed by
a numerically small prefactor~\cite{Laine:2012ht}. This result justifies the neglect
of these nonperturbative effects that is inherent in our use of the HTL self-energy at
momenta of order $g^2T$.

 Although using the HTL self-energy nicely eliminates the infrared problems in
 perturbation theory, 
 we cannot simply use the HTL self-energy throughout our calculation because
 it is not valid for hard external momenta, which in our case
 corresponds to $q_0 \sim T$ (and therefore $q^- \sim T$)
 and $q_\perp \sim T$ in \r{eq:W2Fourier}. The correct procedure is then to use the full or HTL self-energies 
 in the regimes where each is valid, and to match in a region where both are valid.  The
 strategy is illustrated in
  Fig.~\ref{IntegRegion}. The darkest shading illustrates the momentum scales of order $g^2 T$ 
  where we must use the HTL self energies because perturbation theory runs into troubles if 
  we do not do so. 
 In the regions where momenta are of order $T$, the HTL self energies are no longer valid
 and we have to use the full self energies.   We must match from HTL to full self energies
 in a region in which both are valid.
  As illustrated in Fig.~\ref{IntegRegion},
we perform the matching at  $q^-_*$ and $q_\perp^*$ such that $gT < q^-_* < T$ and $gT < q_\perp^* < T$. We find that the matching is smooth at weak
coupling $g\ll 1$, with 
the exact location of the matching scales $q^-_*$ and $q_\perp^*$ not affecting our final results 
as long as the matching is performed in the appropriate region.

Before presenting our results in the next Section, we close this
Section by comparing our approach to perturbation theory, illustrated in the Fig.~\ref{IntegRegion}, to that in some previous field-theoretical analyses of
momentum broadening in weakly coupled quark-gluon 
plasma~\cite{Arnold:2008zu,Arnold:2008vd,CaronHuot:2008ni,Majumder:2009cf}. 
In the soft region ($k_\perp < T$), these authors use the HTL approximation
in their calculations of the probability for momentum broadening. 
In the notation 
of our Eq.~(\ref{eq:W2Fourier}), using the HTL
approximation is well justified when $q_\perp <T$
only over the regime of the $dq^-$ integration in which
$|q^-| < T$, but not over the entire range of the $dq^-$ integration.
We have checked, however, that if we were to use
 the HTL self energies for $q_\perp <T $ over the entire $dq^-$ integration
 the error introduced is quite small. 
Our results therefore agree with theirs in this momentum regime,
for a thin medium. But, only for a thin medium because
once the medium becomes thick one must resum $L$-enhanced
diagrams, as we have done.  
In the hard region
($k_\perp \gg T$) Arnold and Dogan 
correctly use the 
unscreened gluon propagator~\cite{Arnold:2008zu}.
In this regime, resumming $L$-enhanced diagrams does
not modify our results significantly 
(because it is more likely to pick up a very large $k_\perp$ from a single improbable hard kick than
from several less hard but still improbable kicks) and our results therefore
agree with those of Ref.~\cite{Arnold:2008zu} for $k_\perp \gg T$.
We have been quite careful about how we match from the hard region, including
that at $|q^-|\gg T$ at small $q_\perp$, as we have described in Fig.~\ref{IntegRegion}.
This care is unnecessary when $k_\perp \gg T$, and when $k_\perp < T$ it turns
out that doing the matching carefully as we do modifies our results less than
the resummation of $L$-enhanced diagrams does.
So, we shall see in the next section 
that our calculation correctly reproduces the results derived with other techniques
where it should. However, when the medium is thick enough that the effects of the
resummation that we have done become important, we find disagreements with previous 
results in the soft perpendicular momentum region.

\section{Results and discussion}
\label{sec:results}

In Sections II, III and IV we have presented a careful derivation of our 
expression for $P(k_\perp)$ in a weakly coupled plasma and a 
complete description of 
how we shall evaluate it.
The derivation has turned out to be both subtle and technical at various
points, and we therefore promised in Section I that a reader not interested
in subtleties or technical details could skip from the end of Section I to here.
For the benefit of such a reader, we begin here by restating the most salient
points from the previous Sections.
After expanding the probability distribution for transverse
momentum broadening~(\ref{eq:provb}) in the weak-coupling limit and after resumming 
an infinite class of
``length-enhanced'' diagrams that are important if $L$ is large enough that 
$g^2 C_{\mathcal{R}} L \, T$ is not $\ll 1$,
we found that transverse momentum broadening is described
by
\be
P(k_\perp) = \int d^2 x_{\perp} \, e^{-i k_{\perp} \cdot x_{\perp}}\, \exp\left[ \mathcal{W}^{(2)}_{\mathcal{R}}  \right] \ .
\tag{\ref{Pdef}}
\ee
The physical interpretation of resumming  length enhanced diagrams is 
that doing includes the effect of multiple scattering; we show in Appendix~\ref{app:Boltz}
that the same result (\ref{Pdef}) can equally well be derived by solving
a Boltzmann equation for momentum broadening via multiple elastic collisions.
In (\ref{Pdef}),
the properties of the medium enter through
\be
\mathcal{W}^{(2)}_{\mathcal{R}}(x_\perp)= - g^2 C_{\mathcal{R}} L^-   \int \frac{d q^- d^2 q_\perp}{(2 \pi)^3} \, \left[1 - e^{i q_{\perp} \cdot x_{\perp}} \right] D^{>}(q^-, q_\perp) \ .
\tag{\ref{eq:W2Fourier}}
\ee
The Wightman gluon propagator $D^>$ is directly related to the retarded 
gluon propagator by
\be
D^>(Q) = \left[1 + f(q^0)\right] \, 2 \,{\rm Re}\, D_R(Q) \ ,
\tag{\ref{D21def}}
\ee
where $f(q^0)$ is the Bose-Einstein distribution function
and $D_R(Q)$ is given in~(\ref{eq:DRasPi}) in terms of the self-energies that we have 
then computed explicitly in Sec.~\ref{sec:propagator}.  In Section IV we explain where and why
we use the full self energies or the HTL self energies in our evaluation of (\ref{eq:DRasPi}).  
The full and HTL self energies are given
explicitly in Eqs.~(\ref{eq:ReImPiL},\ref{eq:ReImPiT},\ref{FGHTL}).

\subsection{Results for a thin medium, and comparison to previous results}
\label{subsec:5A}
Let us introduce the dimensionless variable
\be
\kappa \equiv \frac{g^2 C_R \,  L T}{2\pi} \ ,
\label{eq:kappa}
\ee
proportional to the thickness of the medium $L$, which determines how
important it is to resum $L$-enhanced diagrams.
We begin by presenting our results for the case where
  $\kappa \ll 1$,  meaning that there is no need  to resum the $L$-enhanced diagrams at all.
In this thin-medium regime,
it is convenient to define the function
\be
P_{\rm thin}(k_\perp) \equiv 
\frac{2\sqrt{2}\pi\,\kappa}{T}\,
\int \frac{d q^- }{2 \pi} D^{>}(q^-, k_\perp) \ ,
\label{eq:Pgdef}
\ee
because
the resummed probability distribution~(\ref{Pdef}) 
reduces to 
\be
P(k_\perp) = P_{\rm thin}(k_\perp) \quad {\rm for} \quad k_\perp \neq 0 \ . 
\ee
This is shown explicitly in Appendix~\ref{app:thinmed}, where we
also explain how to handle subtleties
at $k_\perp=0$ correctly, so as to obtain 
a normalized probability distribution $P(k_\perp)$.

\begin{figure}[t]
\begin{center}
\includegraphics[scale=0.95]{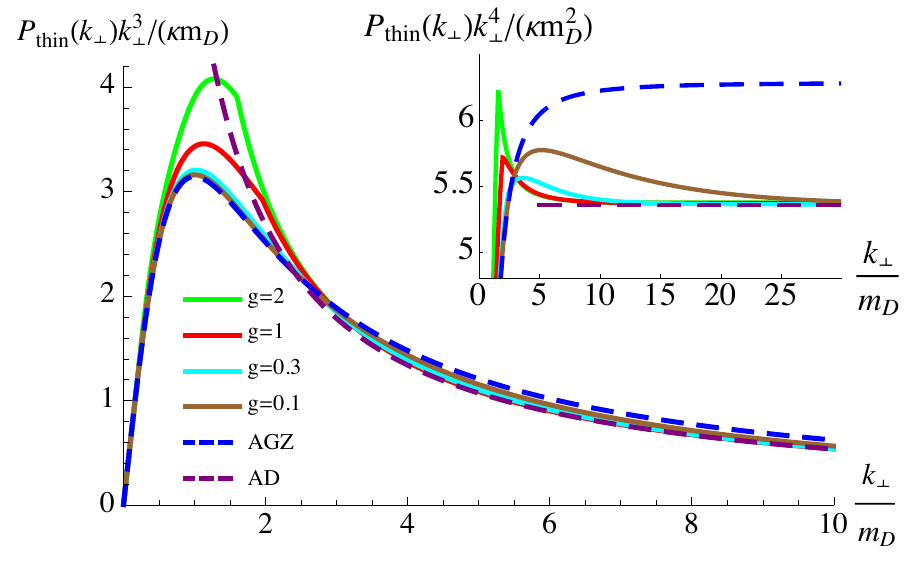}
\caption{The continuous brown, light blue, red and green curves
are the probability distribution $P_{\rm thin}(k_\perp)$
for $g=0.1$, 0.3, 1 and 2 (bottom to top at low $k_\perp$, top to
bottom at high $k_\perp$), multiplied by $k_\perp^3$ and
$k_\perp^4$. 
In the IR, 
$P_{\rm thin}(k_\perp)$ agrees with $P_{\rm thin}^{{\rm AGZ}}(k_\perp)$ (shown as the dashed 
dark blue curves) 
and in the UV, 
$P_{\rm thin}(k_\perp)$ agrees with $P_{\rm thin}^{{\rm AD}}(k_\perp)$ 
(shown as the dashed purple curves).  
The only $L$-dependence in $P_{\rm thin}$ arises from
it being proportional to $\kappa$ meaning that, 
because we have plotted the probability distributions divided by $\kappa$, the quantities
plotted are $L$-independent.  We have scaled both axes by the appropriate
power of the Debye mass $m_D$ to make the quantities plotted dimensionless.
Scaling the plots in this way also ensures that   $P_{\rm thin}^{{\rm AGZ}}(k_\perp)$
and $P_{\rm thin}^{{\rm AD}}(k_\perp)$, shown as the dashed curves, are
independent of $g$.
The kinks in the curves for $g=1$ and $g=2$ are located at the $k_\perp=q_\perp^*$
where we do the matching described in Fig.~\protect{\ref{IntegRegion}}. 
}
\label{fig:P21g01}
\end{center}
\end{figure}

The correct IR and UV behavior 
of the probability distribution $P_{\rm thin}(k_\perp)$ have each been obtained previously:
\begin{itemize}
\item
In the IR region, Aurenche, Gelis and Zaraket showed by explicit 
calculation that (in our notation)~\cite{AGZ}
\be
P_{\rm thin}(k_\perp) =P_{\rm thin}^{{\rm AGZ}} (k_\perp) \quad {\rm for} \quad k_\perp \ll T
\ee
where
\be
P_{\rm thin}^{\rm AGZ}(k_\perp)\equiv \kappa \,  \frac{2 \pi m_D^2}{k_\perp^2 (k_\perp^2 + m_D^2) } \ ,
\label{AGZ}
\ee
with the Debye mass squared as given in~(\ref{eq:mD}). 
\item
In the UV region, $k_\perp \gg T$,  the calculation of Arnold and Dogan
shows that (again in our notation)~\cite{Arnold:2008zu}
\be
P_{\rm thin}(k_\perp) =P_{\rm thin}^{{\rm AD}} (k_\perp) \quad {\rm for} \quad k_\perp \gg T
\ee
where
\be
P_{\rm thin}^{{\rm AD}}(k_\perp) = \kappa \, (4 N_c+ 3 N_f)  \frac{g^2 \zeta(3) T^2}{\pi k_\perp^4}  \ ,
\label{ArnoldDogan}
\ee
with $\zeta(3) \approx 1.202$  the Riemann zeta function. 
\end{itemize}
These expressions can each be obtained from our
$P_{\rm thin}(k_\perp)$, defined in Eq.~(\ref{eq:Pgdef}), by taking
the IR or UV limits. To obtain the IR expression, we use the HTL self-energy everywhere, make 
the additional soft approximation, e.g.~$n_B(q_0) \sim T/q_0$, 
and recover~\r{AGZ}. To take the UV limit,
we use the full self-energy rather than the HTL self-energy and keep only the first order solution to the Dyson equation~(\ref{eq:Dyson}), and recover~(\ref{ArnoldDogan}). 
In Fig.~\ref{fig:P21g01} we plot $P_{\rm thin}(k_\perp)$ multiplied by factors of $k_\perp^3$
and $k_\perp^4$, and show the agreement in the IR with the AGZ result and in the UV with
the AD result.  We see that at both $g=0.1$ and $g=0.3$, the agreement with the AGZ result is
excellent, and extends to values of $k_\perp/m_D$ that are not small at all.  In fact, for $g=0.01$
(which we have not plotted) this agreement extends beyond $k_\perp=10 m_D$.
We also see that although at $g=0.3$ the matching described in Section IV, see
Fig.~\protect{\ref{IntegRegion}}, is smooth, at $g=1$ and $g=2$ it introduces
a kink at  $k_\perp=q_\perp^*$.  This highlights the fact that a weak-coupling
analysis is not quantitatively reliable at these larger values of $g$.
There is a good reason for this: once $g\geq 1$, the separation of
the scales $g^2 T$, $g T$ and $T$ that we discussed in Sec.~\ref{sec:magnetic} and 
used as depicted in Fig.~\ref{IntegRegion} breaks down. 
In order to apply our calculation at $g=1$ and $g=2$, we need a prescription
for how to do the matching described in Fig.~\ref{IntegRegion}, even when the
scales depicted are not  separated.  What he have done is to choose
the matching scale on the horizontal axis of Fig.~\ref{IntegRegion}
as $q^-_*=T$, thinking it would be unreasonable to choose a larger $q^-_*$ even
if it is the case that $g T > T$.  Then we have chosen the matching scale $q_\perp^*$
so as to make the probability distribution $P_{\rm thin}(k_\perp)$ continuous
at  $k_\perp=q_\perp^*$, with a kink there but no discontinuity.
Clearly we could instead have chosen a matching prescription at $g>1$ involving
interpolation over a window in $k_\perp$, but this would have been no less arbitrary,
given that there is a physical reason why the calculation is not under quantitative
control at these large values of $g$.

A leading order expression for $P_{\rm thin}(k_\perp)$ 
for all $k_\perp$ was obtained in Ref.~\cite{Arnold:2008vd}
by interpolating between the small and large $k_\perp$ regimes. 
A next-to-leading-order result was derived in Ref.~\cite{CaronHuot:2008ni}, but within the HTL approximation. The HTL result of Ref.~\cite{CaronHuot:2008ni} was then extended to the $k_\perp > T$ region by making the soft approximation discussed above. Another calculation of $P_{\rm thin}$ 
valid in the IR can be found in Ref.~\cite{Majumder:2009cf}, where the momentum broadening distribution was obtained via a Langevin equation. The solution obtained there using 
the HTL self-energy reproduces the AGZ result.



\subsection{Complete results for the probability distribution $P(k_\perp)$}
\label{subsec:5B}

\begin{figure}[t]
\includegraphics[scale=1.03]{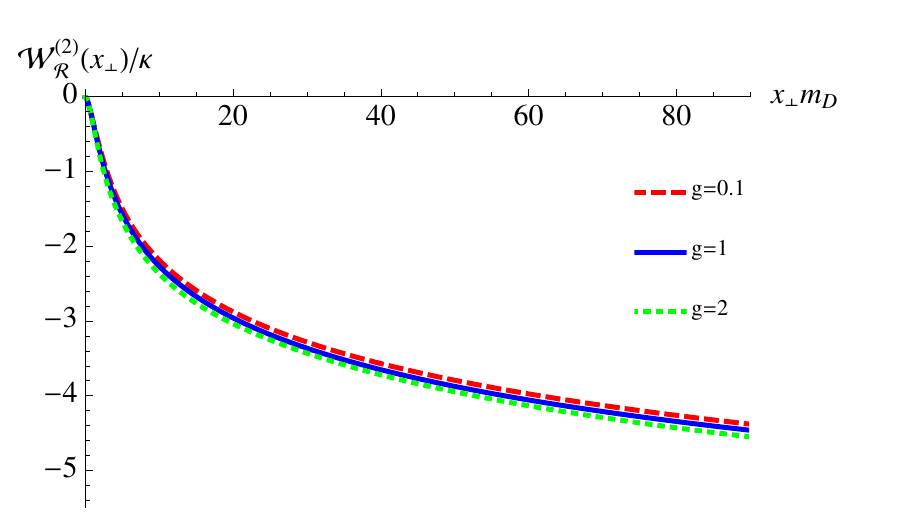}
\caption{$\mathcal{W}^{(2)}_{\mathcal{R}} (x_\perp)/\kappa$ 
for gauge coupling constants $g=0.1$, 1
and  2.  $\mathcal{W}^{(2)}_{\mathcal{R}} (x_\perp)/\kappa$ is independent
of $\kappa$ and, when plotted versus $x_\perp$ in units
of the inverse Debye mass, is almost 
independent of $g$.}
\label{W2_g01}
\end{figure}

The probability distribution $P(k_\perp)$ in~(\ref{eq:provb}) is obtained by Fourier transforming the function $\exp\left[\mathcal{W}^{(2)}_{\mathcal{R}} (x_\perp)\right]$ with
$\mathcal{W}^{(2)}_{\mathcal{R}}$ given by (\ref{eq:W2Fourier}). 
We see therefore that, if the medium being probed is a weakly-coupled plasma, 
$\mathcal{W}^{(2)}_{\mathcal{R}} (x_\perp)$ is the only ``soft function'' 
through which properties of the medium enter
into the probability distribution for momentum broadening.
$\mathcal{W}^{(2)}_{\mathcal{R}} (x_\perp)$ is proportional to $\kappa$ and
depends on the gauge coupling constant $g$, with most of the latter
dependence coming via its dependence on the Debye mass $m_D$ given in (\ref{eq:mD}).
We illustrate this 
in Fig.~\ref{W2_g01}, where we plot $\mathcal{W}^{(2)}_{\mathcal{R}} (x_\perp)/\kappa$
versus $x_\perp m_D$ for several values of $g$.
We have described in detail how we evaluate $\mathcal{W}^{(2)}_{\mathcal{R}} (x_\perp)$ 
 in Sections III and IV.

\begin{figure}[th]
\begin{center}
\includegraphics[scale=0.96]{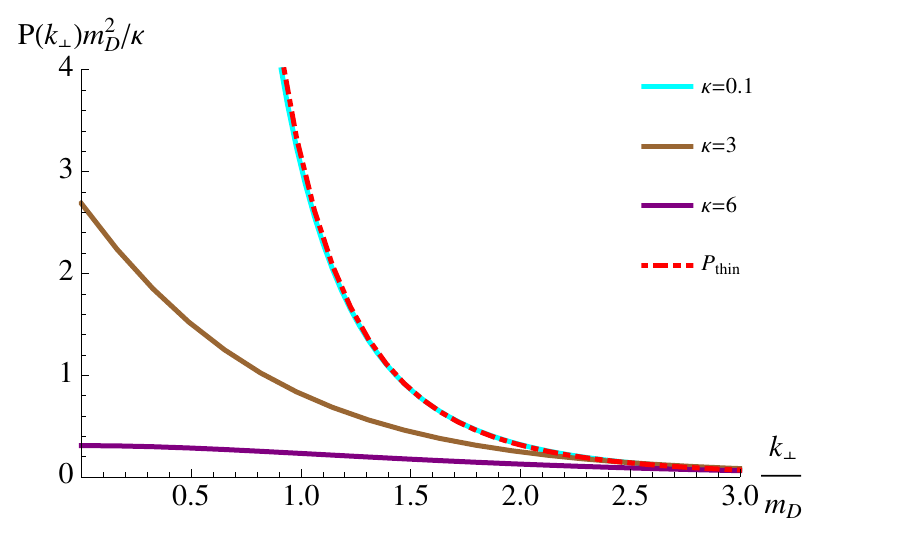} \\
\includegraphics[scale=0.96]{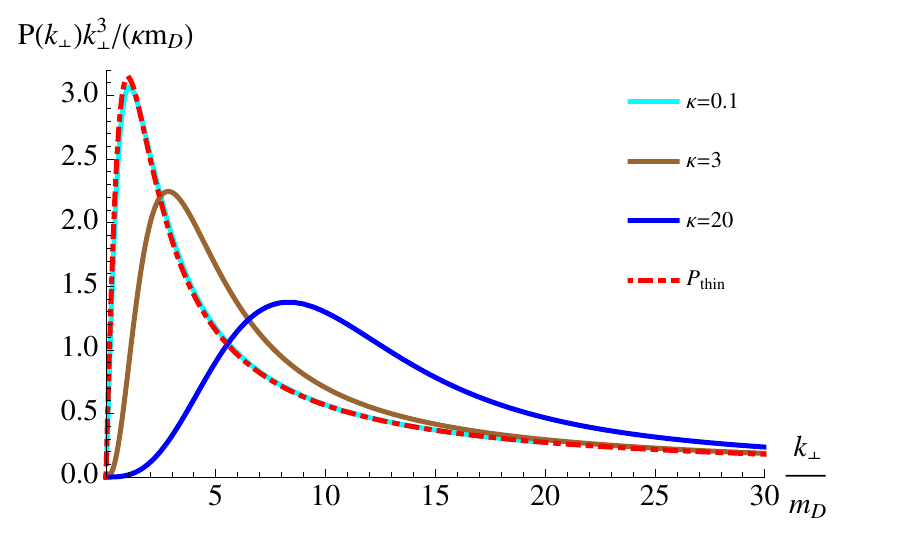} \\
\includegraphics[scale=0.96]{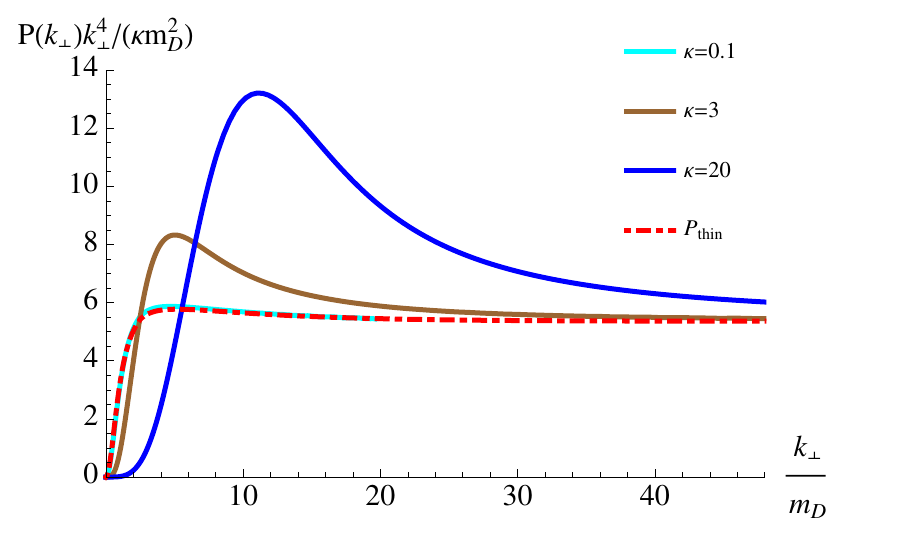} \\
\caption{In the top panel, we plot our full result for the probability distribution
$P(k_\perp)$, after resumming the contributions of $L$-enhanced diagrams.  
We show our results at three different values of $\kappa$ ($\kappa$ increases
from top to bottom at low $k_\perp$), and confirm
that at $\kappa=0.1$ our result agrees with $P_{\rm thin}(k_\perp)$.
In the middle and lower panels, we multiply $P(k_\perp)$ by $k_\perp^3$ and $k_\perp^4$
in order to highlight the behavior at intermediate and large $k_\perp$,
as well as at a larger value of $\kappa$, namely $\kappa=20$.
The gauge coupling constant is  $g=0.1$  throughout.
All the probability distributions are normalized as in (\protect{\ref{eq:ProbabilityNormalization}}).
} 
\label{fig:Pg01}
\end{center}
\end{figure}

In the top panel of Fig.~\ref{fig:Pg01}, we present our numerical 
results for the fully resummed probability distribution $P(k_\perp)$
that describes momentum broadening in a weakly
coupled plasma with gauge coupling constant $g=0.1$.
We show our results for three different values of
the thickness of the medium $L$, meaning three
different values of $\kappa$.  We have also
plotted $P_{\rm thin}(k_\perp)$, and we see that
when $\kappa=0.1$ the medium is so thin that
$P(k_\perp)\simeq P_{\rm thin}(k_\perp)$, 
meaning that there is no need to resum $L$-enhanced diagrams.
Although $P_{\rm thin}(k_\perp) \propto \kappa$, meaning that 
$P_{\rm thin}(k_\perp)/\kappa$
in the Fig.~\ref{fig:Pg01} is $\kappa$-independent,
our full result $P(k_\perp)$ has nontrivial $\kappa$-dependence at small $k_\perp$.
This $\kappa$-dependence is better seen in the middle panel of the Fig.~\ref{fig:Pg01},
where we plot $k_\perp^3 P(k_\perp)$.  Note that the mean value of $k_\perp^2$, which
is proportional to the jet quenching parameter $\hat q$ that we shall discuss in Section V.C,
is given by the area under the curves in this middle panel.  We 
see from the Figure that increasing $\kappa$ steadily shifts probability density away
from small $k_\perp$, pushing it out to larger and larger $k_\perp$.  This makes sense:
as you make the medium thicker, the hard parton spends more time
travelling through the medium, getting kicked, and so can pick up more and more transverse
momentum.
Finally, in the third panel of Fig.~\ref{fig:Pg01} we highlight the behavior of $P(k_\perp)$
at large $k_\perp$.  We see that for any value of $\kappa$ at large enough $k_\perp$
the probability distribution $P(k_\perp)$ approaches $P_{\rm thin}(k_\perp)$,
meaning that resummation of $L$-enhanced diagrams is unnecessary.
This is reasonable on physical grounds:
for any value of $\kappa$ there will be some $k_\perp$ that is so large
that the most probable way of picking up this improbably large $k_\perp$
is via a single scattering, which is described by $P_{\rm thin}(k_\perp)$.

\begin{figure}[t]
\begin{center}
\includegraphics[scale=0.95]{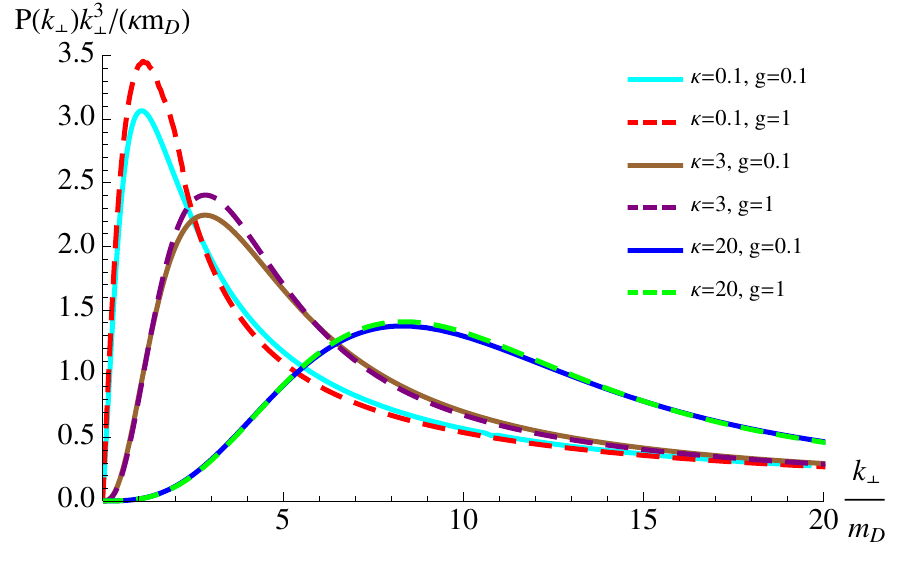}
\caption{
Probability distribution $P(k_\perp)$ at three different values of $\kappa$, multiplied by $k_\perp^3$ and plotted
as in the middle panel of Fig.~\protect{\ref{fig:Pg01}}, but for two different
values of $g$.
}
\label{fig:g_dependence}
\end{center}
\end{figure}

We expect from Fig.~\ref{W2_g01}, and confirm in Fig.~\ref{fig:g_dependence},
that once we plot $P(k_\perp)$ relative to $k_\perp/m_D$, there is
little remaining $g$-dependence. We see from Fig.~\ref{fig:g_dependence} that
the $g$-dependence is at most about 10\% at $\kappa=0.1$, and even much
smaller than that at $\kappa=20$.  Note that what Fig.~\ref{fig:g_dependence} demonstrates
is that, when plotted in this way, our results are insensitive to increasing $g$ {\it at fixed} $\kappa$.
Increasing $g$ while holding $\kappa$ fix requires reducing $L$.    If, instead, we increase
$g$ at fixed $L$, this corresponds to increasing $\kappa$ --- which
(we see in Figs.~\ref{fig:Pg01} and \ref{fig:g_dependence} and below) has a marked effect
on our results.

The $\kappa$-dependence of $P(k_\perp)$ at small $k_\perp$ manifest in Fig.~\ref{fig:Pg01}
is interesting.  For very small $\kappa$, $P(k_\perp)\simeq P_{\rm thin}(k_\perp)$, which
diverges proportional to $1/k_\perp^2$ at small $k_\perp$, as in (\ref{AGZ}).  Plotting our results
at many more values of $\kappa$ than we have shown in Fig.~\ref{fig:Pg01} indicates
that $P(k_\perp)$ diverges as $k_\perp\to 0$ if $\kappa<2$ and is finite at $k_\perp=0$ for 
$\kappa\geq 2$, and indicates that $P(k_\perp)$ is linear in $k_\perp$ at small $k_\perp$
if $\kappa=3$ (as illustrated in Fig.~\ref{fig:Pg01}) and is quadratic in $k_\perp$
at small $k_\perp$ if $\kappa\geq 4$.
All these features of the behavior of $P(k_\perp)$ in the  $k_\perp\to 0$ limit
can be demonstrated analytically, via approximating $P_{\rm thin}(k_\perp)$ 
by $P_{\rm thin}^{\rm AGZ}(k_\perp)$ as is valid for $k_\perp\to 0$, and then
resumming $L$-enhanced diagrams.  We present this analysis in Appendix~\ref{app:AGZ}.
We expect that the $L$-resummed $P^{\rm AGZ}(k_\perp)$ will agree with
the full $P(k_\perp)$ at small $k_\perp$ because $P_{\rm thin}^{\rm AGZ}(k_\perp)$
agrees with $P_{\rm thin}(k_\perp)$ in this regime.  
The physical argument behind this expectation follows. 
For any value of $k_\perp$, resumming the $L$-enhanced
diagrams means taking into account the possibility that the hard parton
could pick up this $k_\perp$ via multiple scatterings, summing over the
infinite number of ways of adding up individual kicks that  yield $k_\perp$ in total. 
If we consider some $k_\perp$ that is small enough that
$P_{\rm thin}(k_\perp) \simeq P_{\rm thin}^{\rm AGZ}(k_\perp)$, then 
getting this $k_\perp$ via multiple kicks that each transfer 
momenta much larger than $k_\perp$
is  improbable.  
The $L$-resummation is therefore dominated by terms in which 
each of the multiple kicks transfers momenta that are comparable to or smaller than 
$k_\perp$, meaning that all the multiple kicks being resummed
are small enough that AGZ is a good approximation. We therefore
expect that, after resummation, $P(k_\perp) \simeq P^{\rm AGZ}(k_\perp)$.
We confirm this by explicit calculation in Appendix \ref{app:AGZ}.
From the analysis in Appendix \ref{app:AGZ} we then 
learn that as $k_\perp\to 0$, the probability distribution $P(k_\perp)$ includes
a term proportional to $k_\perp^2$ as well as a (possibly nonanalytic) term
proportional to $k_\perp^{\kappa-2}$, which dominates at small enough $k_\perp$
if $\kappa<4$.  This term explains the qualitative features that we have described above.

\begin{figure*}[t]
\includegraphics[scale=0.85]{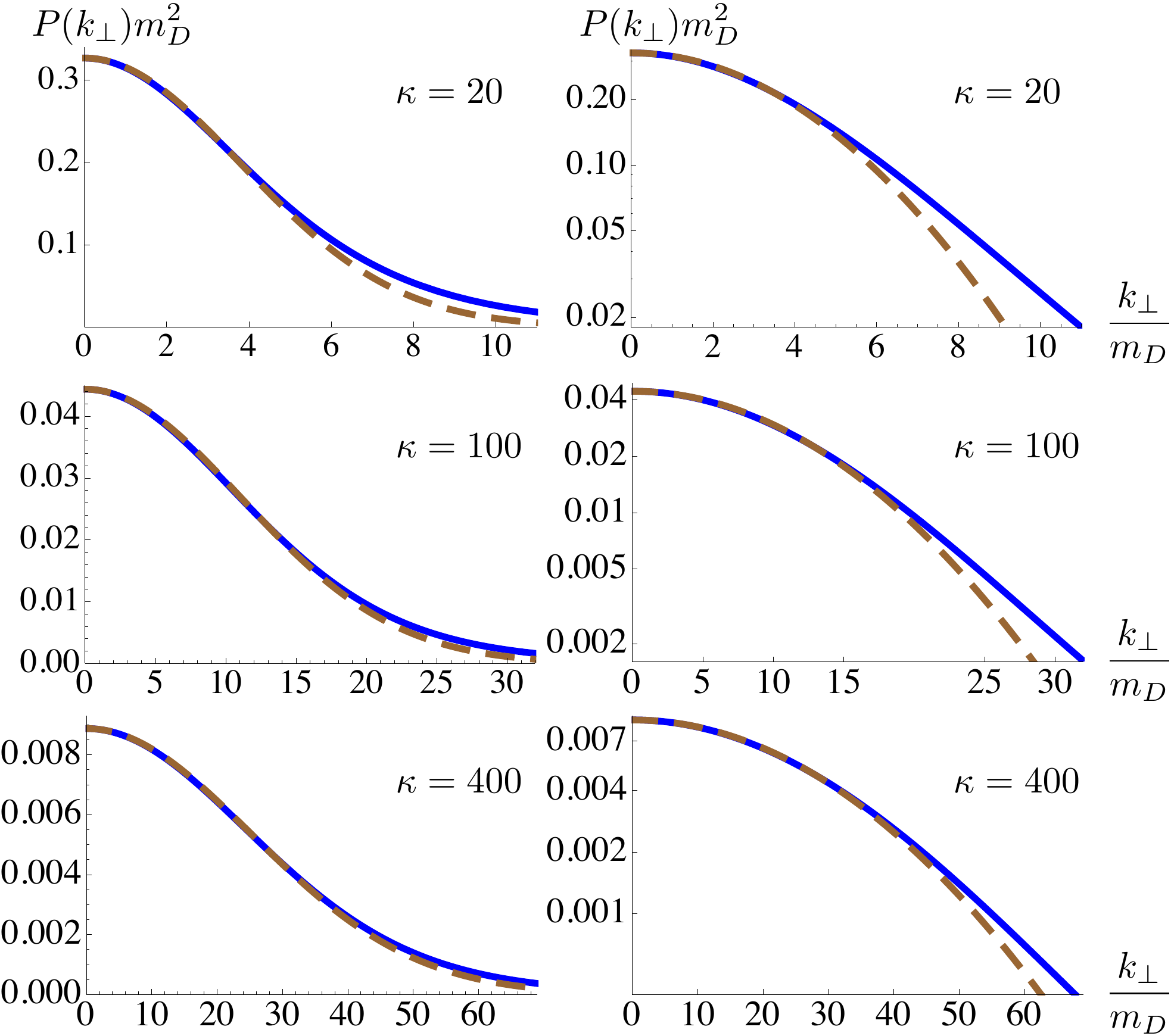}
\caption{The solid blue curves show
the probability distribution $P(k_\perp)$ describing momentum
broadening in a weakly coupled plasma with $g=0.1$ at three different
large values of $\kappa$ plotted on linear (left panels) and log (right panels) scales.
For each $\kappa$, with the brown dashed curves we show a Gaussian fit
of the form $P_{\rm fit}(k_\perp)=A \exp(-a\, k_\perp^2)$
 to $P(k_\perp)$ at low $k_\perp$.  We find $a \,m_D^2 = 0.0345$, $0.00410$ and $ 0.000792$ for 
 $\kappa=20$, $100$ and $400$. 
 We see from the figures that at larger $\kappa$ the 
 Gaussian fit is a good approximation to $P(k_\perp)$ out to larger $k_\perp/m_D$,
 out to larger $k_\perp \sqrt{a}$, and down to lower probabilities. This is the central
 limit theorem in action.  
 To quantify where the Gaussian approximation breaks down, 
 we introduce $k_\perp^G$, defined as the value of $k_\perp$ where the Gaussian fit 
 deviates from the full result by 2\%. 
 We find $k_\perp^G = 4.95 \, m_D=0.92/\sqrt{a}$, $k_\perp^G = 19.51 \, m_D = 1.25/\sqrt{a}$ and 
 $k_\perp^G = 51.44 \, m_D= 1.45/\sqrt{a}$ 
 for $\kappa=20$, 100
 and 400. For any $\kappa$ no matter how large, 
 if $k_\perp$ is large enough --- in particular if $k_\perp\gg k_\perp^G$ ---
 a power-law tail with $P(k_\perp)\propto 1/k_\perp^4$ rises above the Gaussian.}
\label{fig:LargeKappaGaussians}
\end{figure*}

We have seen that for $\kappa\geq 4$, the leading small-$k_\perp$ behavior of $P(k_\perp)$ 
is a constant minus a term quadratic in $k_\perp$.  This immediately makes one think of
a Gaussian.  And, indeed, there is every reason to expect that at large $\kappa$ we should
find a Gaussian probability distribution at small enough $k_\perp$:  large $\kappa$ means
a thick medium, meaning that the hard parton picks up its transverse momentum via the
sum of many kicks.  This means that in the large-$\kappa$ regime we expect that momentum
broadening can be thought of as diffusion in transverse momentum space, with
a Gaussian probability distribution $P(k_\perp)$ arising via the central limit theorem.
In Fig.~\ref{fig:LargeKappaGaussians}, we illustrate $P(k_\perp)$ for 
three large values of $\kappa$, showing that at small enough $k_\perp$ we do indeed
find a Gaussian probability distribution. For each $\kappa$, we 
fit a Gaussian of the form
\be
P_{\rm fit}(k_\perp)=A \exp(-a\, k_\perp^2) 
\label{GaussianFitForm}
\ee
to $P(k_\perp)$ at low $k_\perp$.
We determine $A$ by requiring that 
$P_{\rm fit}(0)=P(0)$ and then determine the width parameter $a$ by fitting
the quadratic dependence of
$\log P(k_\perp)$ around $k_\perp=0$, i.e. by fitting the brown dashed parabolas to the results 
shown as the solid blue curves in the right panels of Fig.~\ref{fig:LargeKappaGaussians}.
(Although $P(k_\perp)$ is a normalized probability distribution,
$P_{\rm fit}$ is not normalized since as we can see in Fig.~\ref{fig:LargeKappaGaussians}
it has less weight in its high-$k_\perp$ tail.) 
In order to gauge the range of $k_\perp$ 
out to which the probability distribution $P(k_\perp)$ is well approximated by the Gaussian
$P_{\rm fit}(k_\perp)$,
we define $k_\perp^G$ as the value of $k_\perp$ where the Gaussian fit function starts to deviate from the actual result by 2\%:
$P(k_\perp^G) - P_{\rm fit}(k_\perp^G) = 0.02\, {P(k_\perp^G)}$.
We see from Fig.~\ref{fig:LargeKappaGaussians} that the
larger $\kappa$ is, the larger is the $k_\perp^G$, both in units 
of $m_D$ and in units of $1/\sqrt{a}$.  More and more of the integrated probability
is found in the regime in which the probability distribution is Gaussian at larger
and larger $\kappa$.
This is the central limit theorem in action.

We can confirm the reliability of the Gaussian fit that we have done by comparing
our results for the Gaussian width parameter $a$ to the value of $a$ for
the 
resummed AGZ distribution $P^{\rm AGZ}(k_\perp)$
that we obtain in App.~\ref{app:AGZ}. 
As we have already emphasized, the resummed AGZ distribution is an excellent approximation to 
the full probability distribution at small $k_\perp$, which is also where we find Gaussian behavior. 
Resumming the AGZ distribution should therefore 
describe the physics well in the region where we perform the fit, with the additional benefit of 
making it possible to do the calculation almost completely analytically. As described in
detail in App.~\ref{app:AGZ}, we fit the Gaussian (\ref{GaussianFitForm}) 
to $P^{\rm AGZ}(k_\perp)$ and find an expression for the width parameter $a$:
\be
a\, m_D^2 = a^{\rm AGZ} \, m_D^2 \equiv  \frac{1}{4} \frac{\int_0^\infty dx \, x^{3 - \kappa} \exp\left[- \kappa \, K_0(x) \right]}{\int_0^\infty dx \, x^{1 - \kappa} \exp\left[- \kappa \, K_0(x) \right]} \ ,
\label{eq:amD2}
\ee 
with $K_0$ the modified Bessel function of the second kind.
The values of $a$ that we obtain from this expression with $\kappa$ set to 20, 100 and 400,
namely $a^{\rm AGZ} \,m_D^2=0.0362$, 0.00410 and 0.000763,
are all within a few percent of those that we have obtained by fitting to our full results
in Fig.~\ref{fig:LargeKappaGaussians}.
Naive application of 
the central limit theorem would suggest that the width
of the Gaussian, $\propto 1/\sqrt{a}$, should increase like $\sqrt{L}$ and hence 
at fixed $g$ it should increase with increasing $\kappa$  
like $\sqrt{\kappa}$.  
Instead we find both from our full
results and from (\ref{eq:amD2}) that $1/a$ grows slightly faster than linearly with $\kappa$, 
apparently
including a $\kappa \log \kappa$ term as well as a term proportional to $\kappa$.

We have also investigated the $g$-dependence of $a$ by repeating
the analysis shown in Fig.~\ref{fig:LargeKappaGaussians} at $g=1$, instead of $g=0.1$ as
in the Figure.  In so doing we find that $a$ decreases by factors of $27000$, $19000$
and $16700$ if we hold $L$ fixed, increasing $\kappa$ from the 20, 100 and 400 in
the Figure to 2000, 10000 and 40000.  
If instead we hold $\kappa$ fixed and increase $g$, we find that to a good
approximation $1/a$ increases like $g^2$, as in (\ref{eq:amD2}).  Our results
for the width parameter $a$ can therefore be summarized by 
writing $1/a = m_D^2 f(\kappa)$ for small values of $g$, with $f(\kappa)$ a function
that grows slightly faster than linearly with $\kappa$ at large $\kappa$.
This means that $1/a$ grows slightly faster than like $g^4$ with increasing $g$ at fixed $L$.

To this point we have focussed on the Gaussian behavior illustrated in 
Fig.~\ref{fig:LargeKappaGaussians} but it is just as important to see in
these plots that
for any $\kappa$, no matter how large,  at large enough $k_\perp$
you see the $P(k_\perp)\propto 1/k_\perp^4$ behavior (\ref{ArnoldDogan}), 
which is the correct form for $P(k_\perp)$ in the asymptotic ultraviolet $k_\perp\to\infty$ 
limit.  This apparent failure of the central limit theorem arises because the
underlying probability distribution for a thin medium
$P_{\rm thin}(k_\perp)$ has a ``fat tail'': its power-law fall-off at large $k_\perp$ ensures that
no matter how many scatterings are added up by resumming $L$-enhanced diagrams,
i.e. no matter how thick the medium is, the behavior of $P(k_\perp)$ at  large $k_\perp$
does not become Gaussian, it remains power-law.  The physics behind this is that 
no matter how thick the medium there is a $k_\perp$ that is so large that the most probable
way of picking up this much transverse momentum is via a single hard scattering.
The ``fat tail'' of the $P_{\rm thin}(k_\perp)$ distribution ensures that,
for large enough $k_\perp$, although such
single hard scatterings are rare they are more probable than picking up
such a large $k_\perp$ from multiple scatterings. 

There are many previous analyses of momentum
broadening via multiple scattering
in a weakly coupled plasma in the literature in which approximations are
made that result in a Gaussian form for $P(k_\perp)$.  (See 
Refs.~\cite{Wiedemann:1999fq,Gyulassy:2002yv,Majumder:2007hx,Liang:2008vz,Ovanesyan:2011xy}, for example,
and see Appendix~\ref{app:Boltz} for further discussion.) 
It is important to realize that altehough at large $\kappa$ 
this is the correct form for ``most of the probability'', i.e. for a region in $k_\perp$ that
contributes the lion's share of the normalization (\ref{eq:ProbabilityNormalization}) 
of the probability density $P(k_\perp)$, at large enough $k_\perp$ the probability density
is not Gaussian but rather a power law. And,
in the $Q\rightarrow\infty$ limit  (high jet-energy limit) in which the calculational framework
within which we are working is controlled it is the large-$k_\perp$ power-law region
of $P(k_\perp)$ that controls the jet quenching parameter $\hat q \propto \langle k_\perp^2 \rangle$,
as we shall discuss in the next subsection.  We shall see there that there
is nevertheless a sense in which $1/(a L)$ can be thought of as a sort of ``soft jet quenching parameter''.
More generally, it is interesting to ask how the width of the Gaussian component of $P(k_\perp)$
manifests itself in aspects of the phenomenology of jet quenching other than
momentum broadening, even in the $Q\rightarrow\infty$
limit.  However, $1/(a L)$ is {\it not} the jet quenching parameter. And, in fact, we shall see that
in the $Q\rightarrow\infty$ limit any dependence of $\hat q$ on $a$ is both subleading
and implicit.


\subsection{Jet quenching parameter}

In this subsection we shall discuss the implications of our results for the jet quenching parameter $\hat q$, defined as 
\beq
\hat q \equiv \frac{\langle k_\perp^2 \rangle}{L} = \frac{1}{L} \int \frac{d^2 k_\perp}{(2\pi)^2} k_\perp^2 P(k_\perp)\ .
\label{qhatdef}
\eeq
Recall that  in the large $k_\perp$ limit,
the resummation discussed in Sec.~\ref{sec:resumming} becomes 
unnecessary, and we recover the limiting behavior
of $P_{\rm thin}(k_\perp)$ given in~\r{ArnoldDogan}. 
We immediately see that 
the ``fat tail'' of the probability distribution, $P(k_\perp)\propto 1/k_\perp^{4}$
at large $k_\perp$,  makes $\hat q$ defined in (\ref{qhatdef}) 
logarithmically divergent 
in the ultraviolet.
Hence, the jet quenching parameter $\hat q$ is not well-defined in a weakly coupled plasma.
This UV divergence has been noted by many authors before us, for example
in Refs.~\cite{Arnold:2008vd,Arnold:2008zu,CaronHuot:2008ni,Majumder:2009cf}.
Nevertheless, the quantity $\hat q$ enters in many calculations of parton
energy loss, even though its definition is based entirely upon momentum
broadening, and we therefore want to compare our results for this
quantity to those in the literature.
To that end, we shall follow standard practice in much of the literature
and regulate the integral in (\ref{qhatdef}) with an ultraviolet cutoff $\Lambda_{UV}$,
which is usually thought of as being a kinematic cutoff of order
$\Lambda_{\rm UV} \sim (Q T)^{1/2}$, with $Q$ the energy of the hard parton.
The thinking behind this conventional choice
is that $\Lambda_{\rm UV}$ should be of order the maximum $k_\perp$
that the hard parton of energy $Q$ can pick up via a single scattering from
a gluon in the medium with momentum of order $T$.
More sophisticated, perhaps process-dependent, 
approaches are also possible~\cite{Arnold:2008zu,Arnold:2008iy,CaronHuot:2008ni,Peshier:2008zz}. 
In the $Q\rightarrow \infty$ limit in which the calculational framework within 
which we are working is controlled, $\Lambda_{\rm UV}$ is much greater than
the $k_\perp^G$ below which $P(k_\perp)$ is Gaussian and it must be in the regime
in which $P(k_\perp)\propto 1/k_\perp^4$.  We shall take this as given initially,
but we shall later consider the possibility that $\Lambda_{\rm UV}$ may
not always be so large for experimentally realizable values of $Q$.

Although our purpose in this subsection is to compare our results for $\hat q$
to those in the literature, we note here that for many purposes 
$\hat q$ may not be the most relevant
parameter with which to characterize $P(k_\perp)$ in a weakly coupled
plasma.  
Because $\hat q$ is ultraviolet-divergent in a weakly coupled plasma,
in this setting it is determined by
$P(k_\perp)$ at asymptotically large $k_\perp$,
in a regime that contributes almost negligibly to the integrated
probability.  We have seen in Fig.~\ref{fig:LargeKappaGaussians} that
although $P(k_\perp)$ has a ``fat tail'' that controls $\hat q$,
most of the probability comes instead from the Gaussian region
at lower $k_\perp$.   So, in the case of a weakly coupled plasma
the jet quenching parameter $\hat q$ describes only the tail, not
the dog itself.
In marked contrast,  we shall see in the next subsection that in a 
plasma that is strongly coupled at
all length scales,
$P(k_\perp)$ is Gaussian at all momentum scales,
and the jet quenching parameter 
$\hat q$
is finite, well-defined, and
 is the only parameter needed in
order to characterize the entire probability 
distribution $P(k_\perp)$~\cite{Liu:2006ug, D'Eramo:2010ak}.

We turn now to the calculation of $\hat q$, with its ultraviolet divergence
regulated in the conventional way.  
In Appendix~\ref{app:qhat} we show that for a probability distribution 
of the form~(\ref{Pdef}) the jet quenching parameter $\hat{q}$ takes the form
\be
\hat q = -\frac{1}{L} \left. \nabla^2 \mathcal{W}^{(2)}_{\mathcal{R}} \right |_{x_\perp = 0} \ ,
\label{qhatnabla2}
\ee
where $\nabla^2$ is the Laplace operator in the transverse plane. 
Before proceeding to apply (\ref{qhatnabla2}), we can make a very
general point.  If we were able to push our calculation through to all orders in perturbation
theory, $\mathcal{W}^{(2)}_{\mathcal{R}}$ in (\ref{Pdef}) would be replaced
by the sum of connected diagrams in the exponent in (\ref{exPne}), which we can
denote $\mathcal{W}^{(c)}_{\mathcal{R}}$. The calculation in Appendix~\ref{app:qhat}
then still goes through, meaning that (\ref{qhatnabla2}) becomes
\be
\hat q = -\frac{1}{L} \left. \nabla^2 \mathcal{W}^{(c)}_{\mathcal{R}} \right |_{x_\perp = 0} \ .
\label{qhatnabla2conn}
\ee
As we have discussed in Section~\ref{sec:resumming}, the contribution of each
connected diagram to $\mathcal{W}^{(c)}_{\mathcal{R}}$ is proportional to $L$.
We then see that (\ref{qhatnabla2conn}) constitutes
a proof that $\hat q$ is independent of $L$ and, equivalently, of $\kappa$
to all orders in perturbation theory.
So, $\hat q$ can only depend on $g$ and $T$ as
well as on $\Lambda_{\rm UV}$.

We now apply (\ref{qhatnabla2}) to our weak-coupling result (\ref{eq:W2Fourier}) for 
$\mathcal{W}^{(2)}_{\mathcal{R}}$ and find 
\be
\hat q = \frac{1}{L} \int \frac{d^2 k_\perp}{(2\pi)^2} k_\perp^2 P_{\rm thin}(k_\perp) \ ,
\label{qhatdef2}
\ee
which is divergent, as we expect.  Even before we regulate the divergence,
we notice the  interesting result that $\hat q$ is the same 
as it would be if the probability distribution for momentum broadening were just
$P_{\rm thin}(k_\perp)$, instead of $P(k_\perp)$.  We have seen in the previous
subsection that the resummation of length-enhanced
diagrams completely changes the shape of $P(k_\perp)$, but
we now see that in a weakly coupled plasma this resummation has no effect
on $\hat q$. This makes sense
since in a weakly coupled plasma
$\langle k_\perp^2 \rangle$  is controlled by the ultraviolet power-law behavior
of $P(k_\perp)$ which is unaffected by the resummation.
In order to actually use the expression (\ref{qhatdef2}), we 
regulate it by introducing an ultraviolet cutoff $\Lambda_{\rm UV}$, as discussed above. 
(As long as $\Lambda_{\rm UV}$ is large enough, introducing the ultraviolet cutoff already in (\ref{qhatdef}) would have been equivalent.)

Because we are assuming 
$\Lambda_{\rm UV} \gg k_\perp^G$, we can present our
results in a semi-analytical form.  
First we divide the 
integral over the magnitude of $k_\perp$ in the expression
(\ref{qhatdef2}) for  $\hat{q}$ into two parts, 
one from 0 to $k'_\perp$ and the other from $k'_\perp$ to $\Lambda_{\rm UV}$,
with $k'_{\perp}$ arbitrary except that it must
satisfy $k'_\perp \ll \Lambda_{\rm UV}$, $k_\perp'\gg k_\perp^G$ and $k_\perp' \gg T$.  We denote the two contributions to $\hat q$
by
\be
\hat q = \hat q_{{\rm IR}}+ \hat q_{{\rm UV}}\ .
\ee
Next, we observe that in the
ultraviolet tail of $P_{\rm thin}(k_\perp)$
the retarded propagator in Eq.~\r{eq:DRasPi} is well approximated by the first order solution to the Dyson equation~\r{eq:Dyson}. This means that  the Wightman propagator $D^>$, expressed in terms of the retarded propagator as in~\r{D21def}, takes the form
\beq
D^{>} (q_0,k_\perp) =  \left(1 + f(q_0)\right)  \frac{{\rm Im} \( \Pi_{R}^L - \Pi_{R}^T \)}{k_\perp^2 \left(k_\perp^2 + q_0^2\right)} \ .
\eeq
We then define the dimensionless coefficient
\be
b \equiv  \int dq_0 \frac{\left(1+f(q_0)\right)}{2 \pi^2 T^3} \frac{k_\perp^2}{\left(k_\perp^2 + q_0^2\right)}
\frac{{\rm Im}\( \Pi_{R}^L - \Pi_{R}^T \)}{g^2} \ .
\label{bdef}
\ee
Given the behavior of the imaginary part of the self-energy in the UV limit, 
the constant $b$ does not depend on either $k_\perp$ or $g$.
We have defined $b$ so as to allow us to write a very compact expression
for the probability distribution $P_{\rm thin}(k_\perp)$ that is valid only in
the ultraviolet limit,
\be
P^{\rm UV}_{\rm thin}(k_\perp) = 2\pi\,  b\, g^4 C_R   \frac{L T^3}{k_\perp^4} \ ,
\ee
and from this we determine that the UV contribution to the jet quenching parameter
is given by
\beq
\hat q_{{\rm UV}} = b\, g^4  C_R\, T^3  \, \int_{k^\prime_\perp}^\Lambda \frac{d k_\perp}{k_\perp} = b\, g^4  C_R \, T^3  \log\frac{\Lambda}{k^\prime_\perp} \ .
\label{eq:qUV}
\eeq
The arbitrary scale $k^\prime_\perp$ must be large enough that
we can safely apply all the UV approximations just discussed. 
From its definition in Eq.~\r{bdef}, we can calculate the value of $b$ explicitly. We find
\be
b = 0.2035 \ .
\label{eq:bus}
\ee
To check our results against those
in the literature, we identify the factor corresponding to $b$ in the result (\ref{ArnoldDogan})
from Ref.~\cite{Arnold:2008zu}, finding
\beq
b^{\rm AD} = \frac{1}{2\pi} \frac{7 \, \zeta(3)}{4 \, \zeta(2)} = 0.2035 \ .
\eeq
Our expression for $\hat q_{{\rm UV}}$ is therefore in 
excellent agreement with that obtained  by Arnold and Dogan.

\begin{figure}[t]
\begin{center}
\includegraphics[scale=0.95]{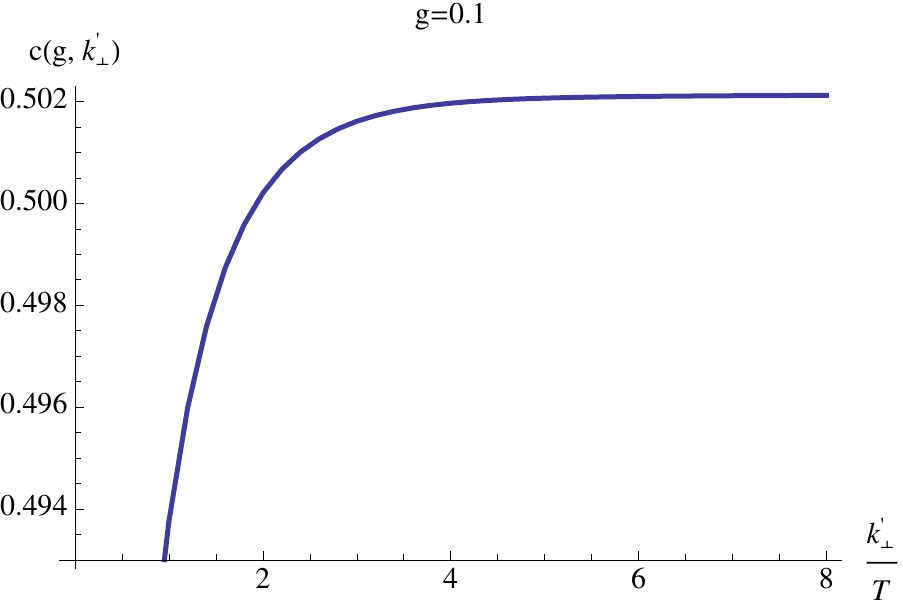}
\caption{The function $c(g, k_\perp^\prime)$ as a function of $k^\prime_\perp$, with the gauge coupling set to $g=0.1$. For large enough $k^\prime_\perp$, $c(g, k_\perp^\prime)$ becomes constant.}
\label{fig:c}
\end{center}
\end{figure}

Turning now to the IR, we have
\beq
\hat q_{{\rm IR}} = \frac{1}{2\pi L} \int_0^{k^\prime_\perp} dk_\perp k_\perp^3 P_{\rm thin}(k_\perp) \ .
\label{eq:qIR}
\eeq
Since we chose  $k^\prime_\perp$  to be well into the UV region, 
the integral in~\r{eq:qIR} 
must depend logarithmically on
$k^\prime_\perp$. 
To isolate this known logarithmic behavior, we 
define the function $c(g, k_\perp^\prime)$ via the expression
\beq
\hat q_{{\rm IR}} = g^4 C_R T^3 \[ c(g, k_\perp^\prime) + b \log \frac{k^\prime_\perp}{T} \] \ ,
\label{eq:qIR2}
\eeq
in which $\hat q_{\rm IR}$ is to be evaluated using (\ref{eq:qIR}) and which therefore serves
to define $c$.
The function $c(g, k_\perp^\prime)$ is plotted in Fig.~\ref{fig:c} as a function of $k_\perp^\prime$ for $g=0.1$. We see that for large enough $k_\perp^\prime$, 
the function $c(g, k_\perp^\prime)$ 
becomes independent of $k_\perp^\prime$.
And, we anyway needed to choose $k_\perp^\prime \gg T$ in order to control the behavior
of $\hat q_{\rm UV}$.  Upon making this choice,
$c(g, k_\perp^\prime) \simeq c(g)$.  The only way that 
properties of the probability distribution $P_{\rm thin}(k_\perp)$ 
in the IR, and indeed in any region of $k_\perp$ except the UV,
enter into the calculation of $\hat q$ is through
the function $c(g)$. In some implicit way, $c(g)$ is related to 
the properties of the $P(k_\perp)$ distribution that we discussed
in Section~\ref{subsec:5B}, like for example the width $1/\sqrt{a}$ 
of its Gaussian component
at large $\kappa$. Notice, however, that $c(g)$ appears only in a subleading
contribution to $\hat q$; the dominant contribution comes from the logarithmic
UV divergence.
We have computed $c(g)$ for several different $g$ values, obtaining
\beq
c(g) = \left \{ 
	\begin{split}
		1.0363 &\text{ for } g=0.01 \\
		0.5023 &\text{ for } g=0.1 \\
		0.0368 &\text{ for } g=1 \\
		-0.0795 &\text{ for } g=2 
	\end{split} \right. \ .
	\label{eq:c}
\eeq

Finally, we combine the two contributions $\hat q_{{\rm UV}}$ in Eq.~({\ref{eq:qUV}}) 
and $\hat q_{{\rm IR}}$ in~Eq.~({\ref{eq:qIR2}}). Summing them, we find
\beq
\hat q = g^4 C_R T^3 \( b \log\frac{\Lambda_{\rm UV}}{T} + c(g) \) \ ,
\label{qhatpar}
\eeq
where the dependence on the arbitrarily chosen $k'_\perp$ has dropped out, as it must.
Thus, at the end of the day we find that the jet quenching parameter is specified
by the constant $b$ and the  function of the coupling constant $c(g)$, as well as
$T$, $\Lambda_{\rm UV}$ and $g$ itself.
Our results for $b$ and $c(g)$ are given in~\r{eq:bus} and ~\r{eq:c}, respectively. 
As a comparison, the AGZ distribution in~\r{AGZ} gives 
\beq
\hat q^{\rm AGZ} =  g^4 C_R T^3 \frac{3}{4\pi} \log \frac{\Lambda_{\rm UV}}{m_D} \ ,
\eeq
from which we can read off the coefficients 
\be
b^{\rm AGZ} = \frac{3}{4\pi} \ , \qquad c^{\rm AGZ}(g) = -\frac{3}{4\pi} \log\(g \sqrt\frac{3}{2} \) \ .
\ee
The coefficients $b$ and $b^{\rm AGZ}$ differ by $\approx 15\%$.
This is the discrepancy between the ultraviolet tails of $P_{\rm thin}(k_\perp)$ and 
$P_{\rm thin}^{\rm AGZ}(k_\perp)$ (recall that $P^{\rm AGZ}_{\rm thin}(k_\perp)$ is
not valid in the ultraviolet) that we have already
seen in
Fig.~\ref{fig:P21g01}.
The values of $c(g)$ and $c^{\rm AGZ}(g)$ are in good agreement for small values of $g$, and differ more for larger values of $g$, where $P_{\rm thin}(k_\perp)$ starts to 
differ from $P_{\rm thin}^{\rm AGZ}(k_\perp)$.

We can even plug in explicit values for the parameters in a range that
is reasonable for jets produced in heavy ion collisions at the LHC. 
We pick the benchmark values $T = 300 \, {\rm MeV}$ and $L = 5 \, {\rm fm}$ for the plasma temperature and thickness, respectively.
We choose the UV cutoff to be $\Lambda_{\rm UV} = 17 \, {\rm GeV}$. (This can
be thought of as the kinematic limit for a 300 GeV parton scattering off a 1 GeV parton from the
medium.)
We consider the momentum
broadening of a hard gluon, meaning that we set $C_{\cal R}=3$,
and find
\beq
\hat q = 0.411 \,g^4 (0.822 + c(g)) \,  {\rm GeV}^2\, {\rm fm}^{-1} \ .
\label{qhatFinalResult}
\eeq
If we then choose $g=2$, corresponding to $\alpha_{\rm QCD}=0.32$, this would correspond to a jet quenching parameter $\hat q \simeq 4.9  \,  {\rm GeV}^2 {\rm fm}^{-1}$.
With these choices of parameters, $\kappa\simeq 14.5$, large enough that the resummation
of length-enhanced diagrams is certainly necessary in order to obtain $P(k_\perp)$.  With
$\kappa\simeq 14.5$, extrapolating from Fig.~\ref{fig:LargeKappaGaussians} 
we estimate $k_\perp^G\sim 4 m_D$, meaning that with the parameters we have chosen it
is around 3 GeV.  So, our assumption that $\Lambda_{\rm UV}\gg k_\perp^G$ is reasonable,
as is our calculation in which $\hat q$ is dominated by the tail of the probability
distribution $P(k_\perp)$.


Although the calculation breaks down if we do so, it is also interesting to speculate
as to how its results differ if we attempt to use parameters more appropriate for the momentum
broadening of hard partons produced
in heavy ion collisions at RHIC.  In the RHIC context, $\Lambda_{\rm UV}$ must be much smaller,
first because the hard partons only have $Q\sim 30-40$ GeV and second because the temperature
is somewhat lower.   Likely, $\Lambda_{\rm UV}$ cannot be much more than 5 GeV.
Also, the relevant values of $g$ must be somewhat larger at RHIC than at the LHC. With
$\kappa$ increasing like $g^2$, according to Fig.~\ref{fig:LargeKappaGaussians} this
will result in a larger $k_\perp^G$.  This means that the assumption 
that $\Lambda_{\rm UV}\gg k_\perp^G$ has broken down, as has our
calculation of (\ref{qhatFinalResult}).  Of course, the whole calculational framework
is breaking down too, for two independent reasons: 
$g$ is becoming uncomfortably large 
and we can no longer trust the $Q\rightarrow\infty$
limit as a guide.

Motivated both by jet quenching at RHIC and by the analysis of the strongly
coupled plasma that we shall discuss in the next subsection,
it is worth asking what happens if $\Lambda_{\rm UV}<k_\perp^G$ even though
as we have just described our weakly coupled calculation is not at all under control
in this regime.
If we nevertheless apply our results, when $\Lambda_{\rm UV}$ lies within the
regime in which $P(k_\perp)$ is well fit by a Gaussian of the form (\ref{GaussianFitForm})
then the power-law tail of the probability distribution is irrelevant in the calculation
of the jet quenching parameter, since the integral in (\ref{qhatdef2}) is cutoff before the
power-law tail makes its appearance.  $\hat q$ is determined entirely from the
Gaussian region,  and is given in terms of the width $1/\sqrt{a}$ of the Gaussian
(\ref{GaussianFitForm}) by 
\be
\hat q_{\rm soft} = \frac{1}{a L} \ , 
\ee
where we have introduced the subscript ``soft'' to remind ourselves that this result
is only valid if $\Lambda_{\rm UV}<k_\perp^G$ and in particular is
not valid in the $Q\rightarrow \infty$ limit where the full
calculation is controlled.
Even though $\hat q_{\rm soft}$ is not the actual jet quenching parameter $\hat q$ (which
is defined in the $Q\rightarrow\infty$ limit and is given by (\ref{qhatpar})) since $\hat q_{\rm soft}$
is determined by the width of the Gaussian that describes the lion's share of the probability
in $P(k_\perp)$ rather than by the power-law tail it could certainly turn out
that $\hat q_{\rm soft}$
is more relevant to the phenomenology of jet quenching than $\hat q$ itself.  Our attempt to
plug numbers into our results suggests that this circumstance is more likely to arise
for jets produced in heavy ion collisions at RHIC and is less likely to arise in the case
of the highest energy jets produced in heavy ion collisions at the LHC.


\subsection{From weak to strong coupling}

We have calculated the probability distribution $P(k_\perp)$ for momentum
broadening in a weakly coupled quark-gluon plasma in QCD.  
We anticipate that our calculation is only quantitatively reliable for $g<1$.  For
example, we argued in Section~\ref{subsec:5B} based upon evidence visible
in  Fig.~\ref{fig:P21g01} that the matching
that we describe in Section IV is not quantitatively reliable at $g=1$ and $g=2$,
although it is gratifying that the small kinks in the $g=1$ and $g=2$ curves in 
Fig.~\ref{fig:P21g01}
have no visible effects in Fig.~\ref{W2_g01}, and therefore no visible effects on
our results, as plotted in 
Figs.~\ref{fig:Pg01}, \ref{fig:g_dependence} and \ref{fig:LargeKappaGaussians}.
Although it is important on theoretical grounds that we have control of the calculation
for $g<1$, this is of little phenomenological interest.  The smallest values of $g$ that
are typically used in comparisons to data from 
heavy ion collisions at RHIC and the LHC are around $g\sim 2$.
(Note that $g=2$ corresponds to $\alpha_{\rm QCD}\simeq 0.32$, 
in many other contexts a weak coupling.)
There are other good reasons beyond the large value of $g$ not to trust
a weakly coupled description of the plasma produced in 
heavy ion collisions at RHIC and the LHC, chief among them being the 
suite of evidence that this plasma is a strongly coupled liquid with a 
shear viscosity that is so small that no description in terms of weakly interacting
quark and gluon quasiparticles can be self-consistent.
That said, it is nevertheless natural to ask, at least at a qualitative level, what
our results suggest for $P(k_\perp)$ at $g\sim 2$ and beyond.
At a qualitative level, the answer is provided by Fig.~\ref{fig:LargeKappaGaussians}.
By far the most important consequence of increasing $g$ at fixed $L$ is the
increase in $\kappa$, and we see from Fig.~\ref{fig:LargeKappaGaussians}
that increasing $\kappa$ makes the probability distribution $P(k_\perp)$ for momentum
broadening ``more and more Gaussian.''  That is, $P(k_\perp)$ is well-approximated 
as Gaussian out to larger and larger $k_\perp$ and for 
a larger and larger fraction of the total probability, pushing the appearance of
the power-law behavior that must be present at asymptotic $k_\perp$ out to
larger and larger $k_\perp$.

The qualitative expectations for the behavior of $P(k_\perp)$ at strong coupling 
that we have gleaned by looking at how our results behave at large $\kappa$
are nicely borne out in the strongly coupled plasma of ${\cal N}=4$ supersymmetric
Yang-Mills (SYM) theory.  Taking advantage of the fact that this
theory, at both zero and nonzero temperature, has a dual 
gravitational description~\cite{AdS/CFT,Rey:1998ik},
holographic calculations of many aspects of its plasma phase
have been used to gain varied insights into the physics
of strongly coupled plasma more generally~\cite{CasalderreySolana:2011us}.
Of interest to us, the
expectation value of the Wilson loop in Eq.~\r{eq:Wseries},
and from it the probability distribution for momentum broadening,
have been calculated for ${\cal N}=4$ SYM theory in the strong coupling
and large-$N_c$ limit~\cite{Liu:2006ug, D'Eramo:2010ak}. 
For a propagating gluon in the adjoint representation, the result reads
\beq
P^{\rm SYM}(k_\perp) = \frac{4 \tilde{a}}{\pi \sqrt{\lambda}\, T^3 L} \exp \left[-\frac{\tilde{a}\, k_\perp^2}{\pi^2 \sqrt{\lambda}\, T^3 L } \right] \ ,
\label{ads}
\eeq
where $\lambda \equiv g^2 N_c$ is the 't Hooft coupling, assumed large, and where
$\tilde{a} \equiv \sqrt{\pi} \, \Gamma(5/4)/\Gamma(3/4) \approx 1.311$.
We see immediately that in this theory, whose plasma is strongly coupled
at all scales, $P(k_\perp)$ is Gaussian out to arbitrarily large $k_\perp$, which
is consistent at a qualitative level with the expectations derived from extending
our calculation for a weakly coupled plasma to larger values of $g$ and hence $\kappa$, as
we have described above.  
At a qualitative level, the lesson from (\ref{ads}) is that in a strongly coupled plasma
momentum broadening should be thought of as diffusion in transverse momentum
space, even though in the calculation behind (\ref{ads}) 
there is
no thin-medium regime,
no analogue of starting with some $P_{\rm thin}$ and resumming
length-enhanced diagrams, and hence no picture of multiple scattering
off quasiparticles building up 
a Gaussian $P(k_\perp)$ via the central limit theorem.  In a strongly coupled plasma,
$P(k_\perp)$ is always Gaussian at any $k_\perp$, for any $L$.  
At a qualitative level, such a regime can be approached starting from a weakly
coupled plasma either by increasing the coupling or by increasing $L$, either of
which corresponds to increasing $\kappa$.


Although the qualitative picture that we have just sketched is 
pleasing, it is important to note that at a quantitative level the Gaussian
probability distribution (\ref{ads}) is quite different from the Gaussian that we obtained by fitting
to our results for a weakly coupled plasma at large $\kappa$ in Fig.~\ref{fig:LargeKappaGaussians}.
In particular, the width of the Gaussian in (\ref{ads}) increases with increasing $g$ only
like $\lambda^{1/4}\sim \sqrt{g}$ while we found in
Section~\ref{subsec:5B} that the width $1/\sqrt{a}$ of the Gaussians
in Fig.~\ref{fig:LargeKappaGaussians} increase with increasing $g$ somewhat faster
than $g^2$.

The result quoted in Eq.~(\ref{ads}) was derived in the $\lambda \rightarrow \infty$ limit. For large but finite 't Hooft coupling $\lambda$, we cannot conclude that $P(k_\perp)$ will be Gaussian at all scales. 
It is reasonable to expect that in this case like at weak coupling
$P(k_\perp)$ will be Gaussian only for $k_\perp$
less than some $k_\perp^G$, with  $k_\perp^G \rightarrow \infty$ as $\lambda\rightarrow\infty$ and
consequently $\kappa \rightarrow\infty$.  The form of $P(k_\perp)$ for 
$k_\perp > k_\perp^G$, where the Gaussian description breaks down, is
not known. However, since there are no quasiparticles off which
hard scattering can occur in the ${\cal N}=4$ SYM theory plasma with large but 
finite $\lambda$, we anticipate that even where $P(k_\perp)$  is not Gaussian 
it will continue to fall off more rapidly than any power of $k_\perp$ at large $k_\perp$.

\begin{figure}[t]
\includegraphics[scale=1.01]{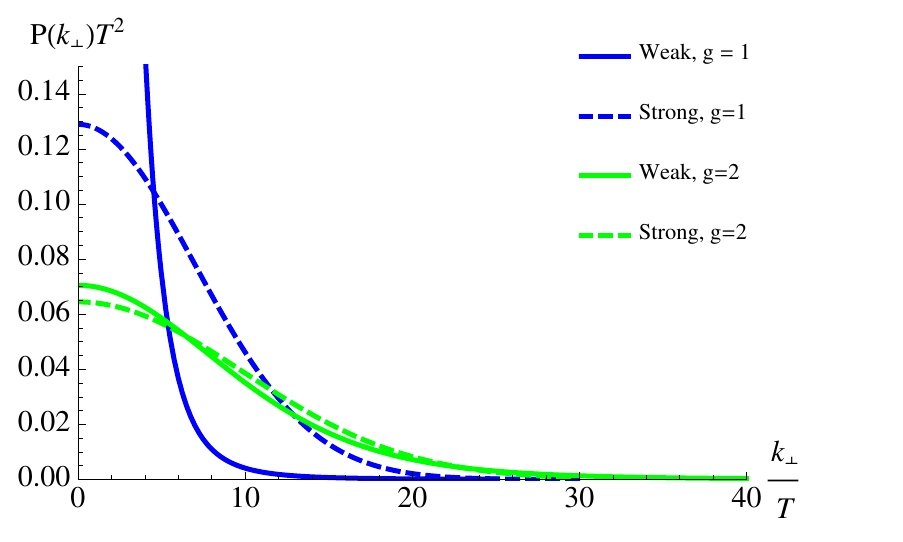}
\includegraphics[scale=1.01]{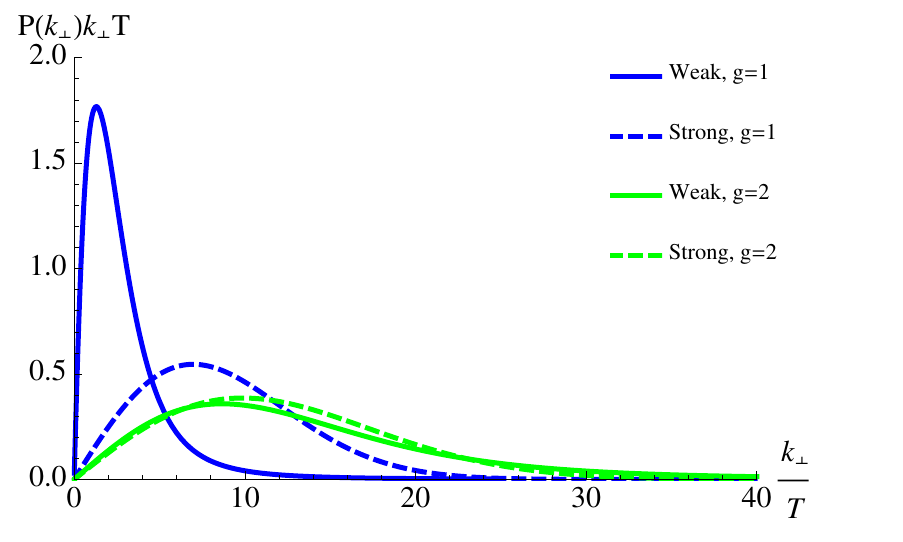}
\includegraphics[scale=1.14]{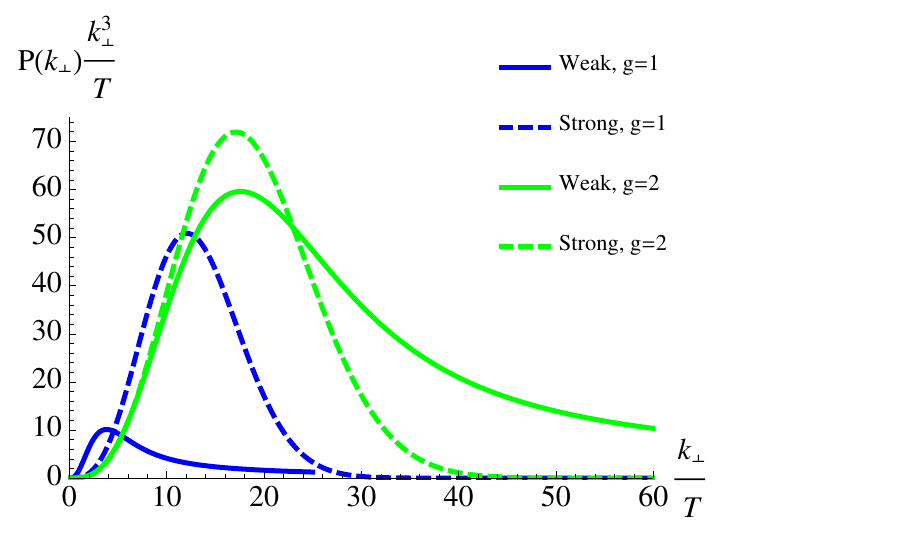}
\caption{Probability distributions for momentum broadening $P(k_\perp)$ 
for weakly coupled QCD plasma
with $g=1$ (continuous blue curve) and $g=2$ (continuous green) 
and $P^{\rm SYM}(k_\perp)$ for strongly coupled
${\cal N}=4$ SYM plasma with $N_c=3$ and $g=1$ (blue dashed) and $g=2$
(green dashed).
The curves are for propagation of a hard gluon  through a region of
plasma with thickness $L = 5$~fm, and temperature $T = 300$~MeV.}
\label{fig:ads}
\end{figure}

We illustrate both the qualitative and the quantitative comparisons that
we have just made between $P(k_\perp)$ in weakly coupled QCD plasma
and $P^{\rm SYM}(k_\perp)$ in strongly coupled ${\cal N}=4$ SYM plasma
in Fig.~\ref{fig:ads} by plotting both for $g=1$ and $g=2$.  In
QCD, $g=1$ and $2$ correspond to $\alpha_{\rm QCD}=0.08$ and 
$0.32$ while in ${\cal N}=4$ SYM theory with $N_c=3$,
these couplings correspond to $\lambda=3$ and $12$.  It is interesting
to note that $g=2$ is in a regime in which $\alpha_{\rm QCD}=0.32$ is considered
a weak coupling in some contexts and $\lambda=12$ is considered a strong
coupling in some contexts.  We perform the comparison for a plasma
with temperature $T = 300 \, {\rm MeV}$ that is $L=5\, {\rm fm}$ thick.
With these choices, $g=1$ and 2 correspond to  
$\kappa = 3.6$ and $\kappa = 14.5$. 
We see in Fig.~\ref{fig:ads} that upon increasing $g$ from 1 to 2, the width
of $P(k_\perp)$ for the weakly coupled QCD plasma increases much
more rapidly than the width of the Gaussian (\ref{ads}) for the strongly
coupled plasma does, as we have described above.
Perhaps the most striking aspect of Fig.~\ref{fig:ads} is just how similar
$P(k_\perp)$ with $g=2$ and $P^{\rm SYM}(k_\perp)$ with $g=2$ and hence 
$\lambda=12$ are,
in particular when plotted as in the top and middle panels.  This is an
indication that at this value of the coupling the strongly coupled and weakly
coupled perspectives yield comparable descriptions of momentum broadening
for low and moderate values of $k_\perp/T$.
We see in the bottom panel of Fig.~\ref{fig:ads}, however, that the physics of
the two descriptions is completely different at large-$k_\perp$, where we see
that the probability distribution $P(k_\perp)$ has a power-law tail only
for the weakly coupled plasma.


\begin{figure}[t]
\includegraphics[scale=1.03]{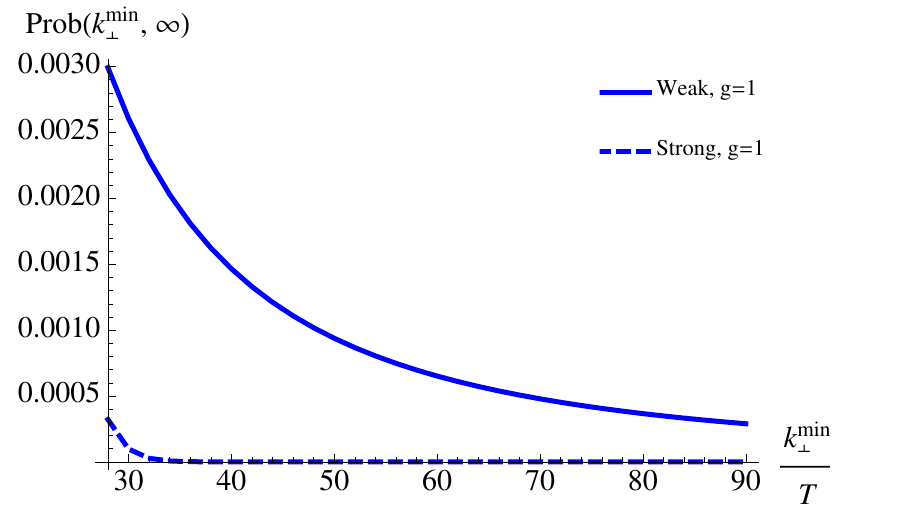}
\includegraphics[scale=1.03]{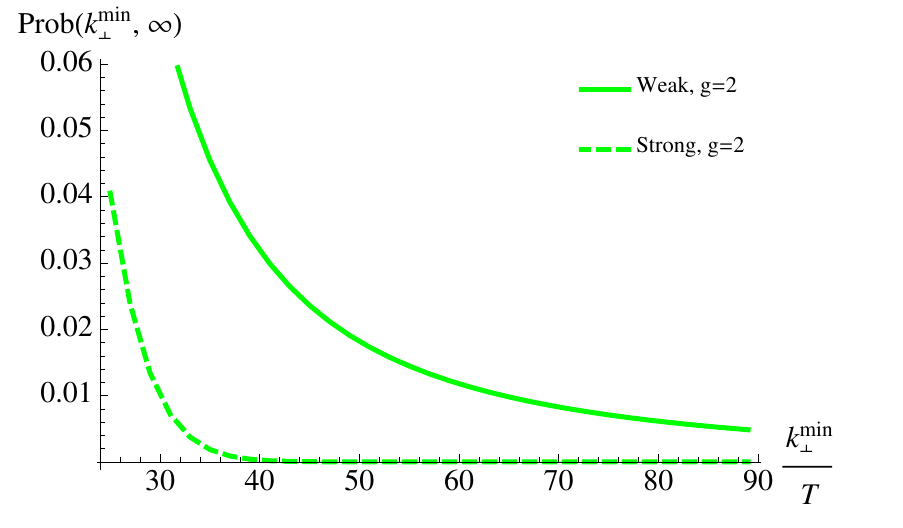}
\caption{Probability that a hard gluon receives 
a transverse momentum greater than $k_\perp^{{\rm min}}$ 
after propagating a distance $L$ through a weakly coupled QCD plasma or
a strongly coupled ${\cal N}=4$ SYM plasma with temperature $T$ and coupling constant
$g$.  Values of $g$, $T$ and $L$ as well as color conventions for the curves are all as in 
Fig.~\protect{\ref{fig:ads}}.}
\label{fig:adsprob}
\end{figure}

We further illustrate the sharp distinction between the behavior of $P(k_\perp)$ 
at large $k_\perp$ in weakly and strongly coupled plasma 
in Fig.~\ref{fig:adsprob}.  
Because the probability distribution for the plasma that is strongly coupled
at all scales is Gaussian whereas that for the weakly coupled plasma
is proportional to $1/k_\perp^4$ at large $k_\perp$, no matter how large
the coupling is there is always a $k_\perp$ beyond which $P(k_\perp)$
is greater in the weakly coupled plasma than in the strongly coupled plasma.
This behavior, which at first hearing may sound counterintuitive, reflects the presence
of point-like quasiparticles in the weakly coupled plasma.
This means that, as Rutherford could have understood, although the probability for
large-angle, large-$k_\perp$, scattering is always low it is much larger in a
plasma containing point-like scatterers than it would be
in a liquid plasma with no quasiparticles at any length-scale 
like the strongly coupled plasma
of ${\cal N}=4$ SYM theory. In
Fig.~\ref{fig:adsprob} we plot the integrated probability
that a hard parton propagating through $L=5$~fm of either the weakly coupled QCD plasma
or the strongly coupled ${\cal N}=4$ SYM plasma with temperature $T=300$~MeV
picks up a transverse momentum kick 
$k_\perp > k_\perp^{{\rm min}}$. 
As we can see, in the strongly coupled plasma with its Gaussian $P^{\rm SYM}(k_\perp)$,
this integrated probability is completely negligible for $k_\perp^{\rm min} \gtrsim 40 \,T$.
In stark contrast, if we assume a weakly coupled QCD plasma and then set $g=2$,
this integrated probability is still more than half a percent for $k_\perp^{\rm min} = 80\, T$.
So, although the two probability distributions are quite similar in the regime of $k_\perp$
which is probable --- indicating that momentum broadening for most partons would
be comparable in these two cases --- rare hard, large-angle, scatterings will be very much more
common if the weakly coupled QCD analysis yields a reasonable approximation.

If we evaluate our results for the momentum broadening of a hard quark rather than a hard
gluon, the conclusions of the above paragraph become even stronger.  
The widths of the Gaussian probability distributions $P^{\rm SYM}(k_\perp)$ describing the momentum
broadening of a hard quark in the strongly coupled ${\cal N}=4$ SYM plasma are half
as wide as those in Fig. \ref{fig:ads}~\cite{Liu:2006ug}, 
meaning that the dashed curves
plotted in Fig.~\ref{fig:adsprob} 
are pushed down so much that they are indistinguishable
from the horizontal axis across the whole range of $k_\perp$ in 
Fig.~\ref{fig:adsprob}.  At weak coupling, in the ultraviolet
$P(k_\perp)$ is proportional to $C_{\cal R}$, meaning that the solid curves in 
 Fig.~\ref{fig:adsprob} get multiplied by a factor of $4/9$ if one treats a hard quark instead
 of a hard gluon. So, in a weakly coupled
 QCD plasma with $g=2$ and $L$ and $T$ as in Fig.~\ref{fig:adsprob} 
  the integrated probability that a hard quark picks up $k_\perp^{\rm min}=30\,T~(60\,T)$ or more
  in transverse momentum is more than two percent  (about half a percent), while either of
  those integrated probabilities
  is completely negligible
   in the strongly coupled plasma with its Gaussian $P^{\rm SYM}(k_\perp)$.
   Because QCD is asymptotically free,  its strongly coupled
   liquid quark-gluon plasma must emerge from 
   weakly coupled quarks and gluons that can be resolved
   at short enough length scales. We therefore expect that
   for large enough $k_\perp$ our weakly coupled QCD analysis yields a reasonable
   approximation to $P(k_\perp)$, meaning that we expect that although large-angle scattering is rare it will be very much more common than it would be if the quark-gluon plasma were a strongly coupled liquid at all length scales.

\section{Outlook}
\label{sec:conclusions}

We have calculated the probability distribution 
$P(k_\perp)$ for an energetic parton that
propagates for a distance $L$ without radiating through
weakly coupled quark-gluon plasma 
with temperature $T$  to pick up transverse momentum $k_\perp$.
Our calculation is built upon Soft Collinear Effective Theory (SCET),
but to date we have not used much of the power of SCET, 
which could in future be brought to bear on the question of
calculating corrections (for example in the ratio of $T$ to the parton
energy) to leading order results like ours.  Before doing this, however,
the most pressing next steps beyond our calculation are to include the
radiation of collinear and soft gluons. 

To date, SCET has been used to relate $P(k_\perp)$ in any medium to
$\mathcal{W}_{\mathcal{R}}(x_\perp)$, the expectation value in that medium
of a  Wilson loop with two long light-like sides separated in the transverse
direction by a distance $x_\perp$.  What we have done here is to use
standard methods from real time thermal field theory, including
Hard Thermal Loop resummation where needed (see Section IV for an explanation
of where it is needed), to calculate the expectation
value $\mathcal{W}_{\mathcal{R}}(x_\perp)$ and from it 
the probability for momentum broadening $P(k_\perp)$, for the case in which the medium is
weakly coupled quark-gluon plasma in thermal equilibrium
at temperature $T$, as in QCD at temperatures
that are sufficiently high that the QCD coupling constant $g$ is less than one.

We first obtained $P_{\rm thin}(k_\perp)$
for a ``thin medium'', in which $\kappa$ (proportional to $g^2 L T$ and defined in (\ref{eq:kappa})) is much less than one.  We have checked that our results in this regime agree with
previous determinations, both at large and small $k_\perp$.
Although it is a stretch to apply a calculation that requires $g<1$ to 
plasma at temperatures such that $g\sim 2$ we do so anyway, since
the lowest estimates of the coupling constant in the plasma produced
in heavy ion collisions at RHIC and the LHC are in this range.
Doing this 
requires consideration of values of $\kappa$ that
are significantly greater than one.  
(A reasonable choice like $L\sim 5$~fm and $T\sim 300$~MeV corresponds
to $\kappa\sim 14$ if $g\sim 2$.)
Handling $\kappa>1$ required us
to resum an infinite class of length-enhanced planar diagrams in order to
obtain the leading behavior of $P(k_\perp)$.  
After resumming the  $L$-enhanced diagrams we obtain results
valid for any value of $\kappa$ including $\kappa\gg 1$.  We find that
in a weakly coupled plasma the properties of the plasma enter the calculation
of $P(k_\perp)$ only through the retarded gluon propagator. We computed
the self-energy therein using real time thermal field theory, including
Hard Thermal Loop resummation in the infrared.
We find
that the form of $P(k_\perp)$ changes qualitatively between $\kappa<2$,
where it diverges at small $k_\perp$, and $\kappa>4$ where it looks Gaussian
at small $k_\perp$.   
Our entire discussion is only valid to leading order in the coupling constant $g$.
It will be interesting in the future to quantify the effects of length-enhancement
at large $\kappa$ for existing next-to-leading order 
calculations~\cite{CaronHuot:2008ni}.

At large $\kappa$, we find that $P(k_\perp)$ takes the form of a Gaussian
at small $k_\perp$ with a power-law tail, proportional to $1/k_\perp^4$, at large
$k_\perp$.  
As $\kappa$ increases, more and more of the integrated probability
resides in the Gaussian, with the power-law tail contributing less and less.
Since $\kappa\to \infty$ can equally well be thought of as the strong
coupling limit as the large-$L$ limit, at a qualitative level our
large-$\kappa$ results are pleasingly consistent with the
known behavior of $P(k_\perp)$ in the plasma of large-$N_c$ ${\cal N}=4$ SYM theory
in the strong coupling limit.  In this theory, $P(k_\perp)$ is precisely Gaussian, with
no power-law tail at all.  Although the widths of the Gaussian in the strongly
coupled ${\cal N}=4$ SYM plasma and of the Gaussian that characterizes 
$P(k_\perp)$ in the weakly coupled QCD plasma for all but improbably large
values of $k_\perp$ do not agree quantitatively, and in particular have different
$g$-dependence, it is interesting to note that they are remarkably similar for $g=2$,
a value of the coupling that can reasonably be thought of as weak since $g^2/(4\pi)$
is small and that can reasonably be thought of as strong since $g^2 N_c$ is large for $N_c=3$.

The most noteworthy difference between momentum broadening in the weakly coupled QCD plasma
and in the strongly coupled ${\cal N}=4$ SYM plasma is at large $k_\perp$.   Because it
has $P(k_\perp)\propto 1/k_\perp^4$ at large $k_\perp$, no matter how
small $\kappa$ or $g$ is there will always be a $k_\perp$ for which
the weakly coupled plasma has a larger $P(k_\perp)$ than that in any strongly
coupled plasma.  In a weakly coupled plasma, the physics of momentum broadening
is different at small $k_\perp$ than at large $k_\perp$.  At small $k_\perp$ and large $\kappa$,
momentum broadening in a weakly coupled plasma can be thought of as diffusion
in $k_\perp$-space, as is the case at all $\kappa$ and all $k_\perp$ in a plasma that
is strongly coupled at all length scales. 
At large enough $k_\perp$, however, momentum broadening
in  a weakly coupled plasma is dominated by single, rare, hard, large-angle scattering
off point-like quasiparticles.  
In the large-$k_\perp$ regime, the power-law tail of $P(k_\perp)$ is unaffected by
resumming $L$-enhanced diagrams precisely because for any given $L$
it arises only at values of $k_\perp$ that are so large that $P(k_\perp)$ describes rare
single hard scattering.  Although rare, these scatterings are more common than
in a plasma that is a strongly coupled liquid
at all scales, like that in strongly coupled ${\cal N}=4$ SYM theory, because
in such a plasma there are no weakly coupled point-like constituents at any length-scale.

Although there are many reasons not to attempt a quantitative comparison 
between our calculation and any results from experiment (for example, we applied
a leading order calculation valid for $g<1$ at $g\sim 2$; for example, we have not yet
included effects of gluon radiation) at a qualitative level it is easy to see what our 
calculation predicts for $P(k_\perp)$ in the QCD plasma at experimentally 
relevant temperatures: as long as observables that bias toward short $L$ are not
used, we should expect $P(k_\perp)$ to be Gaussian at low $k_\perp$ and to have
a power-law tail at large $k_\perp$.  Even if the QCD plasma can be best understood
as a strongly coupled fluid without quasiparticles at its natural length scales of order $1/T$,
since QCD is asymptotically free we know that at short distances the strongly
coupled quark-gluon plasma is made of weakly coupled quarks and gluons and
this means that 
at high enough $k_\perp$ a power-law tail must rise above the Gaussian in $P(k_\perp)$,
as in our calculations in which the plasma is assumed to be weakly coupled at all length scales.

What about the jet quenching parameter $\hat q \equiv \langle k_\perp^2 \rangle/L$?  
Our calculation shows that its role is very different in a 
plasma that is weakly coupled at short distances than in one that is strongly
coupled on all length scales.  In the latter case, if $P(k_\perp)$ is Gaussian,
as in the strongly coupled ${\cal N}=4$ SYM plasma, then $\hat q$ is simply
a measure of the width of the Gaussian probability distribution.  In some other 
plasma that is strongly coupled on all length scales, the same would be true
even if $P(k_\perp)$ were not precisely Gaussian.
On the other hand, if $P(k_\perp)\propto 1/k_\perp^4$ at large $k_\perp$, as in any
plasma that contains weakly coupled constituents at short enough length scales,
then $\hat q$ is ultraviolet divergent and so is 
determined almost completely by the power-law tail of the 
probability distribution $P(k_\perp)$.  In this circumstance, $\hat q$ knows almost
nothing about the (Gaussian) behavior of $P(k_\perp)$ in the regime where almost
all of the probability resides. In this circumstance, therefore,
$\hat q$ tells us almost nothing about the momentum
broadening of almost all hard partons.  In this circumstance, $\hat q$ is almost entirely determined by the improbably small fraction of hard partons that scatter at large angles.

We close with three observations.  First, 
the jet quenching parameter by itself is not sufficient
as a characterization of $P(k_\perp)$.  Depending on the circumstances,
it could be a characterization of the ``bulk'' of the probability distribution
for momentum broadening or it could be telling us only about the 
power-law ``fat tail'' of this distribution, in other words only about 
the physics of those few hard partons that pick up improbably large $k_\perp$.
Knowing how to interpret what a value of $\hat q$ means requires
knowing more than $\hat q$; it requires further characterization of the shape of 
$P(k_\perp)$.  Second, if at some future time it is possible to use data
from heavy ion collision experiments to characterize the
shape of $P(k_\perp)$ over the range of $k_\perp$ where most of the
probability lies, it will take quantitative knowledge of this shape 
to learn about the properties of the plasma --- since we have seen
an example in which a very similar shape arises if the plasma is
assumed weakly coupled or if the plasma is assumed to be infinitely
strongly coupled.  Third, the place to look in order to distinguish
between these cases qualitatively is large $k_\perp$.  Just as 
Rutherford discovered point-like nuclei within what he thought
were liquid-like atoms, we can find evidence for the weakly coupled
quarks and gluons that we know to be the short
length-scale constituents of the strongly coupled plasma
produced in heavy ion collision experiments.  

It remains
to be seen which experimental observables provide access
to $P(k_\perp)$ at large $k_\perp$.  The most direct approach
that we are aware of is to analyze events in which an
initial hard scattering produces an energetic photon back-to-back
with an energetic quark.  The photon tells us the initial energy
and direction of the quark, and we can then hope to determine the
$k_\perp$ relative to this initial direction that the quark has picked up
as it propagates through the plasma.  Fig.~\ref{fig:adsprob} suggests
it would be very interesting to determine whether the quark picks 
up $k_\perp\sim 20$~GeV or more a fraction of a percent of the time,
even a small fraction of a percent of the time,
versus not at all.  
Seeing rare, but large, momentum kicks
would confirm that (as we know) at short enough distance scales
quark-gluon plasma is made of quark and gluon quasiparticles.
It would then become very interesting to study the intermediate $k_\perp$-range
in detail, to start to understand how a strongly coupled liquid emerges from
an asymptotically free gauge theory.
We will be watching with interest as 
early results on jets and energetic hadrons back-to-back with a single
hard photon in heavy ion collisions~\cite{Chatrchyan:2012gt,QMPlenaryTalksOnGammaJetsatLHC,QMTalkWithGammaHadrons}
develop, and as the statistics of such measurements improve.

\begin{acknowledgments}
We 
are particularly grateful to Chris Lee for many very helpful conversations,
in which he taught us about plus distributions and more. We
also acknowledge helpful conversations with Andrea Allais, Peter Arnold,
Simon Caron-Huot, Iain Stewart, Xin-Nian Wang 
and Urs A. Wiedemann. This work is supported by U.S. Department of Energy (D.O.E.) under cooperative research agreement DE-FG0205ER41360. The work of F.D. is supported by the Miller Institute for Basic Research in Science, University of California, Berkeley.
\end{acknowledgments}

\appendix

\section{Schwinger-Keldysh formalism}
\label{app:RTF}
In this Appendix, we briefly review the real-time field theory tools 
that we use in setting up the formalism in Section II and calculating
the self-energy in Section III.
(The Schwinger-Keldysh formalism that we are using is standard.
One of its advantages is that by formally doubling the number
of degrees of freedom, pinch singularities that otherwise plague
real-time quantum field theory are avoided.
For a more complete review, see e.g. Ref.~\cite{Thoma:2000dc}.)
We present the discussion for a scalar field theory to avoid excess notation. The discussion
for gauge bosons is analogous. 

\begin{figure}[t]
\begin{center}
\includegraphics[scale=0.55]{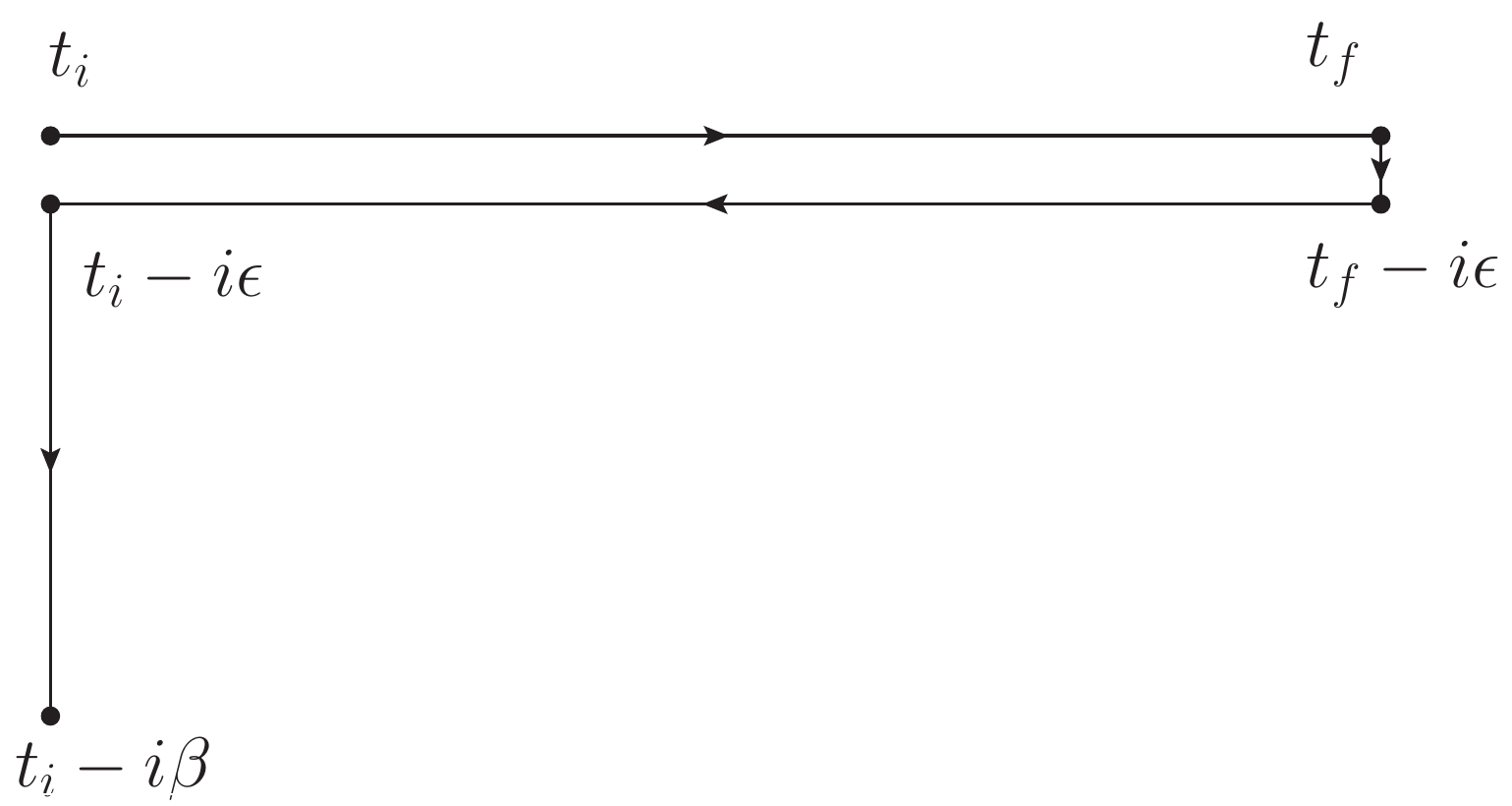}
\end{center}
\caption{Schwinger-Keldysh contour for $\sigma = \epsilon$.}
\label{fig:SKconteps}
\end{figure}

We allow the time coordinate $x^0$ to be complex, and we define thermal Green functions
\begin{equation}
G_C(x_1, x_2, \ldots, x_n) = \left\langle T_C \left\{ \phi(x_1) \phi(x_2) \ldots \phi(x_n) \right\} \right\rangle \ ,
\end{equation}
where the ordering is taken along a path $C$ in the complex plane. For the path $C$
we choose the Schwinger-Keldysh contour in Fig.~\ref{fig:SKconteps}. In the $t_i \rightarrow -\infty$ and $t_f \rightarrow \infty$ limit the vertical pieces of $C$ are irrelevant for the calculation of Green functions. Thus it is convenient to label the field $\phi_i(x)$, where $i = 1$ or $i=2$ 
depending on where the field is evaluated on $C$:
\begin{equation}
\phi_1(x) = \phi_1(x^0,\mathbf{x}) \ , \qquad  \phi_2(x) = \phi_2(x^0-i \epsilon,\mathbf{x}) \ .
\end{equation}
Propagators then become matrices:
\begin{equation}
\begin{split}
& D_{ij}(x - y) =  \\ &
\left(\begin{array}{cc}
\left\langle T\left\{\phi(x^0,\mathbf{x})\phi(y^0,\mathbf{y} )\right\} \right\rangle & \left\langle \phi(y^0 - i\epsilon ,\mathbf{y} )\phi(x^0,\mathbf{x}) \right\rangle \\
\left\langle \phi(x^0-i\epsilon,\mathbf{x})\phi(y^0,\mathbf{y} ) \right\rangle & \left\langle \tilde{T}\left\{\phi(x^0,\mathbf{x})\phi(y^0,\mathbf{y} ) \right\} \right\rangle
\end{array}\right)
\label{eq:Dmatrix}
\end{split}
\end{equation}
where $T$ and $\tilde{T}$ stand for time ordering and anti-time ordering, respectively. 
The generating functional for the free theory can be written as
\be
\begin{split}
Z_C^{\rm free} = & \, \mathcal{N} \exp\left[ - \frac{1}{2} \int_{-\infty}^{\infty} d^4 x \;  \int_{-\infty}^{\infty} d^4 x^\prime \, K(x, x^\prime)  \right] \ , \\
K(x, x^\prime) \equiv & \, J_i(x) D^{\rm free} _{ij}(x-x^\prime) J_j(x^\prime) \ ,
\end{split}
\ee
where the normalization factor $\mathcal{N}$ takes into account the multiplicative constant given by the two vertical pieces in $C$. In order to get the thermal Green functions we have to differentiate the above expression with respect to $J_i(x)$. For $\epsilon \to 0$, the Fourier space free propagator matrix reads
\begin{equation}
\begin{split}
D^{\rm free} _{11}(Q) = &\, \left[\frac{i}{Q^2+i \epsilon}+n(q_{0})\,2\pi\,\delta(Q^2)\right] \ ,  \\   
D^{\rm free} _{12}(q) = &\, \left[\theta(-q_0)+n(q_{0})\right]\,2\pi\,\delta(Q^2)\ ,\\
D^{\rm free} _{21}(Q) = &\, \left[\theta(q_0)+n(q_{0})\right]\,2\pi\,\delta(Q^2)\ ,  \\ 
D^{\rm free} _{22}(q) = &\, \left[\frac{-i}{Q^2- i \epsilon}+n(q_{0})\,2\pi\,\delta(Q^2)\right] \ .
\end{split}
\end{equation}
The $\phi_2(x)$ field induces a modification of the naive Feynman rules. The propagator has off-diagonal elements, meaning that 
it mixes the two fields $1$ and $2$, whereas the vertices have only one type of field, and they do not induce any mixing. In addition we have a minus sign for any vertex connecting fields of type $2$.

The four components of the propagator matrix in (\ref{eq:Dmatrix}) are not independent, since
\begin{equation}
D_{11}(Q) + D_{22}(Q) =  D_{12}(Q) + D_{21}(Q) \ ,
\label{eq:linearcomb}
\end{equation}
and  it is therefore 
more convenient to use a different basis including only three independent propagators. We introduce the Keldysh representation
\begin{equation}
\begin{array}{llll}
{\rm Retarded}: & & D_R(Q) \equiv D_{11}(Q) - D_{12}(Q) \ , \\
{\rm Advanced}: & & D_A(Q) \equiv D_{11}(Q) - D_{21}(Q) \ , \\ 
{\rm Symmetric}: & & D_S(Q) \equiv D_{11}(Q) + D_{22}(Q)  \ .
\end{array}
\end{equation}
From the definitions we can derive the Fourier space expressions for the
three free propagators in this new basis.
\begin{equation}
\begin{array}{llll}
{\rm Retarded}: & & D^{\rm free}_R(Q) = \frac{i}{Q^2 + i\, {\rm sgn}(q_0) \, \epsilon} \ , \\ & & \\
{\rm Advanced}: & & D^{\rm free}_A(Q) = \frac{i}{Q^2  - i \, {\rm sgn}(q_0) \, \epsilon} \ , \\ & & \\
{\rm Symmetric}: & & D^{\rm free}_S(Q) = [1 + 2\, f(q_{0})] \,2\pi\,\delta(Q^2)  \ .
\end{array} 
\label{eq:Keldyshbasis}
\end{equation}
We notice that only the symmetric propagator contains the thermal 
distribution $n(q_0)$ (either the Bose-Einstein distribution $n_B(q_0)$ or the Fermi-Dirac
distribution $n_F(q_0)$ as appropriate)
simplifying the identification of thermal contributions to diagrams.

We can now introduce the quantity that
we calculate in Section III, the self-energy matrix $\Pi_{ij}$ which is
the building block for the propagator $D_{ij}(Q)$, as described by the Dyson equation
\begin{equation}
D_{ij}(Q) = D^{\rm free} _{ij}(Q) + D^{\rm free} _{im}(Q) \, (- i \Pi_{mn}(Q)) \, D_{nj}(Q) \ .
\end{equation}
In order to use the Keldysh representation also for the self-energy we define
\begin{equation}
D_{\alpha}(Q) = D^{\rm free} _{\alpha}(Q) + D^{\rm free} _{\alpha}(Q) \, (- i \Pi_{\alpha}(Q)) \, D_{\alpha}(Q) \ ,
\label{eq:Pidefinition}
\end{equation}
where $ \alpha = R, A, S$. The self-energies $\Pi_{\alpha}(Q)$ introduced above are linear combinations of the self-energies $\Pi_{ij}$. This is shown by plugging the propagators $D_{ij}(Q)$ into the definition (\ref{eq:Pidefinition}), and identifying
\begin{equation}
\begin{array}{llll}
{\rm Retarded}: & & \Pi_R(Q) = \Pi_{11}(Q) + \Pi_{12}(Q) \ , \\
{\rm Advanced}: & & \Pi_A(Q) = \Pi_{11}(Q) + \Pi_{21}(Q) \ , \\ 
{\rm Symmetric}: & & \Pi_S(Q) = \Pi_{11}(Q) + \Pi_{22}(Q)  \ .
\end{array} 
\label{eq:PiRandPi1112}
\end{equation}

\section{A thin medium}
\label{app:thinmed}

In this Appendix we analyze our results in the case of a ``thin medium'', by which
we mean $L$ short enough
such that $\kappa\equiv g^2 C_{\mathcal{R}} L \, T/(2\pi)  \ll 1$.
In this regime, the
resummation of length-enhanced contributions
performed in Sec.~\ref{sec:resumming} is not necessary
meaning that in this regime the full expression (\ref{Pdef}) 
reduces to
\be
P(k_\perp) \simeq \int d^2 x_{\perp} \, e^{-i k_{\perp} \cdot x_{\perp}}\, \left[ 1 + \mathcal{W}^{(2)}_{\mathcal{R}}(x_\perp)  \right]  \ .
\ee
We rewrite this expression in Fourier space, exactly as we did at the end of Sec.~\ref{sec:resumming}, and we find
\beq
P(k_\perp) = (2\pi)^2 \delta^2(k_\perp) \left[1 - \int \frac{d^2q_\perp}{(2\pi)^2} P_{\rm thin}(q_\perp) \right] + P_{\rm thin}(k_\perp) 
\label{Plim}
\eeq
where $P_{\rm thin}(k_\perp)$ is the function defined in Eq.~\r{eq:Pgdef}, a
definition that we reproduce here for convenience:
\be
P_{\rm thin}(k_\perp) \equiv 
\frac{2\sqrt{2}\pi\,\kappa}{T}\,
\int \frac{d q^- }{2 \pi} D^{>}(q^-, k_\perp) \ .
\ee
The probability distribution in Eq.~\r{Plim} consists of two parts.
First, we see a delta function centered at $k_\perp = 0$
whose coefficient has two terms, corresponding to no scattering
and scattering with no transverse momentum exchange.
Second, we see the contribution describing 
scattering with nonzero final transverse momentum, 
given just by $P_{\rm thin}(k_\perp)$. We see that
for a thin medium, and for $k_\perp \neq 0$, we have $P(k_\perp) = P_{\rm thin}(k_\perp)$.

Looking at the probability distribution (\ref{Plim}), we immediately notice that the coefficient of the delta function is not finite. As we have seen in Section V.A, $P_{\rm thin}(q_\perp)$ precisely matches the AGZ distribution~\r{AGZ} in the infrared, meaning that 
it is $\propto q_\perp^{-2}$ for $q_\perp \ll T$. The integral in the delta function coefficient thus blows up. We shall show that, despite this divergence, the probability distribution in Eq.~\r{Plim} is well defined. In order to do that we need to introduce the ``plus distribution'' $[\ldots]_{+}$ for a generic function $g(x)$, defined as~\cite{Ligeti:2008ac}
\be
\begin{split}
\[ \theta (x) g(x) \]_+ \equiv &\, \lim_{\beta \to 0} \frac{d}{dx} [\theta(x-\beta) G(x)]  \text{, with } \notag \\
G(x) = & \, \int_{x_0}^x dx' g(x') \ ,
\label{plusd}
\end{split}
\ee
satisfying the boundary condition $\int_0^{x_0} dx [\theta(x) g(x)]_+ = 0$. Here, $x_0 \neq 0$ when $g(0) = \infty$. Since $P(k_\perp)$ depends only on the absolute value of transverse momentum, it is convenient to use the one-dimensional 
probability distributions $ \tilde P(k_\perp) \equiv 2\pi k_\perp P(k_\perp)$  and $ \tilde P_{\rm thin}(k_\perp) \equiv 2\pi k_\perp P_{\rm thin}(k_\perp)$. Then \r{Plim} becomes
\beq
\tilde P(k_\perp) = (2\pi)^2 \delta(k_\perp) + \tilde P_{\rm thin}(k_\perp) - \delta(k_\perp) \int_0^\infty dq_\perp \tilde P_{\rm thin}( q_\perp) \ ,
\label{tildeP}
\eeq
where the normalization now reads $\int_0^\infty dk_\perp \tilde P(k_\perp) = (2\pi)^2$.

We can 
apply the plus distribution prescription to extract
 the divergent term from $\tilde P_{\rm thin}(k_\perp)$,
obtaining
\be
\begin{split}
& \[ \theta (k_\perp) \tilde P_{\rm thin}(k_\perp) \]_+ = \\ &
 \lim_{\beta \to 0} \left[ \delta(k_\perp- \beta) \int_{k_{\perp 0}}^{k_\perp}  dq_\perp \tilde P_{\rm thin}(q_\perp)   + \theta (k_\perp- \beta) \tilde P_{\rm thin}(k_\perp) \] \ .
\end{split}
\ee
After we take the $\beta \to 0$ limit, and noting that $k_\perp$ is always positive, the above result takes the form
\beq
\[ \tilde P_{\rm thin}(k_\perp) \]_+ = \tilde P_{\rm thin}(k_\perp) - \delta(k_\perp) \int_{k_\perp}^{k_{\perp 0}}  dq_\perp \tilde P_{\rm thin}(q_\perp)  
\label{P>+} \ .
\eeq
This in turns implies the following facts:
\begin{itemize}
\item $\int_0^{k_{\perp 0}} dk_\perp\[ \tilde P_{\rm thin}(k_\perp) \]_+ = 0$ \ ,
\item $\[ \tilde P_{\rm thin}(k_\perp) \]_+ = \tilde P_{\rm thin}(k_\perp)$ for $k_\perp> 0$ \ .
\end{itemize}
Combining~\r{P>+} and~\r{tildeP}, 
we obtain
\begin{align}
\tilde P(k_\perp) & = \delta(k_\perp) \( (2\pi)^2 - \int_{k_{\perp 0}}^\infty  dq_\perp \tilde P_{\rm thin}( q_\perp) \) \notag \\
&+ \[ \tilde P_{\rm thin}(k_\perp) \]_+ ,
\label{Pw+}
\end{align}
where $k_{\perp 0}$ is a free parameter in the range $0<k_{\perp 0}<\infty$.

If we switch back to the two dimensional distributions, drop the tildes, and keep in mind that the plus distribution $\[P(k_\perp)\]_+$ is formally the one dimensional plus 
distribution divided by $2\pi k_\perp$, we find
\be
P(k_\perp) = \delta^2(k_\perp) g(k_{\perp 0}) + \[ P_{\rm thin}(k_\perp) \]^{k_{\perp 0}}_+ \ ,
\label{eq:Pgrplusfin}
\ee
where the factor $g(k_{\perp 0})$ reads
\be
g(k_{\perp 0}) = (2\pi)^2 - 2\pi \int_{k_{\perp 0}}^\infty dq_\perp q_\perp P_{\rm thin}(q_\perp) \ .
\ee
For $k_{\perp 0} > 0$, the coefficient of the 
delta function in Eq.~\r{eq:Pgrplusfin} is finite and positive, and the 
last term in Eq.~(\ref{eq:Pgrplusfin}) is the probability
distribution with its value at $k_\perp = 0$ having been subtracted. Even though both terms depend on $k_{\perp 0}$, the $k_{\perp 0}$ dependence cancels when we sum the two terms. In conclusion, the probability distribution $P(k_\perp)$ in the thin medium limit $g^2 C_{\mathcal{R}} LT \ll 1$ is well defined, and the probability for no scattering plus that for scattering
with no transverse momentum exchange is finite and positive, as it must be.

\section{Exponentiation of $\mathcal{W}^{(2)}_{\mathcal{R}}$} \label{app:exon}

In this Appendix, we show that by summing over all disconnected diagrams which involve only 
$\mathcal{W}^{(2)}_{\mathcal{R}}$, we obtain precisely the exponential~\eqref{eq:W2en}
that we have used in (\ref{Pdef}). 
To get the lowest power in $g$ for a given power of $L$, there is no need to include any connected diagrams
other  than $\mathcal{W}^{(2)}_{\mathcal{R}}$ in the sum. As discussed in the main text, 
if we ignore any factors of $g$ that 
come from within the gluon propagators 
$\mathcal{W}^{(2)}_{\mathcal{R}}$ is of order $g^2 L$, while any other connected diagrams give rise to $g^n L$ with $n > 2$ and thus are of higher order. For example, $\mathcal{W}^{(3)}_{\mathcal{R}}$ is of order $g^3 L$, while the cross diagram in Fig.~\ref{fig:WloopNO} will give a factor $g^4 L$. 

To obtain the $n$-th order term in~\eqref{eq:W2en}, we consider the expansion of $\mathcal{W}_{\mathcal{R}}$ at order $j= 2n$.  For illustration here we demonstrate explicitly the exponentiation of $D^> (y^-, x_\perp)$ in~\eqref{eq:W2} which comes from contractions of  gluons with ends on different Wilson lines.  The story for $D^> (y^-, 0_\perp)$ (gluons with both ends on the same
Wilson line) 
and the cross terms between $D^> (y^-, 0_\perp)$ and $D^> (y^-, x_\perp)$ are similar and we leave it to the reader that they also exponentiate.  
In the notation of Eq.~(\ref{eq:DoubleSeries}), we need only 
consider the term $W^{(n,n)}_{\cal R}$ in $\mathcal{W}^{(2n)}_{\mathcal{R}}$,  
which can be written as  
\begin{widetext}
 \be
  W_{\mathcal{R}}^{(n,n)}
 =\frac{ g^n}{d(\mathcal{R})}
 \int_{x_1^-< x_2^- < \cdots x_n^-} \hspace{-0.5in}
 dx_1^- \ldots dx_n^- \int_{z_1^-< z_2^- < \cdots z_n^-} \hspace{-0.5in} dz_{1}^- \ldots 
 dz_n^- 
 \left \langle {\rm Tr} \left[ 
 A_{\mathcal{R}}^+(x_1^-,x_\perp) \ldots 
 A_{\mathcal{R}}^+(x_n^-,x_\perp)
 A_{\mathcal{R}}^+(z_{n}^-,0_\perp) \ldots 
 A_{\mathcal{R}}^+(z_1^-,0_\perp)
 \right] \right\rangle\ 
 \label{eq:Explicin}
 \ee
 \end{widetext}
where we have denoted the coordinates on the two different 
Wilson lines by $x^-_i$ and $z^-_i$ respectively and made the path ordering explicit. There are many different contractions in~\eqref{eq:Explicin}. The one which is relevant for our purposes here 
is the one in which 
the gluon at $x_i^-$ is contracted with the gluon at $z^-_i$, for each of $i=1, 2, \cdots n$. Other contractions will either give rise to cross terms between $D^> (y^-, x_\perp)$ and $D^> (y^-, 0_\perp)$ or create cross diagrams like that in Fig.~\ref{fig:WloopNO} which are of higher order. 
We then find that 
\begin{widetext}
\be
 W_{\mathcal{R}}^{(n,n)} =  (g^2 C_{\cal R})^n \int_{x_1^-< x_2^- < \cdots x_n^-} \hspace{-0.5in}
 dx_1^- \ldots dx_n^- \int_{z_1^-< z_2^- < \cdots z_n^-} \hspace{-0.5in} dz_{1}^- \ldots 
 dz_n^- \, \prod_{i=1}^n D^> (x_i^--z_i^-, x_\perp)  =  {1 \over n!}  \left( g^2 C_{\cal R} L^-  \int dy^- \, D^> (y^-, x_\perp) \right)^n  \ .
 \ee
\end{widetext}
In the last step we have changed the integration over each pair
of integration variables $(x_i^-, z_i^-)$ to an integration over $Y_i ^-= {x_i^- + z_i^- \over 2}$ and $y_i^- = x_i^- - z_i^-$. The integrations over $y_i^-$ are unconstrained, while the integrations over 
$Y_i^-$, which satisfy the ordering $Y_1^- < Y_2^- < \cdots Y_n^-$, 
give $(L^-)^n/ n!$.  We have proved (\ref{Pdef}). The generalization of the proof of the 
exponentiation of $\mathcal{W}^{(2)}_{\mathcal{R}}$  
that we have presented here to a proof of the exponentiation of the sum of all connected diagrams
is straightforward.

\section{Boltzmann Equation Approach}
\label{app:Boltz}

In this Appendix, we 
provide an explicit alternative
derivation of
the expression (\ref{Pdef}) 
for the probability distribution $P(k_\perp)$
by deriving and solving an appropriate Boltzmann equation for momentum broadening.
The argument can easily be generalized to obtain the
more general result~(\ref{exPne}) that goes beyond leading order in weak coupling.
In this Appendix, it is useful to make the $L$-dependence of the probability distribution 
explicit by denoting it as $P(k_\perp, L)$.

Let us consider the hard parton after it 
has propagated for a distance $L$ in the hot and dense medium. $P(k_\perp,L)$ is the probability distribution 
for its transverse momentum $k_\perp$. 
We want to relate $P(k_\perp, L)$ to $P(k_\perp, L+\Delta L)$, the probability distribution after the parton has traveled a further distance $\Delta L$.
In the $\Delta L \ll L$ limit, modifications to $P(k_\perp, L)$ are only caused by a single scattering event. 
We then introduce the differential collision kernel $C(q_\perp)$, defined as the differential rate for momentum broadening due to elastic collisions
\be
C(q_\perp) \equiv \frac{d^2 \Gamma}{d q_\perp^2} \ ,
\label{eq:collkernel}
\ee 
in terms of which the relation between the probability distributions at $L$ and $L+\Delta L$ can 
be written as
\be
\label{eq:BoltzmannIntegralEquation}
\begin{split}
P(k_\perp, L + \Delta L) = & \, P(k_\perp, L) \left[1 - \Delta L \,  \int  \frac{d^2 q_\perp}{(2 \pi)^2} C(q_\perp) \right] + \\ &
\Delta L  \int  \frac{d^2 q_\perp}{(2 \pi)^2} \, C(q_\perp) \, P(k_\perp - q_\perp, L)  \ .
\end{split}
\ee
The first line describes the probability that 
between $L$ and $L + \Delta L$ there is  no further momentum 
transfer, and the probability distribution is not affected. The second line 
takes into account the possibility that the transverse momentum $k_\perp$ 
at $L+\Delta L$ arises after the hard parton picks up an additional transverse
momentum $q_\perp$ in a scattering event that occurs between $L$ and $L+\Delta L$.
The Boltzmann equation for transverse momentum broadening is obtained by taking 
the $\Delta L \rightarrow 0$ limit in Eq.~\ref{eq:BoltzmannIntegralEquation}, yielding
\be
\frac{d\, P(k_\perp, L)}{d L} = \int \frac{d^2 q_\perp}{(2 \pi)^2} \, C(q_\perp)  \left[ P(k_\perp - q_\perp, L)  - P(k_\perp, L)\right] \ .
\label{eq:BoltzmannDifferentialEquation}
\ee

Solving the Boltzmann equation (\ref{eq:BoltzmannDifferentialEquation}) 
is simpler in coordinate space, where it reads
\be
\frac{d\, P(x_\perp, L)}{d L} = - v \left(x_\perp \right) \, P(x_\perp, L)  \ ,
\label{eq:Boltzcoordinates}
\ee
with 
\be
 v \left(x_\perp \right) \equiv \int \frac{d^2 q_\perp}{(2\pi)^2}  \left[1 - e^{ i q_\perp x_\perp}  \right] C(q_\perp) \ , 
\ee
a quantity often referred to as the dipole cross-section in the literature.
The solution of Eq.~(\ref{eq:Boltzcoordinates}) is just an exponential function, and upon performing the Fourier transform back to momentum space we find 
\be
P(k_\perp, L) = \int d^2 x_\perp \, e^{i k_\perp x_\perp} e^{-  v \left(x_\perp \right) L } \ .
\label{eq:BoltzSol}
\ee
This connection between multiple scattering and the dipole cross section was recognized 
in Refs.~\cite{Baier:1996kr,Baier:1996sk,Zakharov:1997uu,Wiedemann:2000za}. In many calculations, however, 
the dipole cross-section
was approximated as
$v \left(x_\perp \right) \approx C x_\perp^2$, resulting in a probability distribution that is
Gaussian in $k_\perp$. 
As illustrated in Fig.~\ref{fig:LargeKappaGaussians}, we find that this is a good approximation
at small $k_\perp$ but not at large $k_\perp$.

Next, we present  the explicit relationship between the analysis of this Appendix in terms
of a Boltzmann equation and the analysis we follow throughout all other
sections of this paper.
The building block for deriving the Boltzmann equation was the differential elastic collision 
kernel $C(k_\perp)$ defined in Eq.~(\ref{eq:collkernel}). With a view toward making contact with 
results obtained in previous literature via the Boltzmann equation approach,  
we begin by including only elastic collisions in which one gluon exchanged. In the 
framework of the rest of this paper,  
this is obtained from the Wilson line diagrams in Fig.~\ref{fig:Wloop} with only one gluon propagator,
as in Fig.~\ref{fig:W2diag}. We have denoted this contribution by $P_{\rm thin}(k_\perp, L)$, defined in Eq.~(\ref{eq:Pgdef}). The connection between $P_{\rm thin}(k_\perp, L)$ and $C(k_\perp)$ reads
\be
P_{\rm thin}(k_\perp, L) = C(k_\perp) \, L \ .
\ee
With this identification in mind, we revisit the expression for $\mathcal{W}^{(2)}_{\mathcal{R}}$ found in Eq.~(\ref{eq:W2Fourier}), and we recognize
\be
\begin{split}
\mathcal{W}^{(2)}_{\mathcal{R}}(x_\perp) = & \, - \int \frac{d^2 q_\perp}{(2 \pi)^2} \, \left[1 - e^{i q_{\perp} \cdot x_{\perp}} \right] P_{\rm thin}(q_\perp, L) = 
\\ & -  v \left(x_\perp \right) L \ .
\end{split}
\label{eq:vvsW2}
\ee
The solution to the Boltzmann equation found in Eq.~(\ref{eq:BoltzSol}), with $v \left(x_\perp \right)$ connected to $\mathcal{W}^{(2)}_{\mathcal{R}}$ as in Eq.~(\ref{eq:vvsW2}), is therefore 
identical to the expression in Eq.~(\ref{Pdef}) obtained by resumming length-enhanced diagrams.

The manipulations in this Appendix are completely general in the sense that 
they do not rely on any specific form for the collision kernel $C(k_\perp)$ in Eq.~(\ref{eq:collkernel}). 
If we were to include contributions to $C(k_\perp)$ that are higher order in the coupling,
solving the Boltzmann equation would resum these higher order effects as needed when
$L$ is not small.  If we were to include all contributions to $C(k_\perp)$ to arbitrarily high
order in the coupling, we would recover our
general result (\ref{exPne}).

It has long been understood (in the QCD context, since 
Refs.~\cite{Baier:1996sk,Wiedemann:1999fq,Gyulassy:2002yv}; in the context of electrons propagating
through ordinary matter, at least since Refs.~\cite{Moliere}) that if $L$ is long enough that multiple
scattering is important, the dipole cross section 
can to a degree be approximated as 
$v(x_\perp)\propto x_\perp^2$, corresponding to a probability
distribution for transverse momentum broadening (\ref{eq:BoltzSol}) that is Gaussian in $k_\perp$. 
It was also understood by all these authors that, at large enough $k_\perp$,
$P(k_\perp)$ must have a power-law tail corresponding to Rutherford scattering
off a single point-like scatterer.   In the QCD context, this was demonstrated
in a model context by Wiedemann and Gyulassy in Appendix A of 
Ref.~\cite{Wiedemann:1999fq},
where they showed that in the Gyulassy-Wang model~\cite{Gyulassy:1993hr} (in which the medium consists of
static scattering centers which the hard parton sees as Debye-screened 
Yukawa potentials) 
the dipole cross section takes the form $v(x_\perp)\propto x_\perp^2 \log\left(R^2/x_\perp^2\right)$
at small $|x_\perp|$ for some constant $R$. (This form was
known earlier in a different context~\cite{Nikolaev:1990ja}.)
Evaluating (\ref{eq:BoltzSol}), we see that this form for $v(x_\perp)$
corresponds to $P(k_\perp)\propto 1/k_\perp^4$ at large $k_\perp$.

In the calculation we have presented in this paper, we need make no model
assumptions about the nature of the medium. It is a weakly coupled
quantum field theoretical plasma, as in QCD in thermal
equilbrium at asymptotically high  temperatures. And,
as illustrated in Fig.~\ref{fig:LargeKappaGaussians},
when $L$ is large enough that multiple scattering is important we find
by direct calculation both the Gaussian behavior
at small $k_\perp$ and the $1/k_\perp^4$ behavior at large $k_\perp$.

\section{A tensor basis for the self-energy}
\label{app:projectors}
In this Appendix we define 
the projectors used to expand the retarded self-energy in tensor components in Section III
and describe their properties. The presence of the thermal bath
breaks Lorentz invariance by specifying a preferred frame in
which the medium is at rest. We shall work in the medium rest frame,
in which its four-velocity
$U^\mu$ is $U^\mu = (1, 0)$, but in this Appendix we 
keep $U^\mu$ unspecified, defining these projectors for any frame.
  It is convenient to introduce another four-vector $N^\mu$ 
  by projecting the four-velocity $U^\mu$ 
  onto the direction orthogonal to the external momentum $Q^\mu$:
\be
\begin{split}
& N^\mu \equiv P^{\mu\nu} U_\nu = U^\mu - \frac{Q \cdot U}{Q^2} Q^\mu \ ,  \\
& P_{\mu\nu} = g_{\mu\nu} - \frac{Q_{\mu}Q_{\nu}}{Q^2}  \ , \qquad  Q \cdot N = 0 \ .
\end{split}
\ee
The self-energy is a symmetric rank-$2$ tensor. It can therefore be expressed as a linear 
combination of the following tensors:
\begin{equation}
\begin{split}
P_{\mu\nu}^T = &\,  - g_{\mu\nu} + \frac{Q_\mu Q_\nu}{Q^2} + \frac{N_\mu N_\nu}{N^2} \ , \qquad   P_{\mu\nu}^L =  - \frac{N_\mu N_\nu}{N^2} \ , \\
J_{\mu\nu} = & \,   \frac{Q_\mu N_\nu + N_\mu Q_\nu}{\sqrt{-2 N^2 Q^2}}  \ ,  \qquad\qquad\qquad  K_{\mu\nu} =  \frac{Q_\mu Q_\nu}{Q^2}   \ .
\end{split}
\label{eq:tensorbasis}
\end{equation}
It is useful to build a multiplication table for the tensor basis, keeping in mind that the tensors in~(\ref{eq:tensorbasis}) are symmetric in their Lorentz indices. We find
\be
\begin{array}{lllllll}
P^T \cdot P^T = - P^T \ , & & & & P^L \cdot P^L = - P^L \ , \\ 
J \cdot J  = \frac{1}{2} (P^L - K) \ , & & & & K \cdot K = K \ , \\ 
P^T \cdot P^L = 0 \ , & & & &  P^T \cdot J = 0 \ , \\
P^T \cdot K = 0 \ , & & & & P^L \cdot K = 0 \ , \\
{\rm Tr} \,\left(P^L \cdot J \right) = 0 \, & & & & {\rm Tr} \,\left(J \cdot K \right) = 0 \ .
\end{array}
\ee

\section{Resumming AGZ}
\label{app:AGZ}

As we noted in Section V.A, some years ago Aurenche, Gelis and Zaraket derived 
an analytic expression (\ref{AGZ}) for $P_{\rm thin}(k_\perp)$ that is valid in the 
IR region, $k_\perp \ll T$, but not in the UV.  In this Appendix we pursue the
exercise of assuming 
that
$P_{\rm thin}(k_\perp)=P_{\rm thin}^{\rm AGZ}(k_\perp)$
at all $k_\perp$, not just in the IR, and then resumming $L$-enhanced
diagrams to obtain an expression for $P(k_\perp)$ that we 
shall denote by $P^{\rm AGZ}(k_\perp)$.  
Even though $P_{\rm thin}^{\rm AGZ}(k_\perp)$ is not
in fact correct beyond the IR, there are two reasons to pursue this exercise.
First, we anticipate 
that the $P^{\rm AGZ}(k_\perp)$ we obtain by resumming the AGZ expression will agree
with our complete result at small enough $k_\perp$.  
This makes sense on physical grounds, as we have
argued in Section~\ref{subsec:5B}.  Here we confirm this expectation
by explicit calculation.
And, second, starting
with the AGZ expression for $P_{\rm thin}$ allows us to do almost
all of the calculation analytically.
This makes it of value as a benchmark, even 
though the result is not valid beyond the IR.  
And, because the calculation {\it is} valid in the IR 
it will allow us to gain an analytic understanding
of the $\kappa$-dependence of the low-$k_\perp$ behavior of $P(k_\perp)$.


In order to resum $L$-enhanced diagrams upon assuming 
that $P_{\rm thin}(k_\perp)=P_{\rm thin}^{\rm AGZ}(k_\perp)$ we proceed as follows.  We first note from~(\ref{D21def}) and (\ref{eq:Pgdef})
that
\be
\mathcal{W}_{\mathcal{R}}^{(2)} (x_\perp) = - \int \frac{d^2 q_\perp}{(2\pi)^2} 
\left(1- e^{i \,q_\perp \cdot x_\perp} \right) P_{\rm thin}(q_\perp) \ ,
\label{eq:AGZresum1}
\ee
regardless of the form of $P_{\rm thin}$.
In the case where $P_{\rm thin}=P_{\rm thin}^{\rm AGZ}$, given in (\ref{AGZ}), 
the integral in (\ref{eq:AGZresum1}) can be done analytically, yielding
\be
\mathcal{W}_{\mathcal{R}}^{(2)} (x_\perp)
 =\mathcal{W}_{\mathcal{R}}^{(2),{\rm AGZ}} (x_\perp) \equiv 
  -\kappa \left[ \gamma + \log\left(\frac{x}{2}\right) + K_0(x) \right], 
  \label{eq:AGZresum2}
 \ee
where we have defined $x\equiv  x_\perp m_D$ and 
where $K_0(x)$ is the modified Bessel function of the second kind
and $\gamma\approx 0.577$ is the Euler gamma constant.  
 Note that $\mathcal{W}^{(2)}_{\mathcal{R}}(x)\rightarrow 0$ as $x \to 0$ 
since $K_0(x) \to - \gamma - \log(x/2)$ as $x\to 0$. 
The dipole
cross-section $v(x_\perp)$ corresponding to (\ref{eq:AGZresum2}) 
(see Appendix~\ref{app:Boltz}
and in particular (\ref{eq:vvsW2}))
is very similar to that found
in Eq.~(A11) of Ref.~\cite{Wiedemann:1999fq}
for the case of the Gyulassy-Wang model where the medium consists
of static Debye-screened scattering centers.
This similarity is not surprising: when Fourier transformed, 
the screened Yukawa potentials in the Gyulassy-Wang 
model are similar to $P_{\rm thin}^{AGZ}(k_\perp)$ of Eq.~(\ref{AGZ}), 
and in particular they agree at large $k_\perp$ where multiple
scattering is not important.  We therefore
find that (\ref{eq:AGZresum2}) and (A11) of Ref.~\cite{Wiedemann:1999fq}
agree at small $x_\perp$.


Next, we must substitute (\ref{eq:AGZresum2}) into (\ref{Pdef}) in order
to obtain $P^{\rm AGZ}(k_\perp)$.
The angular integral in (\ref{Pdef}) can be performed analytically,
yielding the full, $L$-resummed,  probability distribution
\begin{align}
P(k_\perp)=P^{AGZ}(k_\perp) \equiv \frac{2\pi}{m_D^2} \int_0^\infty dx \, x \, J_0 \(\frac{k_\perp x}{m_D}\) \times \notag \\
\exp \[ - \kappa \( \gamma + \log\( x/2\) + K_0(x) \) \]\ ,
\label{eq:AGZsum}
\end{align}
in which only a single integral remains to be evaluated numerically.

\begin{figure}[t]
\begin{center}
\includegraphics[scale=0.86]{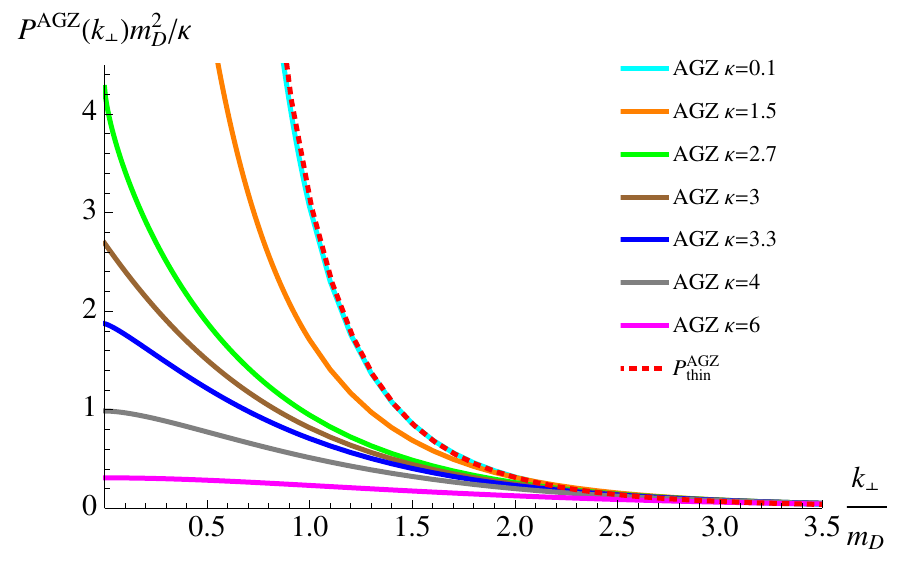}
\includegraphics[scale=0.96]{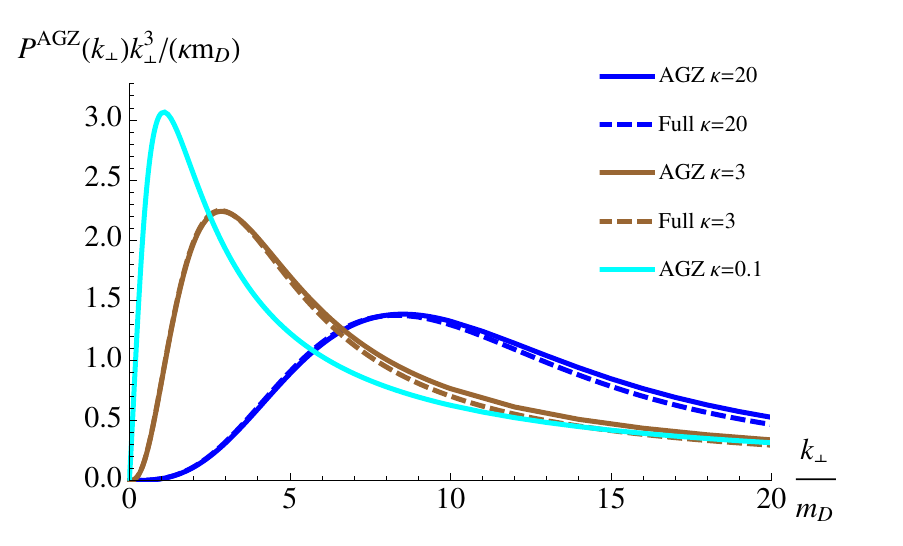}
\includegraphics[scale=0.96]{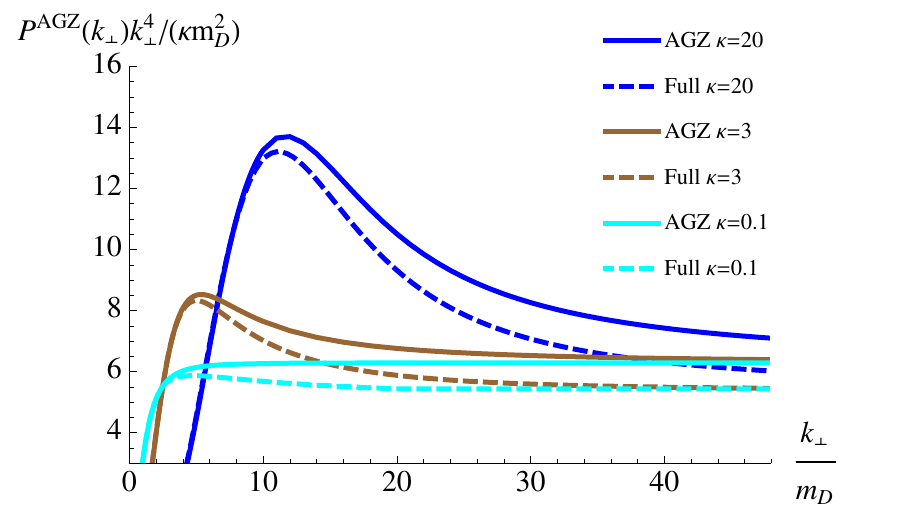}
\caption{The solid curves in the top panel show the $L$-resummed $P^{\rm AGZ}(k_\perp)$
for (top to bottom) $\kappa=0.1$, 1.5, 2.7, 3, 3.3, 4 and 6.  The dashed curve shows $P_{\rm thin}^{\rm AGZ}(k_\perp)$; this curve is  $\kappa$-independent because we have divided by $\kappa$.
We see the $k_\perp \to 0$ behavior of $P^{\rm AGZ}(k_\perp)$ changing as $\kappa$ increases.
In the middle and lower panels, both $P^{\rm AGZ}(k_\perp)$ 
(solid curves) and our full results  for $P(k_\perp)$  from 
the middle and lower panels of Fig.~\ref{W2_g01} (dashed curves) are plotted,
multiplied by $k_\perp^3$  and $k_\perp^4$.
We see that, as expected, the $L$-resummed AGZ distribution agrees with our full
result in the IR, but not in the UV.
All plots are for $g=0.1$.
}
\label{AGZsum1}
\end{center}
\end{figure}

In Fig.~\ref{AGZsum1}, we plot the result (\ref{eq:AGZsum}) for the $L$-resummed
AGZ distribution for various values of $\kappa$.  In the middle and lower panels,
which are analogous to the middle and lower panels of Fig.~\ref{fig:Pg01}, we show
explicitly that the $L$-resummed AGZ distribution agrees with our full result in the IR,
as anticipated.  This confirms for us that when we inspect the $\kappa$-dependence
of $P^{\rm AGZ}(k_\perp)$ as we do in the top panel, we can trust that at small
$k_\perp$ these results agree quantitatively with our full results.  
The first thing that we see in the top panel is that 
at $\kappa=0.1$ the $L$-resummed result agrees with that for a thin medium,
namely $P_{\rm thin}^{\rm AGZ}(k_\perp)$, since when the medium is thin there
is no need to resum $L$-enhanced diagrams.
Next, by plotting (\ref{eq:AGZsum}) for varying $\kappa$  as in this panel
we discover that for $\kappa<2$, the resummed result 
$P^{\rm AGZ}(k_\perp)$ diverges as $\kappa\rightarrow 0$ while
for 
$\kappa \geq 2$, the resummed result 
is finite in this limit.
Careful inspection of the curves with $\kappa>2$ 
shows that for $\kappa=3$ the probability
$P^{\rm AGZ}(k_\perp)$ is linear in $k_\perp$ as $k_\perp\to 0$, for $2<\kappa<3$
it grows faster than linear in this limit, for $3<\kappa<4$ it grows slower than linear
in this limit, and for all $\kappa\geq 4$ the curve is quadratic in $k_\perp$ at small $k_\perp$.
We find the same nontrivial $\kappa$-dependence
of our full results at small $k_\perp$.  The advantage of reproducing these results in the $L$-resummed AGZ distribution 
(\ref{eq:AGZsum})
is that in this simpler setting we can understand them all analytically, as we now explain.

The first step to understanding the $\kappa$-dependence of (\ref{eq:AGZsum}) at small $k_\perp$ is
to divide the $x$-integration in (\ref{eq:AGZsum}) into an integral from $x=0$ to some 
$x=x_0$ with $x_0$ chosen to be very much greater than 1, and an integral from $x=x_0$ to $x=\infty$.
The integral over $0<x<x_0$ yields a result that is analytic as $k_\perp\to 0$, whose leading
small-$k_\perp$ dependence includes a constant term and a term quadratic in $k_\perp$.
The only place where divergent, or other nonanalytic, behavior can arise as $k_\perp\to 0$
is from the
integral over $x_0<x<\infty$.  In the $k_\perp\to 0$ limit,
this integral can be evaluated analytically
in terms of generalized hypergeometric functions, whose asymptotic behavior 
determines that as $k_\perp\to 0$ the probability distribution $P^{\rm AGZ}(k_\perp)$ 
must
include a term with the form 
%
%
\be
-\frac{2\pi}{m_D^\kappa} \(2e^{-\gamma}\)^\kappa f(\kappa) \, k_\perp^{\kappa-2}
\label{AGZIR}
\ee
with
\be
f(\kappa)\equiv  \frac{\kappa}{2^{\kappa} } \frac{\Gamma \(-\frac{\kappa}{2} \) }{\Gamma \(\frac{\kappa}{2} \) }
\ee
for $\kappa \neq 2, 4, 6,\ldots$ and 
\be
f(\kappa)\equiv  \frac{4}{4^{\kappa/2}} \frac{(-1)^{\kappa/2}}{\Gamma(\frac{\kappa}{2})^2} \(\frac{1}{\kappa} + \psi(\frac{\kappa}{2}) \), 
\ee
for $\kappa = 4, 6, 8,\ldots$,
where $\Gamma(n)$ and $\psi(n)$ are gamma and digamma functions respectively. 
The $k_\perp$-dependence of this 
analytic result explains the $k_\perp\to 0$ behavior of all the curves
in the top panel of Fig.~\ref{AGZsum1}.  For $\kappa> 4$, the term (\ref{AGZIR}) is subleading
relative to the regular $k_\perp^2$-dependent term at small $k_\perp$ and so, for $\kappa\geq 4$,
$P^{\rm AGZ}(k_\perp)$ is quadratic at small $k_\perp$.  For $\kappa<4$, (\ref{AGZIR}) dominates
as $k_\perp\rightarrow 0$.  And, it is linear in $k_\perp$ for $\kappa=3$, grows nonanalytically
and slower (faster) than linear for $3<\kappa<4$ ($2<\kappa<3$) and diverges for $\kappa<2$.
So, we have a complete analytic understanding of the nontrivial $\kappa$-dependence
of $P^{\rm AGZ}(k_\perp)$ as $k_\perp\to 0$, and consequently of the same nontrivial
$\kappa$-dependence of $P(k_\perp)$ that we described in Section~\ref{subsec:5B}.

We saw in Section~\ref{subsec:5B}, in particular in Fig.~\ref{fig:LargeKappaGaussians},
that for large enough $\kappa$ the probability distribution $P(k_\perp)$ is well-approximated
as a Gaussian at values of $k_\perp$ that are not too large.  Given how similar
$P^{\rm AGZ}(k_\perp)$ and $P(k_\perp)$ are at small $k_\perp$, see 
Fig.~\ref{AGZsum1}, we can conclude that $P^{\rm AGZ}(k_\perp)$ is also
well-approximated as a Gaussian at small values of $k_\perp$.
Because we have an (almost) analytic understanding of $P^{\rm AGZ}$ in the 
form (\ref{eq:AGZsum}) we can now obtain an (almost) analytic expression for
the width of the Gaussians in Fig.~\ref{fig:LargeKappaGaussians} 
that describe $P(k_\perp)$ at not too large $k_\perp$.


In order to find the Gaussian approximation 
$P^{\rm AGZ}(k_\perp)\simeq A_{\rm AGZ} \exp(-a_{\rm AGZ} \,k_\perp^2)$ 
to $P^{\rm AGZ}(k_\perp)$ at low $k_\perp$, 
we Taylor expand the logarithm of  $P^{\rm AGZ}(k_\perp)$. Recall
that $P^{\rm AGZ}(k_\perp)$ is given by (\ref{eq:AGZsum}),
which
we obtained upon resumming length-enhanced diagrams as is necessary at large $\kappa$.
We write the
expansion as
\be
\log  P^{\rm AGZ}(k_\perp) = \log A_{\rm AGZ}  - a_{\rm AGZ}\, k_\perp^2 + \ldots
\ee
and find
\be
\begin{split}
A_{\rm AGZ} = & \, \int_0^\infty dx \, b_{\rm AGZ}(x)  \ ,  \\  
a_{\rm AGZ} = & \, \frac{1}{4 \, A_{\rm AGZ} \, m_D^2}  \int_0^\infty dx \, x^2 \, b_{\rm AGZ}(x)  \ , 
\end{split}
\label{eq:AGZGaussian}
\ee
where we have defined
\be
b_{\rm AGZ}(x) \equiv \, \frac{2\pi}{m_D^2}  \, x \exp \left[ - \kappa \left( \gamma + \log\left( x/2\right) + K_0(x) \right) \right]  \ .
\ee
The expression for $a_{\rm AGZ}$ in (\ref{eq:AGZGaussian}) can be rewritten as
\be
a_{\rm AGZ} = \frac{1}{4 m_D^2} \frac{\int_0^\infty dx \, x^2 \, b_{\rm AGZ} (x)}{\int_0^\infty dx \, b_{\rm AGZ} (x)} \ ,
\ee
from which we notice that any $x$-independent overall factor in $b_{\rm AGZ}(x)$ 
does not affect the final result for $a_{\rm AGZ}$, i.e.~for  the width of the Gaussian. 
We can therefore simplify the expression for $a_{\rm AGZ}$ further, obtaining
\be
a_{\rm AGZ} \, m_D^2 =  \frac{1}{4} \frac{\int_0^\infty dx \, x^{3 - \kappa} \exp\left[- \kappa \, K_0(x) \right]}{\int_0^\infty dx \, x^{1 - \kappa} \exp\left[- \kappa \, K_0(x) \right]} \ ,
\ee
which is Eq.~(\ref{eq:amD2}).

\section{Simple formula to compute $\hat{q}$}
\label{app:qhat}

In this Appendix, we derive the simple expression (\ref{qhatnabla2}) for the jet quenching parameter $\hat q$, 
valid for any 
probability distribution of the form~\r{Pdef}. We start from the definition of $\hat{q}$ in Eq.~\r{qhatdef} which, with \r{Pdef}, reads
\beq
\hat q= \frac{1}{L} \int \frac{d^2 k_\perp}{(2\pi)^2} k_\perp^2 \int d^2 x_\perp e^{-i k_\perp x_\perp} e^{\mathcal{W}^{(2)}_{\mathcal{R}} } \ .
\label{eq:qhatapp1}
\eeq
By performing the integration over $d^2 k_\perp$, we obtain the two dimensional Laplace operator $\nabla^2$ in the transverse plane, acting on a delta function: 
\beq
\hat q = - \frac{1}{L} \int d^2 x_\perp \[ \nabla^2 \delta^2(\vec x_\perp) \]  e^{ \mathcal{W}^{(2)}_{\mathcal{R}} (x_\perp)}.
\eeq
We can then integrate the above equation by parts, 
drop the vanishing boundary terms in our notation, and express $\hat q$ as 
\be
\hat q = - \frac{1}{L} \int d^2 x_\perp \delta^2(x_\perp) \nabla^2 e^{\mathcal{W}^{(2)}_{\mathcal{R}}} = - \frac{1}{L} \left. \nabla^2 e^{\mathcal{W}^{(2)}_{\mathcal{R}}} \right |_{x_\perp = 0} \ .
\ee
We can further simplify this expression. By explicit calculation, the Laplace operator acting on the exponential reads
\be
\begin{split}
& \nabla^2 e^{\mathcal{W}^{(2)}_{\mathcal{R}}} = e^{\mathcal{W}^{(2)}_{\mathcal{R}}} \[ \nabla^2 \mathcal{W}^{(2)}_{\mathcal{R}} + \left(\partial_x \mathcal{W}^{(2)}_{\mathcal{R}}\right)^2 + \left(\partial_y \mathcal{W}^{(2)}_{\mathcal{R}}\right)^2 \],
\label{lapexp}
\end{split}
\ee
where we have written the vector in the perpendicular plane as $x_\perp = (x,y)$. The last two terms vanish once they are evaluated at $x_\perp = 0$, since the medium is isotropic in the transverse plane, and we therefore find
\beq
\hat q = -\frac{1}{L} \left. \nabla^2 \mathcal{W}^{(2)}_{\mathcal{R}} \right |_{x_\perp = 0} \ ,
\label{qhat>app}
\eeq
which is the result~\r{qhatnabla2} that we used in Sec.~\ref{sec:results}.

As an example of its use, the result (\ref{qhat>app}) can be applied in
a straightforward fashion in any case where the physics of momentum broadening
can be understood as diffusion in $k_\perp$-space, as for example in
the strong-coupling regime that we have discussed in Section V.D.  
In any such context,
\be
\mathcal{W}^{(2)}_{\mathcal{R}} = - D \, x_\perp^2 \ ,
\ee
for some diffusion constant $D$, and applying~\r{qhat>app} yields
\be
\hat q = \frac{4 D}{L} \ ,
\ee
immediately.  
This is of course what we also 
find by applying the definition of $\hat{q}$ in Eq.~\r{eq:qhatapp1}.

Applying the result (\ref{qhat>app}) to the case of a weakly-coupled plasma,
as we analyze in this paper, must come with a further subtlety since, 
as we have discussed in Section V.C, in this case the jet quenching
parameter $\hat q$ is not well defined!
The result of the integration in  Eq.~\r{eq:qhatapp1} is logarithmically divergent, 
meaning that it must be regulated in some way.
Similarly, if we simply 
apply (\ref{qhat>app}) to our weak-coupling result (\ref{eq:W2Fourier}) for 
$\mathcal{W}^{(2)}_{\mathcal{R}}$ we find a divergent result.
To regulate this divergence, we write the result~\r{qhat>app} in Fourier space, and regulate the momentum integral by imposing a UV cutoff $\Lambda_{\rm UV}$, as discussed in Section V.C. 
Upon doing so with $\mathcal{W}^{(2)}_{\mathcal{R}}$ given by~\r{eq:W2Fourier},  we obtain
\be
\hat q = \frac{1}{L} \int^{\Lambda_{\rm UV}} \frac{d^2 k_\perp}{(2\pi)^2} k_\perp^2 P_{\rm thin}(k_\perp) \ ,
\ee
which is the expression~\r{qhatdef2} that we used to evaluate the jet quenching parameter $\hat q$ 
in Section V.C.

\end{document}